\documentclass[manuscript]{acmart}
\usepackage[utf8]{inputenc}

\usepackage{booktabs} 

\usepackage{amsmath}
\usepackage{balance}
\usepackage{euscript}
\usepackage{graphicx}
\usepackage{subcaption}
\usepackage{hyperref}
\usepackage{mdframed}
\usepackage{microtype} 
\usepackage{multirow}
\usepackage[skins,breakable]{tcolorbox}
\usepackage[normalem]{ulem}
\usepackage{xspace}
\usepackage{enumitem}

\newcommand{\nosemic}{\renewcommand{\@endalgocfline}{\relax}}
\newcommand{\dosemic}{\renewcommand{\@endalgocfline}{\algocf@endline}}
\let\oldnl\nl
\newcommand{\nonl}{\renewcommand{\nl}{\let\nl\oldnl}}

\usepackage{amsfonts}

\newcommand{\overbar}[1]{\mkern 1.5mu\overline{\mkern-1.5mu#1\mkern-1.5mu}\mkern 1.5mu}

\def\figcapup{\vspace{-1mm}}
\def\figcapdown{\vspace{-0mm}}

\newcommand{\bm}[1]{\textrm{\boldmath${#1}$}}

\def\eps{\epsilon}

\def\-{\mbox{-}}

\def\lf{\lfloor}
\def\la{\langle}

\def\rf{\rfloor}
\def\ra{\rangle}

\def\Pr{\mathbf{Pr}}

\def\*{\star}

\newcommand{\dmin}{d_{\min}(u,v)}
\newcommand{\dmax}{d_{\max}(u,v)}

\newcommand{\newblue}[1]{\color{black}{#1}}

\usepackage{xcolor}

\definecolor{Mulberry}{rgb}{0.77,0.29,0.55}
\definecolor{CadmiumOrange}{rgb}{0.93,0.53, 0.18}
\definecolor{ForestGreen}{rgb}{0.13, 0.55, 0.13}
\definecolor{WildStrawberry}{rgb}{0.5, 0.7, 0.2}

\newcommand{\core}{\ensuremath{\mathit{core}}}

\def\C{\mathcal{C}}

\def\F{\mathcal{F}}
\def\L{\mathcal{L}}
\def\tO{\tilde{O}}
\def\aC{\C\big(\L_{\eps,\rho}(G),\mu\big)}
\def\cR{\C\big(\mathcal{L}(G), \mu\big)}
\def\newcR{\C\big(\mathcal{L}_{\eps}(G),\mu\big)}
\def\exCplus{\C\big(\L_{(1+\rho)\eps}(G), \mu \big)}
\def\exCminus{\C\big(\L_{(1 - \rho)\eps}(G), \mu \big)}

\def\tsig{\tilde{\sigma}}
\def\tL{{\L}_{\eps, \rho}(G)}

\newcommand{\DT}{\ensuremath{\mathit{DT}}}
\newcommand{\cnt}{\ensuremath{\mathit{s}}}
\newcommand{\DtHeap}{\ensuremath{\mathit{DtHeap}}}

\newcommand{\FindCcID}{\ensuremath{\mathit{FindCcID}}}
\newcommand{\SimCnt}{\ensuremath{\mathit{SimCnt}}}
\newcommand{\ccid}{\ensuremath{\mathit{ccid}}}

\newcommand{\suv}{\sigma(u,v)}
\newcommand{\suvc}{\sigma_c(u,v)}
\newcommand{\du}{d[u]}
\newcommand{\dv}{d[v]}
\def\strclu{\ensuremath{\mathit{StrClu}}}
\def\strcluresult{\ensuremath{\mathit{StrCluResult}}}
\def\dynstr{\mathit{DynStrClu}}
\def\dynelm{\mathit{DynELM}}
\def\scan{\mathit{SCAN}}

\def\pscan{\mathit{pSCAN}}
\def\hscan{\mathit{hSCAN}}
\def\linkscan{\mathit{linkSCAN}}
\def\linkscanstar{\mathit{linkSCAN^*}}

\def\rhorange{[0, \min\{1, \frac{1}{\eps}-1\})} 
\def\strategy{(\frac{1}{2}\rho\eps, \delta)\text{-strategy}}
\def\half{\frac{1}{2}}

\def\figcapup{\vspace{-3mm}}
\def\figcapdown{\vspace{-4mm}}

\newtheorem{observation}{Observation}
\newtheorem{fact}{Fact}

\AtBeginDocument{%
  \providecommand\BibTeX{{%
    \normalfont B\kern-0.5em{\scshape i\kern-0.25em b}\kern-0.8em\TeX}}}

\setcopyright{acmcopyright}
\copyrightyear{2018}
\acmYear{2018}
\acmDOI{10.1145/1122445.1122456}

\acmConference[Woodstock '18]{Woodstock '18: ACM Symposium on Neural
  Gaze Detection}{June 03--05, 2018}{Woodstock, NY}
\acmBooktitle{Woodstock '18: ACM Symposium on Neural Gaze Detection,
  June 03--05, 2018, Woodstock, NY}
\acmPrice{15.00}
\acmISBN{978-1-4503-XXXX-X/18/06}

\title{Dynamic Structural Clustering on Graphs}

\author{Boyu Ruan}
\affiliation{
  \institution{The University of Queensland}
  \city{Brisbane}
  \country{Australia}
}
\email{b.ruan@uq.edu.au}

\author{Junhao Gan}
\affiliation{
  \institution{The University of Melbourne}
  \city{Melbourne}
  \country{Australia}
}
\email{junhao.gan@unimelb.edu.au}

\author{Hao Wu}
\affiliation{
  \institution{The University of Melbourne}
  \city{Melbourne}
  \country{Australia}
}
\email{whw4@student.unimelb.edu.au}

\author{Anthony Wirth}
\affiliation{
  \institution{The University of Melbourne}
  \city{Melbourne}
  \country{Australia}
}
\email{awirth@unimelb.edu.au}

\begin{abstract}
  \textit{Structural Clustering} ($\strclu$) is one of the most popular graph clustering paradigms.
  In this paper, we consider $\strclu$ under {\newblue two commonly adapted similarities, namely Jaccard similarity and cosine similarity} on a \textit{dynamic} graph, $G = \la V, E\ra$, subject to edge insertions and deletions (updates). The goal is to maintain certain information under updates, so that the $\strclu$ clustering result on~$G$ can be retrieved in $O(|V| + |E|)$ time, upon request.
  The state-of-the-art worst-case cost is~$O(|V|)$ per update;
  we improve this update-time bound \textit{significantly} with the $\rho$-approximate notion.
  Specifically,
  for a specified failure probability, $\delta^*$, and \textit{every} sequence of~$M$ updates (no need to know $M$'s value in advance),
  our algorithm, $\dynelm$,
  achieves~$O(\log^2 |V| + \log |V| \cdot \log \frac{M}{\delta^*})$ amortized cost for each update,  \emph{at all times} in linear space.
  Moreover, $\dynelm$
  provides a provable ``sandwich'' guarantee on the clustering quality at all times after \emph{each update} with probability at least $1 - \delta^*$.
  We further develop $\dynelm$ into our ultimate algorithm, $\dynstr$, which also supports \emph{cluster-group-by} queries.
  Given $Q\subseteq V$, this puts the non-empty intersection of $Q$ and each $\strclu$ cluster into a distinct group.
  $\dynstr$ not only achieves \emph{all} the guarantees of $\dynelm$, but also runs \emph{cluster-group-by} queries in~$O(|Q|\cdot \log |V|)$ time.
  We demonstrate the performance of our algorithms via extensive experiments, on 15 real datasets.
  Experimental results confirm that our algorithms are up to \emph{three orders of magnitude} more efficient than state-of-the-art competitors, and still provide quality structural clustering results. {\newblue Furthermore, we study the difference between the two similarities w.r.t. the quality of approximate clustering results.}
\end{abstract}

\keywords{Structural Clustering; Dynamic Graphs; Algorithms}

\begin{document}

\maketitle

\section{Introduction} \label{sec::intro}

\begin{figure*}
	\resizebox{\linewidth}{!}{%
		\begin{tabular}{cccc}
			\includegraphics[height = 32mm]{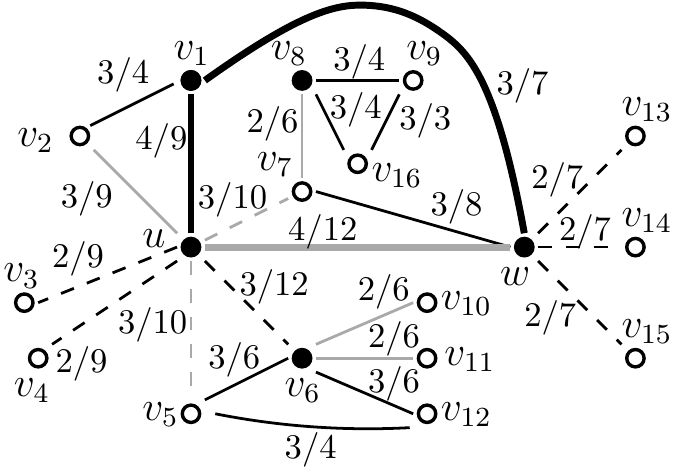}               & \hspace{-4mm}\vline\hspace{1mm}

			\includegraphics[width = 0.15\linewidth]{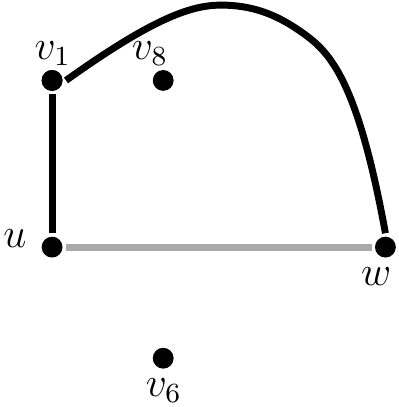} & \hspace{-4mm}\vline\hspace{1mm}

			\includegraphics[height = 28mm]{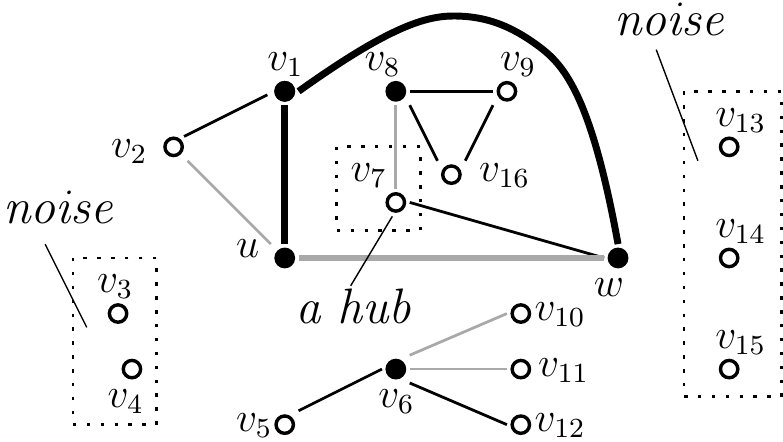}              & \hspace{-10mm}\vline\hspace{3mm}

			\includegraphics[height = 32mm]{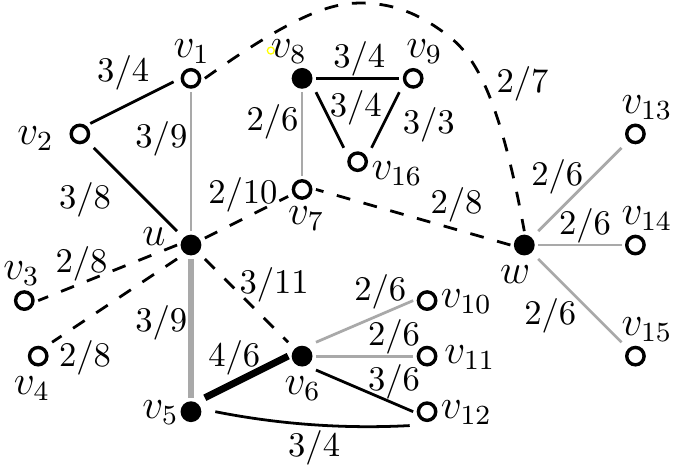}
			\\
			\hspace{-6mm} (a) {\em A graph $G = \la V, E \ra$}                             & (b) {\em The sim-core graph $G_\core$} & \hspace{2mm} (c) {\em The $\strcluresult$ $\cR$} & (d) The result after deleting the edge $(u,w)$ \\[1mm]
		\end{tabular}
	}
	\figcapup
	\caption{
		The $\strcluresult$ with~$\eps = 1/3$ and~$\mu =3$. In (a) and (d), the Jaccard Similarities between the endpoints are shown beside the edges.
		Similar edges are shown as solid lines and dissimilar edges appear as dashed lines.
		The core and non-core vertices are in black and white, respectively.
		All the {\em sim-core} edges are further highlighted in bold.
		Moreover, the edges whose labels are {\em doest-not-matter} under the $\rho$-approximate notion with $\rho= 1/6$ are shown in gray.
	}
	\label{fig:strclu}
\end{figure*}

Clustering on graphs is a fundamental and highly applicable data-mining task~\cite{s-csr-2007}.
The graph vertices are assigned to clusters so that similar vertices are put into the same cluster, while dissimilar vertices are separated from each other, in different clusters.
While most clustering approaches~\cite{bbc-ml-2004, gvw-isaac-2018, pdfv-nature-2005, srs-ijcai-2015, mcre-sdm-2009, gfbs-kis-2014, dhzgs-icdm-2001,wxsw-icde-2014, ggvwz-mfcs-2019} aim to {partition} the vertices into {\em disjoint} sets,
{\em Structural Clustering} (aka $\strclu$, the focus of this paper)~\cite{xyfs-kdd-2007, yxs-icss-2008, sfo-pvldb-2015, clqzy-tkde-2017, wqzcl-vldbj-2019} allows clusters to overlap with each other, so that a vertex might be assigned to multiple clusters.
In particular, in $\strclu$,
there are different roles each vertex might play in the clustering.
Some vertices are \emph{core} members of the clusters, while others may be \emph{noise} (i.e., belonging to no cluster), and still others might be \emph{hubs} (i.e., being members of, and hence bridging, multiple clusters).

\vspace{1mm}
\noindent
\underline{\em Structural Clustering.}
The success of structural clustering arises directly from a family of careful definitions, governing the roles of the edges and vertices.
Given an undirected graph $G = \la V, E\ra$, the neighbourhood of a vertex~$u$ is defined as the set of vertices adjacent to~$u$ in~$G$, plus~$u$ itself. In $\strclu$, each edge in~$E$
has a {\em label}: either {\em similar} or {\em dissimilar}.
An edge~$(u,v)$ is labelled as similar if and only if the {\em structural similarity} (e.g., the {\em Jaccard similarity} or {\newblue\em cosine similarity}) between the {\em neighbourhoods} of~$u$ and~$v$ is at least a certain specified {\em similarity threshold},~$\eps$.
If at least a specified integer~$\mu$ of similar edges are incident on vertex~$u$, then~$u$ is a {\em core} vertex:
consequently, all vertices that share a similar edge with~$u$ are put in the same cluster as~$u$.
Of course, another core vertex~$v$ might be ``similar-adjacent'' to~$u$, and thus in its cluster: repeating this principle, all vertices similar-adjacent to~$v$ are in the same cluster as~$v$ and hence as~$u$.
This chain effect continues until no more vertices are added to~$u$'s cluster, resulting in a $\strclu$ cluster.
We repeat the above process until all the core vertices are in some cluster and return the $\strclu$ result.
As some non-core vertex $w$ may be ``similar-adjacent'' to multiple core vertices that are in different clusters,
$w$ is thus included in multiple clusters, as a {\em hub}  bridging these clusters.
On the other hand, a non-core vertex also possibly belongs to no cluster, and thus becomes {\em noise} (i.e., an outlier).
These different roles information of the vertices and edges have greatly enriched the structural information of a graph.
The power of $\strclu$ arises from these specific vertex roles.

\vspace{1mm}
\noindent
\underline{\em An Example.}
Figure~\ref{fig:strclu}(a) shows an example of Structural Clustering {\newblue under Jaccard similarity}, where all the similar (respectively., dissimilar)  edges appear as solid (respectively., dashed) edges;
and all the core (respectively., non-core) vertices are coloured black (respectively., white).
Since~$u$ is a core, each vertex in~$u$'s neighbourhood sharing a similar edge with~$u$, i.e., $\{v_1, v_2, w\}$, is added to the same cluster as~$u$.
Due to the core vertex~$w$, the non-core vertex~$v_7$ is further added to the cluster.
Thus, $C_1 = \{u, w, v_1, v_2, v_7\}$ is a $\strclu$ cluster.
Likewise, $C_2 = \{v_8, v_9, v_{16}, v_{7}\}$ and $C_3 = \{v_6, v_5, v_{10}, v_{11}, v_{12}\}$ are the other two $\strclu$ clusters, whereas $v_3, v_4, v_{13}, v_{14}, v_{15}$ are {\em noise}.
Finally, observe that~$v_7$ is a {\em hub} vertex, belonging to both clusters~$C_1$ and~$C_2$.

\vspace{1mm}
\noindent
\underline{\em Applications.}
Before continuing with our technical description of the problems in this paper, we first consider the significance of~$\strclu$.
Since its introduction~\cite{xyfs-kdd-2007} in 2007, $\strclu$ has not only attracted significant follow-up work\footnote{Over 850 citations in Google Scholar, as at March 2021.}, but has also served a wide range of real-world applications.
For example, $\strclu$ is an essential component of the {\em atBioNet}, developed by the U.S. Food and Drug Administration's (FDA) National Center for Toxicological Research (NCTR)~\cite{DCL-BMC-2012, MTX-BMC-2008}. {\newblue atBioNet is a web-based tool for genomic and proteomic data, that can perform network analysis follower by biological interpretation for a list of seed proteins or genes (i.e., proteins or genes provided by user).}
Here, $\strclu$ supports identifying functional modules in protein-protein interaction networks and enrichment analysis.
	{\newblue Another example is in community detection~\cite{PKV-DMKD-2012}, where the users are modelled as vertices and the following relationships are modelled as edges. Each cluster in the clustering results of $\strclu$ can be regarded as a community in the social network. Given a collection of tagged photos, $\strclu$ can be utilized to identify landmarks and events~\cite{PZKV-IEEE-2010} by applying $\strclu$ on a {\em hybrid similarity image graph}, constructed by taking both visual similarity and tag similarity into consideration. One interesting application of $\strclu$ is detecting frauds on blockchain data~\cite{c-cmbda-2019}. The noise information of $\strclu$ result is deployed on a graph constructed using the features extracted from blockchain data. The outliers in the $\strclu$ result are regarded as frauds that need to be paid attention to.}


\vspace{1mm}
\noindent
\underline{\em The Dynamic Scenario.}
While the importance and usefulness of $\strclu$ is witnessed from its various applications,
new challenges arise from the {\em dynamic} nature (subject to updates) of contemporary graph data.
The significance and application of $\strclu$ is only increased by considering dynamic graphs, where edges might be added or deleted.
The consequent imperative research is to design highly efficient algorithms that can handle updates to the graph~$G$ so that queries
on the $\strclu$ result are answered efficiently, i.e.,
without re-computing from scratch.
By maintaining the edge labels under updates,
the $\strclu$ result can be obtained in~$O(n + m)$ time,
where~$n$ and $m$ are the {\em current} numbers of vertices and edges in~$G$.
We refer to this linear-time problem as {\em dynamic $\strclu$}.

\vspace{1mm}
\noindent
\underline{\em The Challenges.}
For dynamic $\strclu$, there are two state-of-the-art
algorithms,
$\pscan$~\cite{clqzy-tkde-2017} and $\hscan$~\cite{wqzcl-vldbj-2019}, which can process each update within~$O(n)$ and~$O(n\log n)$ time, respectively.
While these bounds are unfortunately too high for an update (on large graphs with millions of vertices),
this~$O(n)$ bound seems unlikely to be improved.
To see this, when an edge~$(u,w)$ is inserted, labelling~$(u,w)$ may require computing the structural similarity between the neighbourhoods of~$u$ and~$w$,
and hence, may require~$\Omega(n)$ time in the worst case.
Worse still, as shown in Figure~\ref{fig:strclu}(a), after removing the edge~$(u,w)$, the label of every edge incident on~$w$ flips (as shown in Figure~\ref{fig:strclu}(d)).
When this edge is re-inserted in Figure~\ref{fig:strclu}(d), those edge labels revert.
Consequently, the maintenance on these labels already takes up to~$O(n)$ time when the degree of~$w$ is large.

\vspace{1mm}
\noindent
\underline{\em Our Solutions.}
We can, however, reduce the running time of updates from the bound of~$O(n)$ to roughly~$O(\log^2 n)$ {\newblue under both Jaccard similarity and cosine similarity}.
There are two (reasonable) trade-offs: (i) an approximate notion of the edge labels, and (ii)
amortized rather than worst-case analysis of the update time.
Specifically, we adapt the $\rho$-approximate notion to edge labels in $\strclu$: this notion circumvents computational hardness in  other key clustering problems~\cite{gt-tods-2017, gt-sigmod-2017, gt-sigmod-2018}.
Experiments confirm that our approach loses only a little in clustering quality but provides a dramatic improvement (i.e., $1000$-fold) in update efficiency.

\vspace{1mm}
\noindent
\underline{\em Cluster-group-by Queries.}
In addition to simply returning entire clustering results, we further enhance our solution to support the so-called {\em cluster-group-by} queries~\cite{gt-sigmod-2017}, which are examples of traditional group-by queries in database systems.
Given a subset $Q\subseteq V$, the cluster-group-by query of~$Q$ asks to group the vertices in~$Q$ by the {\em identifiers} of the clusters (if any) containing them.
Consider Figure~\ref{fig:strclu}(a), for $Q = \{u, v_7, v_9,  v_{14}\}$,
the cluster-group-by query returns two groups: $\{u, v_7\}$ and $\{v_7, v_9\}$.
This is because both $u$ and $v_7$ belong to $C_1$, meanwhile, both $v_7$ and $v_9$ belong to $C_2$, and $v_{14}$ is noise.
As a special case, when~$Q = V$, the cluster-group-by query is equivalent to retrieving the whole $\strclu$ result, where each group is exactly a cluster.
Thus, the cluster-group-by query is a more general form of  clustering query.
Furthermore, since~$|Q|$ is typically much smaller than~$|V|$,
an efficient algorithm should be able to answer the query much faster, with an output cost depending on~$|Q|$ rather than~$|V|$.
We note that cluster-group-by queries are
not only applicable in \emph{all} general clustering applications,
but also more favourable in  scenarios where users are focused on  the part of clustering results \emph{corresponding to a certain set of vertices}.

\vspace{1mm}
\noindent
\underline{\em Our Contributions.}
This article delivers the following contributions:
\begin{itemize}[leftmargin = *]
	\item {\newblue We first adapt Jaccard similarity as our definition of structural similarity.} Under the $\rho$-approximate notion,
	      we propose the {\em dynamic edge labelling maintenance} ($\dynelm$) algorithm for maintaining edge labels under updates.
	      Specifically,
	      for a specified failure probability,~$\delta^*$, for {\em every} sequence of~$M$ updates, where~$M$'s value need not be known in advance,
	      $\dynelm$ guarantees:
	      \begin{itemize}
		      \item the amortized update cost is~$O(\log^2 n + \log n \cdot \log \frac{M}{\delta^*})$, significantly better than the state-of-the-art~$O(n)$ bound;
		      \item with probability at least $ 1 - \delta^*$, the edge labels are correct (under the $\rho$-approximate notion) {\em all the time}, i.e., the clustering result is correct after {\em each} update;
		      \item the space consumption is~$O(n + m)$, linear in the graph size;
		      \item on request, the $\strclu$ result  can be retrieved in~$O(n + m)$ time.
	      \end{itemize}

	\item
	      To support fast cluster-group-by queries, we introduce our ultimate algorithm: the {\em Dynamic $\strclu$} algorithm ($\dynstr$).
	      Although $\dynstr$ maintains some extra data structures, it not only achieves all the guarantees provided by $\dynelm$, but also answers every cluster-group-query~$Q$ in~$O(|Q|\cdot \log n)$ time, substantially less than~$O(n + m)$ when $|Q|$ is far smaller than $n$.

	\item
	      We conduct extensive experiments on~{15} real datasets, where the largest dataset, \emph{Twitter}, contains up to~$1.2$ billion edges.
	      The experimental results confirm that our $\dynelm$ and $\dynstr$ algorithms are up to {\em three-orders-of-magnitude} (i.e.,
	      1000$\times$) faster on updates than the state-of-the-art algorithms,
	      while still returning a high-quality $\strcluresult$ {\newblue under Jaccard similarity}.
	\item
	      {\newblue We extend our $\dynelm$ and $\dynstr$ algorithms to support cosine similarity. With non-trivial analysis, we prove that the amortized time cost and space consumption remains the same as these algorithms under Jaccard similarity.}
	\item
	      {\newblue We visualise the clustering results under both Jaccard similarity and cosine similarity using {\em Gephi}~\cite{ICWSM09154}, and confirm that the clustering results of Structural Clustering are meaningful and human understandable.}
	\item
	      {\newblue Finally, we conduct extra experiments on 5 representative datasets. The experimental results confirm that our $\dynelm$ and $\dynstr$ algorithms still have outstanding performance and return a high-quality $\strcluresult$ under cosine similarity.}
\end{itemize}

\noindent
\textbf{Paper Organization.}
Section~\ref{sec:preliminaries} introduces the preliminaries. Section~\ref{sec:prob-def} defines the problems.
In Section~\ref{sec:estimator}, we design a similarity estimator. In Sections~\ref{sec:maintenance} and~\ref{sec:dynelm}, we propose the $\dynelm$ algorithm and prove its theoretical guarantees. We discuss $\dynstr$ in Section~\ref{sec:ultimate-algo}. {\newblue Section~\ref{sec:ext} presents the extension of our $\dynelm$ and $\dynstr$ algorithms under cosine similarity and theoretical analysis.} Section~\ref{sec:exp} shows experimental results {\newblue under both Jaccard similarity and cosine similarity}. Finally, Section~\ref{sec:conclusion} concludes the paper.

\section{Preliminaries}\label{sec:preliminaries}
\subsection{Structural Clustering Setup} \label{sec:prelim-scan}

\noindent
\underline{\em Basic Definitions.}
Consider an undirected graph~$G = \la V, E \ra$ with~$n = |V|$ vertices and~$m = |E|$ edges.
Two vertices $u,v\in V$ are {\em neighbors} of each other if they share an edge, $(u, v)\in E$.
For a vertex~$v\in V$,  the {\em neighborhood} of~$v$, denoted by~$N[v]$, is defined as the set of all neighbors of~$v$, as well as $v$ itself, i.e., $N[v]=\{u\in V \mid (u, v)\in E\} \cup \{v\}$.
The {\em degree} of~$v$, denoted by $d[v]$, is the number of neighbors of~$v$ and hence, $d[v] = |N[v]|-1$.
Each edge $(u,v)\in E$ can be labelled as either {\em similar} or {\em dissimilar}, and $(u,v)$ is called a {\em similar} or {\em dissimilar} edge, respectively.
We define an {\em edge labelling} of~$G$, denoted by~$\mathcal{L}(G)$, as a function: $V\times V \rightarrow \{\text{similar}, \text{dissimilar}\}$, specifying a label for each edge.

Given a {\em constant} integer {\em core threshold},~$\mu \geq 1$, and an edge labelling,~$\mathcal{L}(G)$, we introduce the following definitions.

\vspace{1mm}
\noindent\underline{\em Similar Neighbors.} If an edge~$(u,v)\in E$ is similar, then each of~$u$ and~$v$ is a {\em similar neighbor} to the other.

\vspace{1mm}
\noindent \underline{\em Core Vertex.} A vertex~$u \in V$ is a {\em core vertex} if~$u$ has at least~$\mu$ similar neighbors.
Otherwise,~$u$ is a {\em non-core vertex}.

\vspace{1mm}
\noindent \underline{\em Sim-Core Edge.} An edge~$(u,v) \in E$ is called a {\em sim-core edge} if~$(u,v)$ is similar and both~$u$ and~$v$ are core vertices,
e.g., $(u,w)$ is a sim-core edge, while $(u,v_2)$ and $(u,v_6)$ are not because $v_2$ is a non-core and $(u, v_6)$ is dissimilar.

\vspace{1mm}
\noindent \underline{\em Sim-Core Graph.} The {\em sim-core graph} $G_{\core}$ consists of all the core vertices and all the sim-core edges, e.g.,
Figure~\ref{fig:strclu}(b).

\vspace{1mm}
\noindent \underline{\em $\strclu$ Cluster.} There is a one-to-one mapping between the connected components (CCs) of $G_{\core}$ and the $\strclu$ clusters.
For each CC of $G_{\core}$, its corresponding~$\strclu$ cluster is the set of all the vertices in $G$ that are similar to some (core) vertex in this CC.
The $G_\core$ in Figure~\ref{fig:strclu}(b) has three connected components: $\{u, w, v_1\}$, $\{v_8\}$ and $\{v_6\}$.
The three corresponding $\strclu$ clusters are: $C_1 = \{u, w, v_1, v_2, v_7\}$, $C_2 = \{v_8, v_9, v_{16}, v_7\}$ and $C_3 = \{v_6, v_5, v_{10}, v_{11}, v_{12}\}$, shown in Figure~\ref{fig:strclu}(c).
In particular, $v_7$ is a {\em hub} belonging to both the clusters $C_1$ and $C_2$, while vertices $v_3$, $v_4$, $v_{13}$, $v_{14}$ and $v_{15}$ are {\em noise}, belonging to no cluster.

\vspace{1mm}
\noindent \underline{\em $\strcluresult$.}  The {\em $\strcluresult$}
on~$G$, for the parameter ~$\mu$, and the labelling~$\mathcal{L}(G)$, is the set of all $\strclu$ clusters, denoted by~$\cR$.

\begin{fact}\label{fact:uniqueness}
	Given~$\mu$ and $\L(G)$, the $\strcluresult$~$\cR$ is {\em unique} and can be computed in~$O(n + m)$ time.
\end{fact}

\vspace{1mm}
\noindent
\underline{\em The $\strclu$ Problem.}
Denote the similarity between a pair of vertices $u, v \in V$ by~$\sigma(u,v)$.
If~$u$ and $v$ are not adjacent, then $\sigma(u, v)=0$. Otherwise, $\sigma(u,v)$ is measured by
$$
	\sigma(u, v) = \frac{|N[v]\cap N[u]|}{|N[v]\cup N[u]|},
$$
{\newblue under {\em Jaccard similarity} between their neighbourhoods; or
$$
	\sigma(u,v) = \frac{|N[v]\cap N[u]|}{\sqrt{d[u]\cdot d[v]}},
$$
under {\em cosine similarity} between their neighbourhoods. In the following contents, we first focus on Jaccard similarity and use Jaccard similarity as our definition of structural similarity due to its simpler form. We defer the discussion on cosine similarity to Section~\ref{sec:ext}.}

Given a {\em constant} {\em similarity threshold} $\eps \in (0, 1]$, $\mu \geq 1$ and~$G$, the $\strclu$ problem computes the $\strcluresult$,~$\newcR$, with respect to a \emph{valid} edge labelling~$\mathcal{L}_{\eps}(G)$, viz.
\begin{definition}[Valid Edge Labelling]\label{def:similar-edge}
	Edge labelling $\mathcal{L}_{\eps}(G)$ is {\em valid} if and only if for every edge~$(u,v)$,
	$\mathcal{L}_{\eps}(G)(u,v)=\text{similar} \iff \sigma(u,v) \geq \eps$.
\end{definition}


\subsection{Related Work}
The $\strclu$ problem was first proposed in~\cite{xyfs-kdd-2007} and the {\em SCAN} algorithm was proposed for solving the problem.
$\scan$ computes the similarities between the endpoints of each edge in $G$, and thus, its worst-case running time is bounded by
$O(m^{1.5})$. While there are a considerable number of follow-up works on $\strclu$, such as $\scan\text{++}$~\cite{sfo-pvldb-2015},
$\pscan$~\cite{clqzy-tkde-2017} and $\hscan$~\cite{wqzcl-vldbj-2019}, they are all heuristic and none of them is able to break the
$O(m^{1.5})$-time barrier.

Lim et al. proposed a related algorithm, called $\linkscan$, and its approximate version $\linkscanstar$~\cite{lrkjl-icde-2014}.
The (structural) clustering results computed by these algorithms are on a transformation of the original graph (i.e., not on the original graph).
Thus, there is no guaranteed connection to their clustering results and those of $\scan$. Moreover, the running time of these two algorithms is $O(n\cdot d^2)$, where~$d$ is the average degree of all the vertices, and hence, in the worst case, the bound still degenerates to~$O(n^3)$.

The problem becomes more challenging when the graph is subject to edge insertions and deletions. $\pscan$ and $\hscan$ are two state-of-the-art algorithms that can support updates and can return the $\strcluresult$ in $O(n + m)$ time upon request, where $m$ is the current number of edges in the graph.
For an update (either insertion or deletion) with an edge $(u,w)$, both $\pscan$ and $\hscan$ need to retrieve $N[u]$ and $N[w]$ in the worst case.
The update cost of $\pscan$ is bounded by $O(n)$, while $\hscan$ requires $O(n\log n)$ time, as it aims for a more general purpose, of supporting $\strcluresult$ reporting in $O(n + m)$ time with $\eps$ and $\mu$ given on the fly.

\subsection{The $\rho$-Approximate Notion}
\label{sec:rho-approximate}
The notion of $\rho$-approximation was initially proposed to circumvent the computational hardness in some clustering problems~\cite{gt-tods-2017, gt-sigmod-2017, gt-sigmod-2018}.
We adopt this notion to relax slightly the validity requirement (Definition~\ref{def:similar-edge}) for edge labellings.
We introduce an additional {\em constant} parameter, $\rho \in \rhorange$, where the value range of $\rho$ is intentionally set to ensure: (i) $(1 - \rho)\eps > 0$ and (ii) $(1 + \rho)\eps \leq 1$.

\begin{definition}[Valid $\rho$-Apprxomiate Edge Labelling]\label{def:approximate-similar-edge}
	Given $\rho \in \rhorange$, an edge labelling $\L_{\eps, \rho}(G)$ is a valid $\rho$-approximate edge labelling, if for every  edge $(u,v)\in E$,
	\begin{itemize}[leftmargin = *]
		\item if $\sigma(u,v) \geq (1 + \rho) \eps$, $(u,v)$ {\em must be} labelled as similar;
		\item if $\sigma(u,v) < (1 - \rho) \eps$, $(u, v)$ {\em must be} labelled as dissimilar;
		\item for every other~$\sigma(u,v)$ value, between~$(1 - \rho) \eps$ and~$(1 + \rho)  \eps$,
		      the label of~$(u, v)$ {\em does not matter}, namely, it is allowed to be either similar or dissimilar.
		      As shown in Figure~\ref{fig:strclu}, the edges in color gray fall in this does-not-matter case.
	\end{itemize}
\end{definition}

Two observations are worth noticing here.
First, when~$\rho > 0$, due to the ``does-not-matter'' case, there may exist multiple valid $\rho$-approximate edge labellings, i.e., multiple valid $\L_{\eps,\rho}(G)$'s.
Nonetheless,
given~$\L_{\eps, \rho}(G)$,
by Fact~\ref{fact:uniqueness}, the $\strcluresult$ $\aC$ is still uniquely defined.
Second, only the edges with $\sigma(u,v) \in [ (1 - \rho) \eps, (1 + \rho) \eps]$ fall into the does-not-matter case.
When~$\rho$ is small (e.g., $\rho = 0.01$),~$\L_{\eps, \rho}(G)$ and~$\L_{\eps}(G)$ are usually close,
and their $\strcluresult$s would not differ much.
This intuition is confirmed on~15 real datasets in our experiments (see Section~\ref{sec:exp}) and formalized by the theorem below.

\begin{theorem}[Sandwich Guarantee]\label{thm:sandwich}
	Given $\eps$, $\mu$ and $\rho$,
	let $\L_{(1+\rho)\eps}(G)$ and $\L_{(1-\rho)\eps}(G)$ be the valid edge labellings
	with respect to similarity thresholds $(1 + \rho)\eps$ and $(1 - \rho)\eps$, respectively.
	For an {\em arbitrary} valid $\rho$-approximate edge labelling, $\mathcal{L}_{\eps, \rho}(G)$, we have:
	\begin{itemize}[leftmargin = *]
		\item for every cluster~$C_+ \in \exCplus$, there exists a cluster~$\tilde{C} \in \aC$ such that $C_+ \subseteq \tilde{C}$;
		\item for every cluster~$\tilde{C} \in \aC$, there exists a cluster~$C_- \in \exCminus$ such that~$\tilde{C} \subseteq C_-$.
	\end{itemize}
\end{theorem}

\begin{proof}
	Let $E_-$, $E_+$ and $E_\rho$ be the sets of {\em similar} edges labelled in $\L_{(1-\rho)\eps}(G)$, $\L_{(1+\rho)\eps}(G)$ and $\tL$, respectively.
		{\newblue Consider an edge $(u,v)\in E_+$, it must satisfy $\sigma(u,v)\geq (1+\rho)\eps$. By the definition of $\rho$-approximate edge labelling, $(u,v)$ must be labelled as similar under $\tL$, i.e., $(u,v)\in E_\rho$. Thus it can be verified that $E_+\subseteq E_\rho$. On the other hand, for a edge $(u,v)\in E_\rho$, by the definition, it must satisfy $\sigma(u,v)\geq (1-\rho)\eps$. Therefore, it must be labelled as similar under $\L_{(1-\rho)\eps}(G)$, i.e., $(u,v)\in E_-$. In summarisation, $E_-$, $E_+$ and $E_\rho$ satisfy the relationship $E_+\subseteq E_\rho \subseteq E_-$.

			Consider a cluster $C_+ \in \exCplus$, the similar edges inside $C_+$ all belong to $E_+$ and thus all belong to $E_\rho$. Therefore, for any vertex $v\in C_+$, either core or non-core vertex, will belong to the same cluster $\tilde{C} \in \aC$. As a result, $C_+\subseteq \tilde{C}$. The first bullet is proven. The second bullet can be proven symmetrically.}
\end{proof}

\subsection{Distributed Tracking} \label{sec::prelim-dt}
Our final preliminary is the {\em Distributed Tracking} (DT) problem \cite{kcr-sigmod-2006, gsk-ta-2011, hyz-algorithmica-2019} and its solutions.
The DT problem is defined in a {\em distributed} environment: there are~$h$ participants,~$s_1, s_2, \ldots, s_h$,  and a  {\em coordinator},~$q$.
Each participant~$s_i$
has a two-way communication channel with the coordinator~$q$,
while direct communications between participants are prohibited.
Furthermore, each~$s_i$ has an integer {\em counter}~$c_i$, which is initially~$0$.
At each time stamp, {\em at most one} (possibly \emph{none}) of these~$h$ counters is incremented, by~$1$.
Given an integer threshold $\tau > 0$, the job of the coordinator~$q$ is to report (immediately) the ``\emph{maturity}''
of the condition
$\sum_{i = 1}^h c_i = \tau$.
The goal is to minimize the {\em communication cost}, measured by the total number of \emph{messages} (each of $O(1)$ words) sent and received by $q$.

A straightforward solution is for each participant to inform~$q$ whenever its counter is incremented.
The total communication cost of this approach is clearly~$\tau$ messages,
which can be expensive if~$\tau$ is large.
The DT problem actually admits an algorithm~\cite{hyz-algorithmica-2019} with~$O(h \log (\tau / h))$ messages.
The algorithm performs in {\em rounds} and in each round, it works as follows:
\begin{itemize}[leftmargin = *]
	\item If $\tau \leq 4h$, use the straightforward algorithm with~$O(h)$ messages.
	\item If $\tau > 4h$, $q$ sends to each~$s_i$ a {\em slack}
	      $\lambda = \lf {\tau}/(2h) \rf$.
	      \begin{itemize}
		      \item Define $\hat{c}_i$ as the {\em checkpoint} value, indicating when $s_i$ next needs to check in with $q$.
		            Initially, $\hat{c}_i = \lambda$.
		      \item As soon as $c_i = \hat{c}_i$, $s_i$ sends a signal to~$q$, and then~$\hat{c}_i$ is increased by~$\lambda$, indicating the next check-in time~of $s_i$.
		      \item When~$q$ receives the $h^{\text{th}}$ signal in this round,~$q$ obtains the precise value of~$c_i$ from each~$s_i$ and computes
		            $\tau' = \tau - \sum_{i=1}^{h} c_i$.
		            If $\tau' = 0$,~$q$ reports maturity. Otherwise,~$q$ starts a new round with $\tau \leftarrow \tau'$, from scratch
		            with the new threshold~$\tau$.
	      \end{itemize}
\end{itemize}

\noindent
\textbf{Analysis.} In each round,~$q$ sends~$h$ slacks, receives~$h$ signals and collects~$h$ counters from the participants.
The communication cost in each round is bounded by~$O(h)$ messages. Furthermore, it can be verified that at the end of each round,~$\tau' \leq 3/4 \cdot \tau$.
Referring to the original~$\tau$, there are at most~$O(\log (\tau / h))$ rounds. The overall communication cost is bounded by $O(h \log (\tau/ h))$ messages.

\section{Problem Formulation \& Rationale}\label{sec:prob-def}
In this paper, we consider the $\strclu$ problem in a dynamic scenario, where the graph $G= \la V, E \ra$ is subject to updates.
Each update is either an insertion of a {\em new} edge or a deletion of an {\em existing} edge. 

\begin{definition}[Basic Dynamic $\strclu$ Problem]
	For a specified valid edge labelling definition (e.g., Definitions~\ref{def:similar-edge} and ~\ref{def:approximate-similar-edge}),
	given $\eps\in(0, 1]$ and $\mu \geq 1$, the goal of the {\em Basic Dynamic $\strclu$ Problem} is to maintain a valid edge labelling~$\L(G)$.
\end{definition}

By Fact~\ref{fact:uniqueness}, with a valid edge labelling~$\L(G)$ {being maintained} at hand, the $\strcluresult$~$\cR$ is uniquely defined and can be returned in~$O(n + m)$ time, where $n$ and~$m$ are the \emph{current} numbers of vertices and edges in~$G$, respectively. 

	As mentioned earlier, we significantly reduce the state-of-the-art $O(n)$ update cost to~$O(\log^2 n + \log n \cdot \log \frac{M}{\delta^*})$ amortized for every sequence of $M$ updates, and with probability at least $1 - \delta^*$, the clustering result is correct at all times under the $\rho$-approximate notion.

	\vspace{1mm}
	\noindent
	\underline{\em The Rationale in Our Solution.}
	Observe that the does-not-matter case in the $\rho$-approximate notion essentially provides a {\em leeway} 
	for maintaining edge labels.
	Each edge can:  
	(i) be labelled with approximate similarity,
	and (ii) ``afford'' a certain number of updates without needing to flip the label. 
	Our solution is designed based on these two crucial properties.
	First, we propose a sampling-based method (in Section~\ref{sec:estimator}) to estimate the Jaccard similarity, by which the cost of labelling an edge is reduced to poly-logarithmic.
	Second, we show that each edge can afford $k$ {\em affected} updates (formally defined in Section~\ref{sec:maintenance}) without needing to check its label.
	As such, we deploy a DT instance to track the moment of the $(k+1)^\text{th}$ affected update for each edge, at which moment, the edge needs to be relabelled.
	However, 
	these two ideas alone are not sufficient to beat the~$O(n)$ update bound.
	To complete the design of our solution, we further need to organize the DT instances carefully with heaps.
	Finally, by performing a non-trivial amortized analysis (in Section~\ref{sec:dynelm}), 
	our solution to the basic problem is thus obtained.

Embarking from this solution, we further study a more challenging problem to support {\em cluster-group-by queries}:

\begin{definition}[Cluster-Group-By Query]
	Consider an edge labelling $\L(G)$;
	for an arbitrary subset $Q \subseteq V$, on a {\em cluster-group-by} query of~$Q$ on~$G$, we return $C_i \cap Q$ as a distinct group (with a unique identifier),
	for every cluster~$C_i \in \cR$ satisfying $C_i \cap Q \neq \varnothing$.
	
\end{definition}
\noindent
Finally, the ultimate problem is defined as follows:

\begin{definition}[Ultimate Dynamic $\strclu$ Problem]
	In addition to maintaining a valid edge labelling~$\L(G)$, 
	the goal of the Ultimate Dynamic $\strclu$ Problem is to further maintain certain data structures, by which every cluster-group-by query $Q\subseteq V$, with respect to~$\L(G)$, can be answered in~$O(|Q|\cdot \log n)$ time. 
\end{definition}

\section{Estimating Jaccard Similarity}\label{sec:estimator}
In this section, we propose a {\em sampling-based} method for estimating the similarity coefficient.

\vspace{1mm}
\noindent
\underline{\em The Sampling-Estimator.}
Consider an arbitrary edge $(u,v)\in E$; let~$a = |N[u] \cap N[v]|$ and~$b = |N[u] \cup N[v]|$.
Our sampling technique relies on a {\em biased} estimator.
First, we define a random variable~$X$, generated by the following steps.

\begin{itemize}[leftmargin = *]
	\item Flip a coin~$z$, where~$z=1$ with probability~$\frac{|N[u]|}{a + b}$ and~$z = 0$ with probability~$\frac{|N[v]|}{a + b}$.
	\item If $z = 1$, then uniformly at-random pick a vertex from~$N[u]$; otherwise, uniformly at-random pick a vertex from~$N[v]$. Denote the vertex picked by~$w$.
	\item If $w \in N[u] \cap N[v]$, then~$X = 1$; otherwise,~$X = 0$.
\end{itemize}

According to the above generation procedure,
\begin{align*}
	\Pr[X = 1] & = \Pr[X = 1 \wedge z=1] + \Pr[X = 1 \wedge z = 0]                                           \\
	           & = \frac{|N[u]|}{a + b} \cdot \frac{a}{|N[u]|} + \frac{|N[v]|}{a + b} \cdot \frac{a}{|N[v]|}
	=\frac{2a}{a + b}\,.\nonumber
\end{align*}
\\
Let $X_1, X_2, \ldots, X_L$ be $L$ independent instances of~$X$, and define $\bar{X} = \frac{1}{L} \sum_{i=1}^L X_i$.
We have
\begin{equation}\label{equation:estimator}
	E[\bar{X}] = \frac{2a}{a + b} = \frac{2a /b}{1 + a/b} = \frac{2\sigma(u,v)}{1 + \sigma(u,v)} \Leftrightarrow \sigma(u,v) = \frac{E[\bar{X}]}{2 - E[\bar{X}]}\,.
\end{equation}

\begin{theorem}\label{thm:L}
	Define $\tsig(u,v) = \frac{\bar{X}}{2 - \bar{X}}$.
	By setting
	$L = \frac{2}{\Delta^2}\ln \frac{2}{\delta}$, we have
	$\Pr[|\tsig(u,v) - \sigma(u,v)|> \Delta] \leq \delta\,.$
\end{theorem}

\begin{proof}
	Observe that
	\begin{align*}
		 & \Pr[|\tsig(u,v) - \sigma(u,v)| > \Delta] = \Pr[|\frac{\bar{X}}{2 - \bar{X}} - \frac{E[\bar{X}]}{2 - E[\bar{X}]}| > \Delta] \\
		 & = \Pr[\frac{2 \cdot | \bar{X} - E[\bar{X}]|}{(2 - \bar{X})(2-E[\bar{X}])} > \Delta]
		\leq \Pr[|\bar{X} - E[\bar{X}]| > \frac{\Delta}{2}]\,,
	\end{align*}
	\\
	where the inequality is from~$\bar{X}$ and~$E[\bar{X}]$ being in~$[0, 1]$, and hence, $(2 - \bar{X})(2-E[\bar{X}]) \geq 1$.
	By the Hoeffding Bound~\cite{hoeffding-1994}, the probability such that $|\bar{X}-E[\bar{X}]|\geq t$ is bounded by $2e^{-2Lt^2}$.
	Thus, when $L = \frac{1}{2\cdot (\Delta/2)^2} \ln \frac{2}{\delta}$, we have $\Pr[|\bar{X} - E[\bar{X}]| > \frac{\Delta}{2}]\leq \delta$.
\end{proof}

We call the estimator, $\tsig(u,v) = \frac{\bar{X}}{2 - \bar{X}}$, a {\em $(\Delta, \delta)$-similarity-estimator}, with which we label edges by the following strategy.
\begin{definition}[The $(\Delta, \delta)$-Strategy]
	Every edge~$(u,v)\in E$, is labelled as similar if and only if~$\tsig(u,v) \geq \eps$.
\end{definition}

\begin{lemma}\label{lmm:approximate-correctness}
	With~$\Delta \leq \rho\eps$,
	with probability at least $1 - \delta$, the~$(\Delta, \delta)$-strategy labelling is $\rho$-approximate valid.
\end{lemma}
\begin{proof}
	The correctness of the labels in~$\tL$ follows from the fact that
	for any edge~$(u,v)\in E$, with probability at least $1 - \delta$,
	$|\sigma(u,v) - \tsig(u,v)| \leq \Delta \leq \rho\eps$. Thus, with the same probability,
	\begin{itemize}
		\item if $\sigma(u,v) \geq (1 +\rho)\eps$, we have~$\tsig(u,v) \geq \eps$;
		\item if $\sigma(u,v) < (1 - \rho) \eps$, we have~$\tsig(u,v) < \eps$.
	\end{itemize}
	In either of these cases,~$(u,v)$ must be labelled correctly under the $\rho$-approximate notion.
	For all other cases, the label of~$(u,v)$ does not matter.
\end{proof}

\noindent
\textbf{\newblue Remark.}
{\newblue
	An important superiority of our sampling-estimator over Min-Hash~\cite{b-sequences-1997} is that it allows us to compute~$\tsig(u,v)$ in~$O(L)$ time in an {\em ad hoc manner}.
	That is, need not maintain any data structures {\newblue (e.g., min-hash signatures)}, and thus, a $O(n + m)$ overall space consumption suffices.
	In the dynamic scenario,
	this feature saves substantially on maintenance costs.}

\section{Maintaining the Edge Labelling}\label{sec:maintenance}
Next, we reveal the details of the main tools behind the $\dynelm$ algorithm.
This algorithm maintains a valid~$\rho$-approximate edge labelling,~$\tL$, and is detailed in Section~\ref{sec:dynelm}.
In particular, we adopt the $(\Delta, \delta)$-strategy, with~$\Delta = \half\rho\eps$ and with~$\delta$ to be set later, aka, the $\strategy$,
to determine edge labels.

\subsection{Update Affordability}\label{subsec:affordability}
Observe that when an update~$(u,w)$ occurs, the \emph{affected} similarity values
are those between~$u$ and its neighbors, and those between~$w$ and its neighbors.
These edges are called the {\em affected edges} of~$(u,w)$, while~$(u,w)$ is an {\em affecting update} for each of these affected edges.

\begin{observation}\label{obs:affected-similarity}
	Consider an update (either an insertion or a deletion) of~$(u,w)$;
	if~$(u, v)$ is an arbitrary affected edge incident on~$u$, with $a =| N[u] \cap N[v]|$ and $b = |N[u] \cup N[v]|$ immediately before the update, the effects of the update are:
	\begin{itemize}[leftmargin = *]
		\item Case~1: an insertion of~$(u,w)$,
		      \begin{itemize}
			      \item if $w \in N[v]$, $\sigma(u, v)$ increases to~${(a + 1)}/{b}$;
			      \item if $w \not \in N[v]$, $\sigma(u, v)$ decreases to~${a}/{(b + 1)}$.
		      \end{itemize}
		\item Case 2: a deletion of~$(u,w)$,
		      \begin{itemize}
			      \item if $w \in N[v]$, $\sigma(u, v)$ decreases to~${(a - 1)}/{b}$;
			      \item if $w \not \in N[v]$, $\sigma(u, v)$ increases to~${a}/{(b - 1)}$.
		      \end{itemize}
	\end{itemize}
	Symmetric changes occur with each edge~$(v, w)$ incident on~$w$.
\end{observation}
\vspace{-1mm}

Our crucial observation is that the \emph{does-not-matter case} in the $\rho$-approximate notion,
affords each edge a certain number of updates that do not instigate a label change.

\begin{lemma}\label{lmm:dis-sim}
	If an edge $(u,v)$ is labelled as dissimilar by the $\strategy$, then, {with probability $\geq 1 - \delta$}, $(u,v)$ can afford at least $k = \lfloor \half \rho\eps \cdot d_{max}(u,v) \rfloor$ affecting updates before its label flips (from dissimilar to similar), where $d_{max}(u,v) = \max\{d[u], d[v]\}$.
\end{lemma}
\begin{proof}
	Let $a = |N[u] \cap N[v]|$ and $b = |N[u]\cup N[v]|$, initially.
	Edge~$(u,v)$ being labelled dissimilar by the $\strategy$ implies $\tsig(u,v) < \eps$, and thus, with probability $\geq 1 - \delta$,
	$\sigma(u,v) = a/ b \leq (1 + \half\rho)\eps < (1 + \rho) \eps \leq 1 \Rightarrow a < b$.
	Since both~$a$ and~$b$ are integers, we have $ a + 1 \leq b$, and hence $(a + 1) / b \geq a / (b -1)$.
	Therefore, in considering the minimum number of affecting updates to cause~$(u,v)$'s label flip (from dissimilar to similar), we focus on the edge insertions that increase~$\sigma(u,v)$.
	After~$k = \lfloor \half \rho\eps \cdot d_{max}(u,v) \rfloor \leq \half \rho\eps b$ such updates,
	by the first bullet of Case 1 in Observation~\ref{obs:affected-similarity},
	\begin{equation*}\label{eq:dis-sim}
		\sigma(u, v) = \frac{a}{b} + \frac{k}{b} \leq (1 + \frac{1}{2}\rho) \eps  + \frac{\frac{1}{2}\rho\eps b}{b} = (1 + \rho) \eps\,.
	\end{equation*}
	Therefore, after~$k$ arbitrary affecting updates, the dissimilar label of~$(u,v)$ remains valid with probability $\geq 1 - \delta$.
\end{proof}

\begin{lemma}\label{lmm:sim-dis}
	If edge~$(u,v)$ is labelled as similar by the $\strategy$, then, {with probability $\geq 1 - \delta$}, $(u,v)$ can afford at least $k = \lfloor \half\rho\eps \cdot d_{\max}(u,v)\rfloor$ affecting updates before its label flips.
\end{lemma}
\begin{proof}
	The proof is analogous to that of Lemma~\ref{lmm:dis-sim}.
\end{proof}

\subsection{Distributed Tracking on Updates}

\begin{figure*}
	\resizebox{\linewidth}{!}{%
		\begin{tabular}{ccc}
			\hspace{-10mm}
			\includegraphics[height = 35mm]{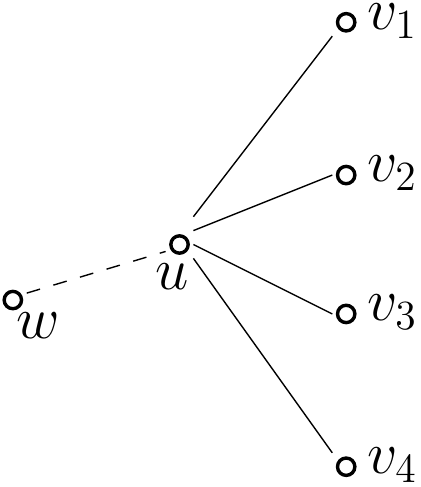} & \hspace{-1mm}\vline\hspace{1mm}
			\includegraphics[height = 35mm]{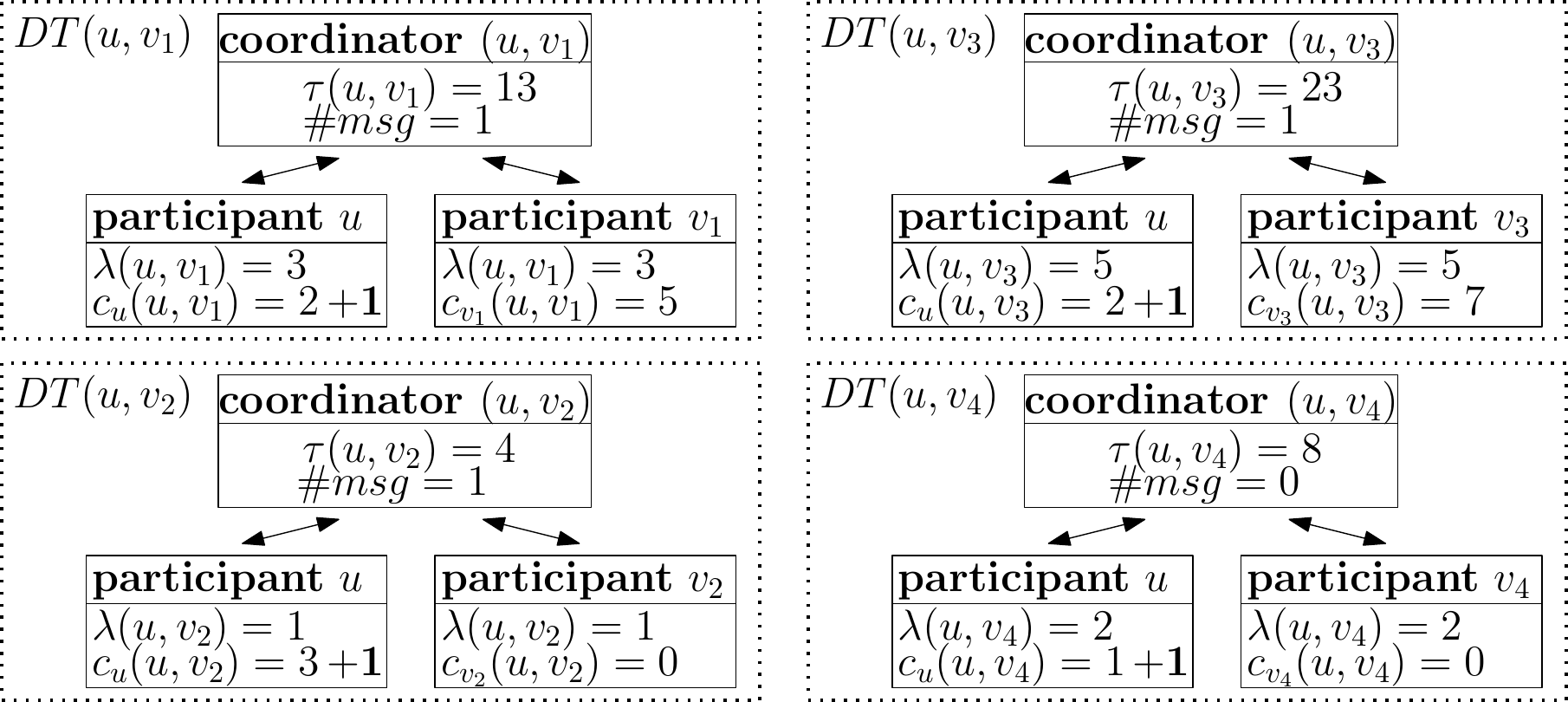} & \hspace{-1mm}\vline\hspace{1mm}
			\includegraphics[height = 35mm]{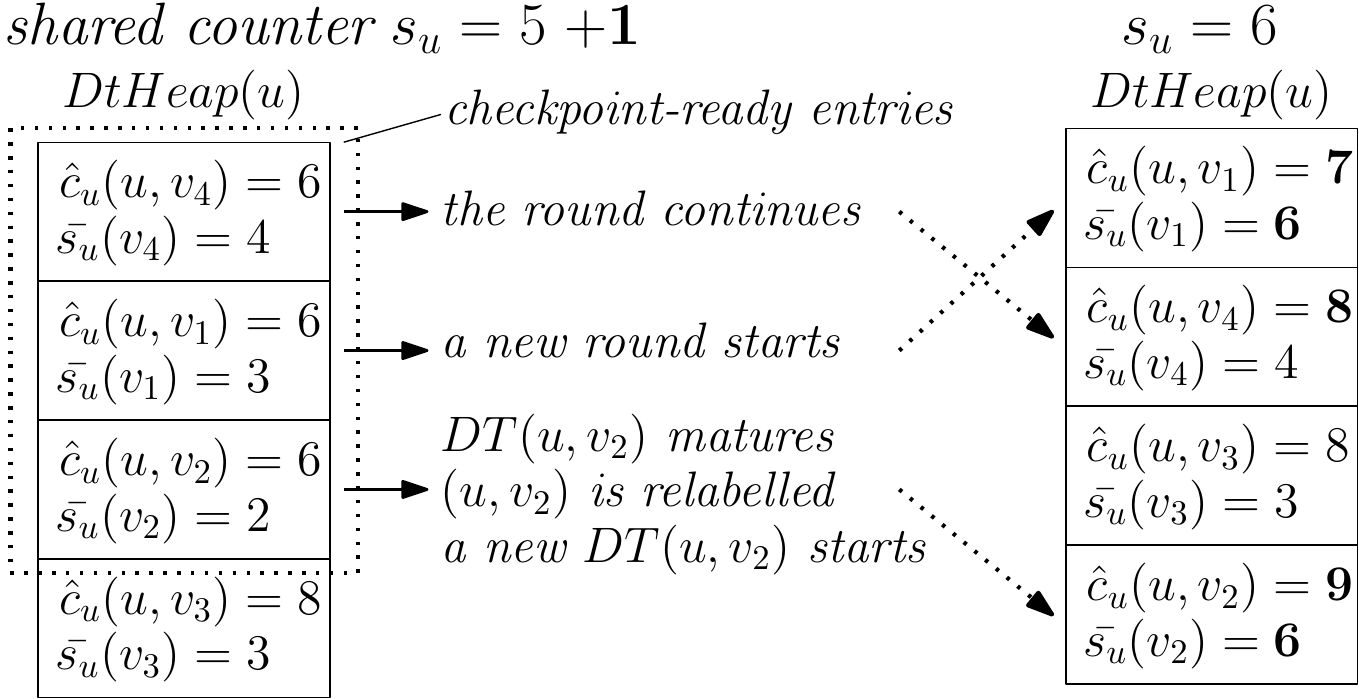}
			\\
			(a) {\em An update of $(u,w)$}                                       & \hspace{6mm} (b) {\em Maintain the DT instances individually} & \hspace{2mm} (c) {\em Organize the DT instances with heaps} \\[1mm]
		\end{tabular}
	}
	\figcapup
	\caption{ An example of handling an update of $(u,w)$ with DT instances on the $u$ side (the update process on the $w$ side is symmetric and thus ommited), where the un-affected edges incident on $v_i$ (for $i = 1, \ldots, 4$) are omitted.
	}
	\label{fig:dt-explanation}
\end{figure*}

Lemmas~\ref{lmm:dis-sim} and~\ref{lmm:sim-dis}
together show that an edge~$(u,v)$  labelled by the $\strategy$ can afford at least $ k = \lfloor \frac{1}{2} \rho\eps \cdot d_{\max}(u,v) \rfloor$
affecting updates without its label flipping.
It thus suffices to
check its label upon the $(k + 1)^{\text{th}}$ affecting update since it was last labelled.

\vspace{1mm}
\noindent
\textbf{Creating DT Instances for Edges.}
To achieve this purpose, we adopt distributed tracking (DT) to {\em track}
the number of affecting updates
for each edge.
Specifically, for each edge~$(u,v)$, we {\em simulate} a DT instance,
denoted by~$\DT(u,v)$, in a single thread in main memory.
The edge~$(u,v)$ itself is the coordinator, with its
endpoints~$u$ and~$v$ the participants;
and the tracking threshold is set to
\begin{equation}
	\tau(u,v) = \lfloor \half\rho\eps\cdot d_{\max}(u,v) \rfloor + 1\,.
\end{equation}

The counter $c_u(u,v)$ (resp., $c_v(u,v)$) is the current number of affecting updates of~$(u,v)$
incident on~$u$ (resp., $v$).
As soon as $c_u(u,v) + c_v(u,v) = \tau(u,v)$, the coordinator at~$(u,v)$ reports {\em maturity}.
Since the label~$(u,v)$ could be invalid,
we {\em relabel}~$(u,v)$ with the~$\strategy$.
After then, a new~$\DT(u,v)$ instance, with a new threshold~$\tau(u,v)$ (based on the new $d_{\max}(u,v)$) is instantiated.

However, as we are simulating DT in main memory and counting running time only,
in addition to the simulated $O(\log \tau(u,v))$ communication cost, each counter increment costs~$O(1)$ time.
Incrementing $c_u(u,v)$ for all neighbors~$v\in N[u]$ still leads to a $\Omega(d[u]) = \Omega(n)$ cost for an update.
Figures~\ref{fig:dt-explanation}(a) and~\ref{fig:dt-explanation}(b) show an example; for an update of $(u, w)$, one needs to {\em individually} increase the counter of the participant $u$ in each of the DT instances.

\vspace{1mm}
\noindent
\textbf{Organizing DT Instances by Heaps.}
The key to address this issue is to
maintain a {\em shared} common counter, instead of increasing~$c_u(u,v)$ for each $v\in N[u]$ individually.
Let every vertex~$u$ have, instead of~$c_u(u,v)$, a single counter~$\cnt_u$ (shared among all~$\DT(u,v)$), initially set to~$0$, recording the number of affecting
updates of edges incident on~$u$.
The crucial observation here is that
the checkpoint value~$\hat{c}_u(u,v)$ is only updated when
there would have
been~$\lambda(u,v)$ (the slack value in $DT(u,v)$) affecting updates incident on~$u$.
Thus, the number of increments (i.e., $\lambda(u,v)$) is important, rather than the value of $\hat{c}_u(u,v)$.
Therefore, we can \emph{shift} the checkpoint value~$\hat{c}_u(u,v)$ by the value~$\cnt_u$ at the time the checkpoint is set.
After shifting,
for each vertex~$u$, we set up a \emph{min-heap}, denoted by $\DtHeap(u)$,
with the shifted checkpoints,~$\hat{c}_u(u,v)$, as keys.
For each~$(u,v)$,
we maintain an entry in $\DtHeap(u)$ associated with:
\begin{itemize}[leftmargin = *]
	\item $\overbar{\cnt_u}(v)$: the value of~$\cnt_u$ when the current round in $DT(u,v)$ starts.
	      With $\overbar{\cnt_u}(v)$,
	      the {\em unshifted} counter value in this participant can be computed, by $\cnt_u - \overbar{\cnt_u}(v)$,  when the coordinator needs it;
	\item $\hat{c}_u(u,v)$: the key of the entry, initialized as $\hat{c}_u(u,v) = \overbar{\cnt_u}(v) + \lambda(u,v)$, where~$\lambda(u,v)$ is the slack value in the current round of~$\DT(u,v)$.
\end{itemize}
When an affecting update arrives,~$u$ only needs to inform its coordinators
if there is some entry with a key equal to~$\cnt_u$. Each of these entries is called a {\em checkpoint-ready} entry.
Therefore, in this way,
we no longer have to scan the whole neighbourhood of a vertex.

\vspace{1mm}
\noindent
\underline{\em A Running Example.} Figure~\ref{fig:dt-explanation}(c) shows an example, where the shared counter $s_u = 5$ before the update, and the entries corresponding to the DT instances of $(u,v_4)$, $(u, v_1)$ and $(u,v_2)$ are at the {\em top} of $\DtHeap(u)$ as they have the same smallest key values, i.e., $\hat{c}_u = 6$.
When the update of $(u,w)$ is performed, $s_u$ is increased to $s_u = 6$, and hence, the three entries at the heap top become checkpoint-ready.
For the entry of $\DT(u,v_4)$, the participant $u$ only needs to notify the coordinator $(u, v_4)$ and the current round continues
(as shown in Figure~\ref{fig:dt-explanation}(b), this is the first notification in the round):
$\overbar{\cnt_u}(v_4)$ does not change and $\hat{c}_u(u, v_4)$ is increased by $\lambda(u,v_4) = 2$ indicating that when $s_u$ reaches $8$, this entry will become checkpoint-ready again.
For the entry of $\DT(u,v_1)$,
after the notification sent for this entry, the current round ends: $\tau(u,v_1) \leftarrow (13 - 3) - 5 = 5$ and $\lambda(u,v_1)\leftarrow \lfloor \frac{5}{2 \times 2} \rfloor  = 1$.
Thus, $\overbar{\cnt_u}(v_1)\leftarrow s_u = 6$ and $\hat{c}_u(u,v_1) = 6 + 1 = 7$.

Finally, for the entry of $\DT(u,v_2)$, the DT instance is matured. Hence, the edge $(u,v_2)$ is relabelled by the $\strategy$ and a new DT instance with respect to the new $\tau(u,v_2)$ is started.

\section{The $\dynelm$ Algorithm}
\label{sec:dynelm}
An outline of the $\dynelm$ algorithm for handling an update, of edge~$(u,w)$, is as follows:
\begin{itemize} [leftmargin = *]
	\item {\bf Step 1.} Initialize the set of label-flipping edges $\F \leftarrow \varnothing$; and increment
	      $\cnt_u$ and $\cnt_w$ (by~$1$), respectively.
	\item {\bf Step 2.} There are two cases:
	      \begin{itemize}[leftmargin = *]
		      \item \underline{\em Case 1: this update is an insertion.} Insert~$(u, w)$ into~$G$ and
		            label it by the $\strategy$. If~$(u,w)$ is labelled as similar, add~$(u,w)$ to~$\F$. Moreover, create $DT(u,w)$ with $\tau(u,w)$.
		      \item \underline{\em Case 2: this update is a deletion.} If $(u,w)$ is labelled as similar, add $(u,w)$ to $\F$. Delete $(u,w)$ from $G$; and delete $DT(u,w)$.
	      \end{itemize}
	\item {\bf Step 3.} While there is a checkpoint-ready entry in $\DtHeap(u)$, pop the entry (from the top). Let $\DT(u,v)$ be the DT instance corresponding to this entry. Instruct~$u$ to inform the
	      coordinator~$(u,v)$.
	      When~$DT(u,v)$ is mature, relabel~$(u,v)$ by the $\strategy$. If its label flipped, add~$(u,v)$
	      to~$\F$. Remove its entry from~$\DtHeap(v)$, and restart the DT with a {\em new }~$\tau(u,v)$.
	      Repeat until there is no checkpoint-ready entry in $\DtHeap(u)$.
	\item {\bf Step 4.} Perform a symmetric process of Step~3 for~$w$.
	\item {\bf Step 5.} Return~$\F$, the set of edges whose labels flipped.
\end{itemize}

\subsection{Theoretical Analysis}\label{sec:basic-analysis}
In this subsection, we prove the following theorem:
\begin{theorem}\label{thm:basic-algo}
	Given a specified failure probability $\delta^*$,
	for every sequence of $M$ updates (the value of~$M$ value need not be known in advance),
	there exists an implementation of the $\dynelm$ algorithm that achieves the following guarantees:
	\begin{itemize}[leftmargin = *]
		\item the amortized cost of each update is~$O(\log^2 n + \log n \cdot \log \frac{M}{\delta^*})$;
		\item the space consumption is always linear in the size of $G$, i.e., $O(n + m)$;
		\item with probability at least $1 - \delta^*$, the $\rho$-approximate edge labelling~$\tL$ maintained is always valid. Hence, with the same probability, the clustering result,~$\aC$, is always correct (under the $\rho$-approximate notion).
	\end{itemize}
\end{theorem}

\begin{corollary}
	If the number of vertices, $n$, is fixed over the whole update sequence and the total number of updates $M$ is bounded by $O(n^c)$ for some constant $c$, e.g., $M = n^{99}$, by setting $\delta^* = 1 / n$, the amortized update bound can be simplified to $O(\log^2 n)$.
\end{corollary}

We consider the following implementation of $\dynelm$.
\begin{itemize}[leftmargin = *]
	\item The neighborhood $N[u]$ of each vertex $u$ is maintained by a binary search tree: each neighbor insertion, deletion and search can be performed in $O(\log d[u]) = O(\log n)$ time.
	\item The $\DtHeap(u)$ of each vertex $u$ is implemented with a binary heap: each heap operation takes $O(\log d[u]) = O(\log n)$ time.
	      Moreover, according to the DT algorithm, there are at most $O(\log \tau(u,v)) = O(\log n)$ rounds (before its maturity) for each $\DT(u,v)$,
	      and each round takes $O(h) = O(1)$ (because only $h = 2$ participants in the instance) operations in the DT heaps of $u$ and $v$.
	      The overall cost of each $\DT(u,v)$ is bounded by $O(\log^2 n)$.

	\item For the $i^{\text{th}}$ invocation of the $\strategy$,  the parameter $\delta$ is set to
	      \begin{equation}
		      \delta_i = \frac{\delta^*}{ i \cdot (i + 1)}\,,\text{
			      where $i = 1, 2, \ldots$.}
	      \end{equation}

	      Let $\kappa$ be the total number of invocations of the strategy.
	      According to Theorem~\ref{thm:L}, the required sample size for the $i^{\text{th}}$ invocation is
	      \begin{equation}
		      L_i = \frac{2}{(\frac{1}{2}\rho\eps)^2}\ln \frac{2}{\delta_i} = O(\log \frac{\kappa}{\delta^*})\,.
	      \end{equation}

	      Therefore, the cost of each invocation of $\strategy$ is bounded by $O(L_i \cdot \log n) = O(\log n \cdot \log \frac{\kappa}{\delta^*})$.
\end{itemize}
Next, we prove the three bullets in Theorem~\ref{thm:basic-algo} one by one.

\vspace{1mm}
\noindent
\textbf{Bullet 1 in Theorem~\ref{thm:basic-algo}: Amortized Update Cost.}
We analyse the amortized update cost of $\dynelm$ step by step.
Clearly, the running time of Step 1 in the $\dynelm$ algorithm is $O(1)$, and
Step 2 can be performed in $O(\log n \cdot \log \frac{\kappa}{\delta^*})$ time in the worst case.
It remains to bound the amortized cost of Step 3 and Step 4.

Consider an update of $(u,w)$.
A crucial observation is that
the update of $(u,w)$ can {\em only} contribute (via a counter increment) to the maturity of the DT instances of its affected edges,
which exist at the {\em current moment}.
Let $\DT(u,v^*)$ be the instance with the {\em smallest} threshold value $\tau(u,v^*)$ among all the affected DT instances at the current moment, and $d'[u]$ the degree of $u$ when $\DT(u,v^*)$ was created.
We claim that the degree of $u$, $d[u]$, at the current moment is at most $d'[u] + \tau(u,v^*)$.
This is because since the creation of $\DT(u,v^*)$, there can be at most $\tau(u,v^*)$ insertions adjacent on $u$; otherwise, $\DT(u,v^*)$ must have matured and hence, would not exist at the current moment.
Therefore,
\begin{align*}
	d[u] \leq d'[u] + \tau(u,v^*) \leq \frac{\tau(u,v^*)}{\rho\eps/2} + \tau(u,v^*) = O(1) \cdot \tau(u,v^*)\,.
\end{align*}

Furthermore, as each of the affected $\DT(u,v)$ requires at least $\tau(u,v)$ affecting updates to mature,
the current update of $(u,w)$ is actually accounted for only $\frac{1}{\tau(u,v)}$ of the cost of the DT maturity as well as the following edge re-labelling cost, i.e., $\frac{1}{\tau(u,v)}\cdot O(\log^2 n + \log n \cdot \log \frac{\kappa}{\delta^*})$.
Summing up over all the neighbors of $u$, the amortized cost instigated by an update of $(u,w)$ in Step 3 is:
\begin{align*}
	  & \sum_{v \in N[u]} \frac{O(\log^2 n + \log n \cdot \log \frac{\kappa}{\delta^*})}{\tau(u,v)}
	\leq  \sum_{v \in N[u]} \frac{O(\log^2 n + \log n \cdot \log \frac{\kappa}{\delta^*})}{\tau(u,v^*)}                                                    \\
	= & \frac{d[u]}{\tau(u,v^*)} \cdot O(\log^2 n + \log n \cdot \log \frac{\kappa}{\delta^*})= O(\log^2 n + \log n \cdot \log \frac{\kappa}{\delta^*})\,.
\end{align*}
\\
By symmetry, the update of $(u,w)$ is also charged a $O(\log^2 n + \log n \cdot \log \frac{\kappa}{\delta^*})$ cost from  the affected DT instances of the vertex $w$.
Therefore, combining the costs of all the four steps, the amortized cost of each update is bounded by $O(\log^2 n + \log n \cdot \log \frac{\kappa}{\delta^*})$.

To complete our analysis, we claim that each update of $(u,w)$ can instigate at most {\em $O(1)$  amortized invocations} of the $\strategy$,
where one is for labelling $(u,w)$ when the update is an insertion, the others are for the $O(1)$ DT maturity charged to the current update.
Thus, the total number of the invocations of the strategy, $\kappa$,  is at most $O(1)$ times of the number of updates, i.e., $\kappa = O(M)$, and the amortized update cost follows.

\begin{lemma}\label{lmm:amortized-bound}
	For any sequence of $M$ updates, the amortized update cost of $\dynelm$ is bounded by $O(\log^2 n + \log n \cdot \log \frac{M}{\delta^*})$.
\end{lemma}

\noindent
\textbf{Bullet 2 in Theorem~\ref{thm:basic-algo}: Overall Space Consumption.}
Based on the aforementioned implementation of $\dynelm$,
for each vertex~$u$, a binary search tree and a DT heap on the neighborhood~$N[u]$ are maintained,
the space consumption of each vertex is bounded by $O(1 + d[u])$.
Summing up over all the vertices,
the overall space consumption of the $\dynelm$ algorithm is bounded by~$O(n + m)$.
\begin{lemma}\label{lmm:space}
	At all times, $\dynelm$ consumes~$O(n + m)$ space.
\end{lemma}

\noindent
\textbf{Bullet 3 in Theorem~\ref{thm:basic-algo}: Correctness and Failure Probability.}
The correctness of the approximate edge labelling $\tL$ maintained by $\dynelm$ follows immediately from the correctness of the $\strategy$ (Lemma~\ref{lmm:approximate-correctness}) and the update affordability (Lemmas~\ref{lmm:dis-sim} and~\ref{lmm:sim-dis}).
It remains to bound the failure probability.
According to our implementation and by Union Bound, the failure probability is bounded by:
\begin{equation*}
	\sum_{i = 1}^\kappa \delta_i = \sum_{i = 1}^\kappa \frac{\delta^*}{i \cdot (i + 1)} = \delta^* \cdot \sum_{i=1}^\kappa \left(\frac{1}{i} - \frac{1}{i+1}\right) = \delta^*\cdot (1 - \frac{1}{\kappa + 1}) \leq \delta^*\,.
\end{equation*}
\begin{lemma}\label{lmm:probability-bound}
	With probability at least $1 - \delta^*$, the $\rho$-approximate edge labelling maintained by $\dynelm$ is always valid.
\end{lemma}

\section{The Ultimate Algorithm}\label{sec:ultimate-algo}
We round out our algorithm development by designing the $\dynstr$ algorithm for solving the \emph{ultimate} dynamic $\strclu$ problem. Specifically, we prove the following theorem.

\begin{theorem}\label{thm:ultimate-algo}
The $\dynstr$ algorithm both
\begin{itemize}[leftmargin = *]
	\item admits all the same guarantees as $\dynelm$ in Theorem~\ref{thm:basic-algo}; and
	\item answers every cluster-group-by query in time linear-polylog in the query size, i.e., for~$Q \subseteq V$, in~$O(|Q|\cdot \log n)$ time. 
\end{itemize}
\end{theorem}

\vspace{1mm}
\noindent
\textbf{The Algorithm Framework.}
The $\dynstr$ algorithm  mainly consists of the following three modules:
\begin{itemize}[leftmargin= *]
	\item \underline{\em Edge Label Manager} (ELM): This module invokes the $\dynelm$ algorithm as a black box, to maintain~$\tL$. 
	\item \underline{\em Vertex Auxiliary Information} (vAuxInfo): For each vertex~$u\in V$, we maintain  auxiliary information:
		\begin{itemize}
			\item a counter, $\SimCnt(u)$, for recording the current number of similar neighbors of~$u$;
			\item a partition of~$u$'s neighbors, which partitions~$u$'s neighbors into three self-explanatory categories:
			(i) sim-core neighbors,
			(ii) sim-non-core neighbors, and
			(iii) dissimilar neighbors. 
		\end{itemize}
		$\SimCnt(u)$ can be updated in~$O(1)$ time; and moving a neighbor from one category to another also takes~$O(1)$ time, given that the labelling has been done in ELM.
	\item \underline{\em CC Structure of $G_{\core}$}:
		In this module, we maintain the connected components in~$G_{\core}$. In particular, we maintain a data structure, denoted by {CC-Str$(G_{\core})$}, to support the following operations:
	\begin{itemize}
		\item \emph{Insert} a sim-core edge~$(u,v)$ into~$G_{\core}$.
		\item \emph{Remove} from~$G_{\core}$ an edge~$(u,v)$.
		\item $\FindCcID(u)$: Return the identifier of the connected component in~$G_{\core}$ that contains the (core) vertex~$u$.
	\end{itemize}
\end{itemize}

\begin{fact}[\cite{hlt-acm-2001,t-stoc-2000}]\label{fact:cc-str}
	There exists a $O(n + m)$-space data structure that implements CC-Str$(G_{\core})$ and can support: (i) each edge insertion or deletion~in~$O(\log^2 n)$ amortized time, and
	(ii) each $\FindCcID$ operation in $O(\log n)$ worst-case time.
\end{fact}

\vspace{1mm}
\noindent
\textbf{The $\dynstr$ Algorithm Steps.}
To process an update, the $\dynstr$ algorithm maintains~$\tL$, with the $\dynelm$ algorithm;
this returns the set~$\F$ of edges whose labels have flipped due to the update.
Given this flipped set~$\F$, $\dynstr$ maintains the two other modules as follows.  

\vspace{1mm}
\noindent
\underline{\em Maintaining vAuxInfo.} For each edge~$(u,v) \in \mathcal{F}$,
\begin{itemize}[leftmargin= *]
	\item update~$\SimCnt(u)$ and~$\SimCnt(v)$ in constant time: if the label of $(u,v)$ is flipped to similar, both $\SimCnt(u)$ and $\SimCnt(v)$ are increased by 1; otherwise, they are decreased by 1, respectively.
	\item if necessary, flip~$u$'s (resp.,~$v$'s) {\em core status},
	and hence change the neighbor category of~$u$ for its {\em similar} neighbors.
\end{itemize}
Let~$V'$ comprise every vertex whose core status has flipped, while~$\mathcal{F'}$ is the set of all the edges whose {sim-core status} have flipped {between {\em sim-core} and {\em non-sim-core}.}

\vspace{1mm}
\noindent
\underline{\em Maintaining $G_{\core}$.} 
For each edge~$(u,v)\in \mathcal{F'}$, if the status of~$(u,v)$ flipped from {\em non-sim-core} to {\em sim-core}, insert~$(u,v)$ into~$G_{\core}$.
Otherwise, remove~$(u,v)$ from $G_{\core}$.
Furthermore, for each vertex~$u \in V'$, if~$u$ flipped from {\em non-core} to {\em core}, 
insert $u$ to $G_{\core}$ by {\em conceptually} inserting to~$G_{\core}$ a self-loop edge
~$(u, u)$, which does not necessarily physically exist.
Otherwise, remove~$u$ from $G_{\core}$ by {\em conceptually} removing the self-loop edge~$(u, u)$,
in which case,~$u$ must be a singleton vertex in~$G_{\core}$ with no incident edge other than the self-loop edge. This is because, all its incident edges in~$G_{\core}$ have been removed when processing the edges in~$\F'$.
All these operations on~$G_{\core}$ can be performed via CC-Str$(G_{\core})$.

\begin{figure}[t]
\begin{center}
	\includegraphics[width = \linewidth]{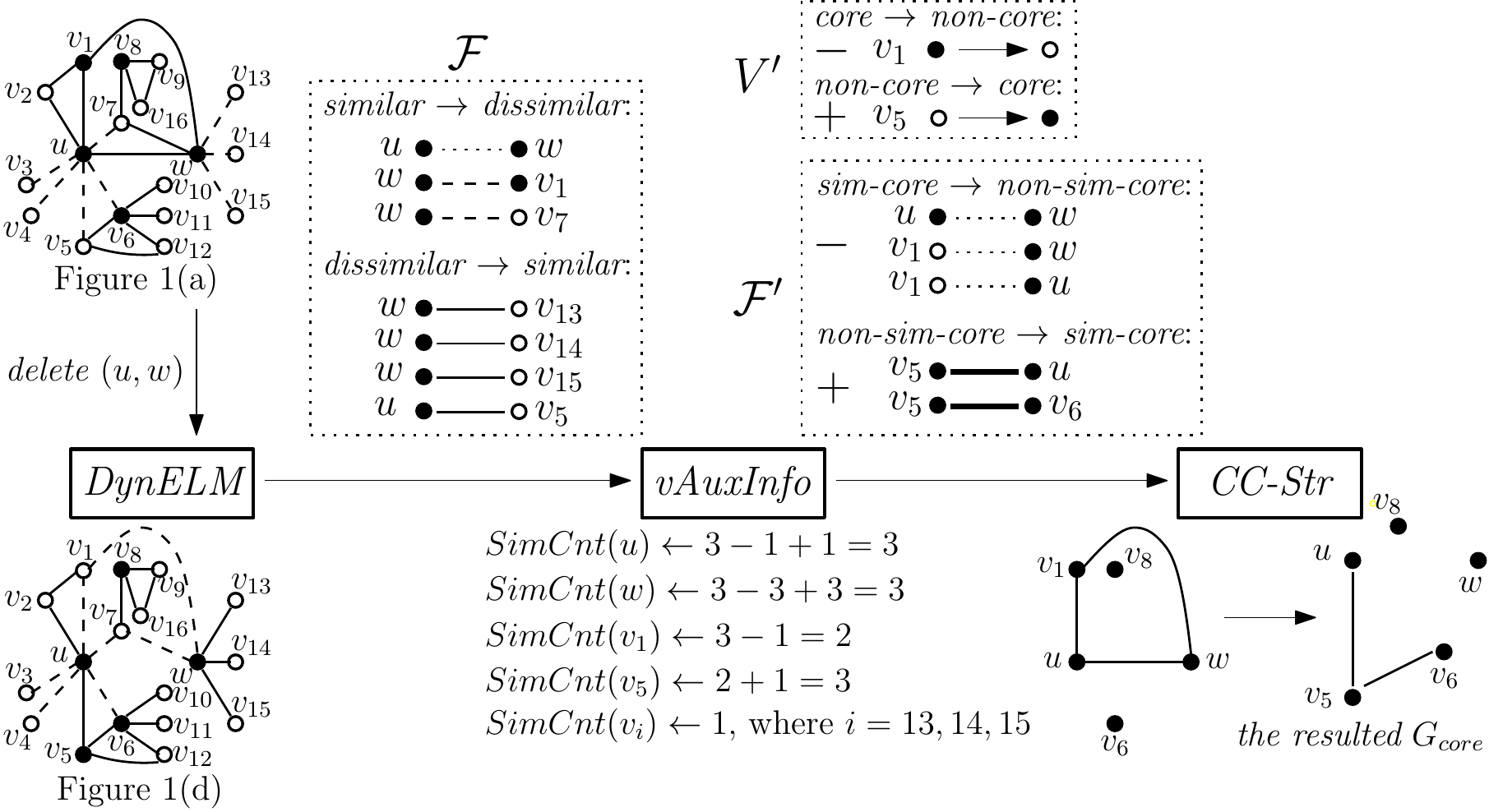}
	\caption{
		The process of $\dynstr$ handling a deletion of~$(u,w)$ from Figure~\ref{fig:strclu}(a),
		with $\eps = 1/3$, $\mu = 3$ and $\rho = 0.01$. 
		The resulted state of after this update is shown in Figure~\ref{fig:strclu}(d).
	}
\label{fig:dynstrclu-process}
\end{center}
\end{figure}

\vspace{1mm}
\noindent
\textbf{A Running Example.} Figure~\ref{fig:dynstrclu-process} shows the maintenance process of $\dynstr$ for deleting the edge~$(u,w)$ from Figure~\ref{fig:strclu}(a), where
the resulted state is as shown in Figure~\ref{fig:strclu}(d).
To process the deletion of~$(u,w)$, $\dynstr$ invokes $\dynelm$ to maintain the edge labelling~$\tL$.
The returned set~$\F$ of edges with labels flipped is shown in the figure. 
In particular, since~$(u,w)$ is a similar edge getting deleted, its label is treated as flipping from similar to dissimilar.
Next, the~{\em vAuxInfo} module updates the $\SimCnt$ information for the endpoints of the edges in~$\F$;
details are shown in the figure.
Since 
$\SimCnt(v_1)$ is decreased from $3$ to $2$ and $\SimCnt(v_5)$ is increased from $2$ to $3$,
the core status of $v_1$ is flipped from core to non-core, while $v_5$'s is from non-core to core.
Thus, $V' = \{v_1, v_5\}$, the set of all the vertices with core status flipped. 
Furthermore, with $\F$ and $V'$, the set $\F'$ of the edges whose {\em sim-core} status are flipped can be easily obtained. 
For example, the sim-core status of edges $(u,w)$ and $(v_1, w)$ are flipped from sim-core to non-sim-core for different reasons.
While the flip of $(u,w)$ is because of its label flipping to dissimilar, the flip of $(v_1, w)$ is cased by $v_1$ becoming non-core.
Likewise, since $v_5$ becomes a core vertex, the similar edges $(v_5, u)$ and $(v_5, v_6)$ become sim-core.
Finally, the CC-Str maintains~$G_{\core}$ with~$V'$ and~$\F'$ by: (i) removing~$v_1$ and adding~$v_5$; (ii) removing all the edges in~$\F'$ turning into non-sim-core, while adding those edges becoming sim-core.
The resulting~$G_{\core}$ is as shown in Figure~\ref{fig:dynstrclu-process}.
 
\vspace{1mm}
\noindent
\textbf{Theoretical Analysis.} 
We analyse the overall maintenance cost on the above two modules. 
\begin{lemma}\label{lmm:VF}
	$|V'| = O(|\F|)$ and $|\F'| = O(|\F|)$.
\end{lemma}
\begin{proof}
	Observe that the core status of vertex~$u$ is flipped only if~$\SimCnt(u)$ changes: at least one edge incident on~$u$ has its label flipped.
	As such edge must be in $\F$, $|V'| \leq 2\cdot |\F| = O(|\F|)$ holds.

	Next , we bound~$|\F'|$.
	We define {\em persistently similar edges} as those edges that remain similar after the update.
	There are only {two} possibilities for edge~$(u,v)$ to {be in~$\F'$:}
	(i) the label of~$(u,v)$ is flipped, or
	(ii)~$(u,v)$ is persistently similar and has at least one endpoint with core status flipped.
	Clearly, there are at most~$|\F \cap \F'| \leq |\F|$ edges added to~$\F'$ due to the first case.
	For the edges in~$\F' \setminus \F$, 
	they must belong to the second case.
	Thus, this is at most the number of persistently similar edges incident on some vertex in~$V'$.  
	For each $u\in V'$, there can be at most~$\mu -1 = O(1)$ persistently similar edges incident on~$u$.
	Because otherwise,
	there is a contradiction with the fact that~$u$'s core status has been flipped. 
		For example, in Figure~\ref{fig:dynstrclu-process}, $v_1\in V'$ has one persistent similar edge $(v_1, v_2)$ and $v_5 \in V'$ has two: $(v_5, v_6)$ and $(v_5, v_{12})$.
		Both of these numbers are at most $2$.
		Otherwise, 
		$v_1$ would not become a non-core
		and $v_5$ would have been a core before the update.
	Thus, there can be at most~$O(|V'|) = O(|\F|)$ persistently similar edges incident on the vertices in~$V'$.
	Therefore,~$|\F' \setminus \F| = O(|\F|)$, and hence~$|\F'| = O(|\F|)$.
\end{proof}

\begin{lemma}\label{lmm:framework-cost} 
	The cost of maintaining vAuxInfo and~$G_{\core}$ is bounded by~$O(|\F| \cdot \log^2 n)$.
\end{lemma}
\begin{proof}
	We bound the maintenance cost of {\em vAuxInfo} first.
	Since only the endpoints of the edges in~$\F$ can have~$\SimCnt$ changed, 
	the cost of maintaining~$\SimCnt$ is clearly bounded by~$O(|\F|)$.	
	{As for the neighbor category, there are only two possible types of changes:} (i) between similar and dissimilar neighbors, or (ii) between sim-core and sim-non-core neighbors. 
	While the former is caused by edge-label flips, the latter is due to sim-core status change.
	The total number of neighbor category alternations caused by this update is thus at most~$|\F| + |\F'|$.
	As each such alternation takes~$O(1)$ time, the maintenance cost is~$O(|\F| + |\F'|)$.
	Therefore, by Lemma~\ref{lmm:VF}, the per-update maintenance cost for {\em vAuxInfo} is~$O(|\F|)$.

	As for the maintenance of~$G_{\core}$, it is clear that there are $|V'| + |\F'| = O(|\F|)$ operations with CC-Str$(G_{\core})$. 
		By Fact~\ref{fact:cc-str}, each such operation incurs a~$O(\log^2 n)$ amortized cost. 
		The~$O(|\F|\cdot \log^2 n)$  per-update maintenance cost of $G_{\core}$ thus follows.
	
\end{proof}

\begin{theorem}\label{thm:framework-cost2}
		The $\dynstr$ Algorithm admits all the same guarantees as $\dynelm$ in Theorem~\ref{thm:basic-algo}
\end{theorem}
\begin{proof}
	First, to analyse the amortized update cost, observe that 
	an edge can be added to $\F$ only when it is relabelled.
   Thus, we can amortize the $O(|\F|\cdot \log^2 n)$ cost over all these edges in $\F$.
   Hence, each of such edges is charged for an extra $O(\log^2 n)$ cost when it is relabelled;
   this charging increases the relabelling cost bound to $O(\log^2 n + \log n \cdot \log \frac{M}{\delta^*})$.
   Therefore, the amortized update cost bound remains the same as that of $\dynelm$.
	Second, by Fact~\ref{fact:cc-str}, the space consumption of $\dynstr$ is also bounded by $O(n + m)$.
	Finally, as the maintenance for {\em vAuxInfo} and CC-Str$(G_{\core})$ is deterministic, the correctness probability remains the same as $\dynelm$. 
\end{proof}

\vspace{1mm}
\noindent
\textbf{The Cluster-Group-By Query Algorithm.} Given $Q\subseteq V$, the cluster-group-by query algorithm is as follows:
\begin{itemize}[leftmargin = *]
	\item Initialize an empty (vertex, ccid)-pair set:~$P \leftarrow \varnothing$. 
	\item For each~$u \in Q$: 
	\begin{itemize}
		\item If~$u$ is core, obtain the ID of the CC containing~$u$ in $G_{\core}$, denoted by~$\ccid(u)$.
		Add the pair~$(u, \ccid(u))$ to~$P$.
	\item If~$u$ is non-core, for each sim-core neighbor~$v$ of~$u$ (possibly none exists),
		add a pair~$(u, \ccid(v))$ to~$P$. 
	\end{itemize}
\item Sort the pairs in~$P$ by the~$ccid(\cdot)$ keys.
Put all vertices with the same~$\ccid(\cdot)$ into the same group and output the resulted groups.
\end{itemize}

\begin{lemma}\label{lmm:query-cost}
	The running time complexity of this Cluster-Group-By Query Algorithm is~$O(|Q|\cdot \log n)$.
\end{lemma}

\begin{proof}
	While each core vertex in~$Q$ produces exactly one pair, each non-core vertex in~$Q$ can produce at most~$\mu-1$ pairs. 
	The size of~$P$ is thus~$O(|Q|)$. Furthermore, by Fact~\ref{fact:cc-str}, the $\ccid(\cdot)$ of each pair can be obtained with CC-Str$(G_{\core})$ in~$O(\log n)$ time.
Combining the sorting cost, the total running time is~$O(|Q|\cdot \log n)$.
\end{proof}

\vspace{1mm}
\noindent
\textbf{Proof of Theorem~\ref{thm:ultimate-algo}.} The theorem follows immediately from Theorem~\ref{thm:framework-cost2} and Lemma~\ref{lmm:query-cost}.

\vspace{1mm}
\noindent 
\textbf{Remark.} The amortized update cost bound in Theorem~\ref{thm:ultimate-algo} (and hence, in Theorem~\ref{thm:basic-algo}) is general enough for ``hot-start'' cases, where a graph $G$ with $m_0$ edges is given at the beginning. To handle this case, one can first insert each of these
$m_0$ edges one by one,  with a total cost~$\tO(m_0)$, and then charge this~$\tO(m_0)$ cost to the next~$\Omega(m_0)$ updates. There is just a constant factor blow-up in the amortized update cost.

\section{Extension to Cosine Similarity}\label{sec:ext}
{\newblue
In this section, we introduce our extension work that adopts cosine similarity as the definition of structural similarity.

The cosine similarity~$\sigma_c(u,v)$ between two vertices~$u$ and~$v$ is defined as in~\cite{xyfs-kdd-2007}:
\begin{itemize}
  \item if~$(u,v)\in E$, then~$$\sigma_c(u,v)=\frac{|N[u]\cap N[v]|}{\sqrt{d[u]\cdot d[v]}};$$
  \item if~$(u,v)\notin E$, then~$$\sigma_c(u,v)=0.$$
\end{itemize}

To see how it follows the definition of cosine similarity, we can construct a~$n$-dimension vector~$\vec p$ and let~$\vec p[i]=1$ if the~$i$-th vertex is in~$N[u]$ and~$p[i]=0$ otherwise. Another~$n$-dimension vector~$\vec q$ can also be constructed in the same way based on~$N[v]$. Then~$|N[u]\cap N[v]|=\vec p\cdot \vec q$,~$d[u] = ||\vec p||$, and~$d[v]=||\vec q||$. Therefore,~$\sigma_c(u,v)$ equals the cosine similarity between these two vectors~$\vec p$ and~$\vec q$.

For cosine similarity, we have obtained similar results as Jaccard similarity. Specifically, we extend~$\dynelm$ and~$\dynstr$ algorithms {\em under cosine similarity} and their performances are guaranteed in the following theorem.
\begin{theorem}\label{thm:cosine-algo}
  The~$\dynelm$ algorithm admits all the same guarantees in Theorem~\ref{thm:basic-algo} {\em under cosine similarity}, and the~$\dynstr$ algorithm admits all the same guarantees in Theorem~\ref{thm:ultimate-algo} {\em under cosine similarity}.
\end{theorem}

We also adopt~$\rho$-approximate notion for approximate edge labeling. To complete the proof of Theorem~\ref{thm:cosine-algo}, we need to extend three main components in our algorithm, namely: {\em similarity estimator}, {\em update affordability}, and {\em distributed tracking on updates}. We proof similar results on these three components in the following subsections.

\subsection{Estimating Cosine Similarity}
In this subsection, we proof that the {\em sampling-based} method also works for estimating cosine similarity between two vertices. Let~$X_1,X_2,\dots,X_L$ be~$L$ independent instances of the random variable~$X$ defined in Section~\ref{sec:estimator}, and define~$\bar{X}=\sum_{i=1}^L X_i$. Recall in Equation~\ref{equation:estimator}, we have:
\begin{equation*}
  \sigma(u,v)=\frac{|N[u]\cap N[v]|}{|N[u]\cup N[v]|}=\frac{E[\bar{X}]}{2-E[\bar{X}]}.
\end{equation*}
Combining the fact that:
\begin{equation*}
  |N[u]\cap N[v]|=d[u]+d[v]-|N[u]\cup N[v]|.
\end{equation*}
We can compute that:
\begin{equation*}
  \begin{split}
    |N[u]\cap N[v]|&=d[u]+d[v]-|N[u]\cup N[v]| \\
    &=d[u]+d[v]-\frac{|N[u]\cap N[v]|}{\suv} \\
    &=d[u]+d[v]-\frac{(2-E[\bar{X}])|N[u]\cap N[v]|}{E[\bar{X}]}.
  \end{split}
\end{equation*}
As a result, we have:
\begin{equation}
  |N[u]\cap N[v]| = \frac{1}{2}(\du + \dv)E[\bar{X}].
\end{equation}
Thus,~$\frac{1}{2}(\du + \dv)\bar{X}$ can serve as an estimator for the size of intersection between two neighbourhoods. To compute the cosine similarity~$\sigma_c(u,v)$, we have:
\begin{equation}
  \sigma_c(u,v)=\frac{|N[u]\cap N[v]|}{\sqrt{\du\cdot \dv}}=\frac{(\du+\dv)E[\bar{X}]}{2\sqrt{\du\cdot\dv}}.
\end{equation}

Let~$d_{\min}(u,v)=\min\{\du,\dv\}$ and~$d_{\max}(u,v)=\max\{\du,\dv\}$. Then, we estimate the cosine similarity in the following manners:
\begin{itemize}
  \item if $d_{\min}(u,v)<\eps^2 d_{\max}(u,v)$, edge~$(u,v)$ can directly be labelled as dissimilar;
  \item otherwise, define~$\tilde{\sigma}_c(u,v)=\frac{(\du+\dv)\bar{X}}{2\sqrt{\du\cdot\dv}}$ and use it as an estimator for~$\sigma_c(u,v)$.
\end{itemize}

For the first case, we prove its correctness with the following lemma:
\begin{lemma}\label{lemma:eps2}
  For an edge~$(u,v)$, if~$d_{\min}(u,v)<\eps^2 d_{\max}(u,v)$, then~$\sigma_c(u,v)<\eps$.
\end{lemma}
\begin{proof}
  The cosine similarity between~$u$ and~$v$ can be upper bounded by:
  \begin{equation*}
    \begin{split}
      \sigma_c(u,v)&=\frac{|N[u]\cap N[v]|}{\sqrt{d[u]\cdot d[v]}} \\
      &\leq \frac{d_{\min}(u,v)}{\sqrt{d_{\min}(u,v)\cdot d_{\max}(u,v)}} \\
      &=\sqrt{\frac{d_{\min}(u,v)}{d_{\max}(u,v)}} < \eps.
    \end{split}
  \end{equation*}
\end{proof}

For the second case, we can guarantee the quality of our estimator by the theorem below.
\begin{theorem}
  Suppose $\eps^2 d_{\max}(u,v)\leq d_{\min}(u,v)\leq d_{\max}(u,v)$, by setting~$L=\frac{(\eps^2+1)^2}{8\eps^2\Delta^2}\ln(\frac{2}{\delta})$, we have~$\Pr[|\tilde{\sigma}_c(u,v)-\sigma_c(u,v)|>\Delta] < \delta$.
\end{theorem}
\begin{proof}
  In this case, we can result in the following bound:
  \begin{equation*}
    \begin{split}
      \frac{\du+\dv}{\sqrt{\du\cdot\dv}}&=\frac{d_{\min}(u,v)+d_{\max}(u,v)}{\sqrt{d_{\min}(u,v)\cdot d_{\max}(u,v)}} \\
      &=\sqrt{\frac{d_{\min}(u,v)}{d_{\max}(u,v)}} + \sqrt{\frac{d_{\max}(u,v)}{d_{\min}(u,v)}} \\
      &\leq \eps + \frac{1}{\eps}.
    \end{split}
  \end{equation*}
  The final inequality results from the fact that~$1\leq\sqrt{\frac{d_{\max}(u,v)}{d_{\min}(u,v)}}\leq\frac{1}{\eps}$.

  Observe that
  \begin{equation*}
    \begin{split}
      \Pr[|\tilde{\sigma}_c(u,v)-\sigma_c(u,v)|>\Delta]&=\Pr[\frac{\du+\dv}{2\sqrt{\du\cdot\dv}}|\bar{X}-E[\bar{X}]|>\Delta] \\
      &\leq\Pr[|\bar{X}-E[\bar{X}]|>\frac{2\Delta}{\eps + \frac{1}{\eps}}].
    \end{split}
  \end{equation*}
  By the Hoeffding Bound~\cite{hoeffding-1994}, when $L = \frac{(\epsilon + \frac{1}{\epsilon})^2}{8\Delta^2}\ln (\frac{2}{\delta})$, we have $\Pr [ |\bar{X} - E[\bar{X}]| > \frac{2\Delta}{\epsilon + \frac{1}{\epsilon}} ]\leq \delta$.
\end{proof}

Based on the above estimation for cosine similarity, we can utilize the same~$(\Delta,\delta)$-strategy to label edges. Specifically, in the extension, we adopt $\strategy$ to determine edge labels as in the original work.

\subsection{Update Affordability}
When an update~$(u,w)$ occurs, the cosine similarity of the affected edges will be affected in a similar way to Jaccard similarity as in Section~\ref{subsec:affordability}.

\begin{observation}\label{obs:cosine}
  Consider an update of edge~$(u,w)$, and an arbitrary affected edge~$(u,v)$,
  let~$\sigma_c(u,v)=\frac{|N[u]\cap N[v]|}{\sqrt{\du\dv}}=
    \frac{a}{\sqrt{\du\dv}}$
  be the structural similarity between~$u$ and~$v$ before the update of~$(u,w)$, where~$a=||N[u]\cap N[v]|$.
  The effect of such update $(u,w)$ can be:
  \begin{itemize}
    \item Case~$1$: an insertion of~$(u,w)$,
          \begin{itemize}
            \item if~$w\in N[v]$,~$\suvc$ is increased to~$(a + 1)/\sqrt{(\du+1)\dv}$;
            \item if~$w\notin N[v]$,~$\suvc$ is decreased to~$a/\sqrt{(\du+1)\dv}$.
          \end{itemize}
    \item Case~$2$: an deletion of~$(u,w)$,
          \begin{itemize}
            \item if~$w\in N[v]$,~$\suvc$ is decreased to~$(a - 1)/\sqrt{(\du-1)\dv}$;
            \item if~$w\notin N[v]$,~$\suvc$ is increased to~$a/\sqrt{(\du-1)\dv}$.
          \end{itemize}
  \end{itemize}
  Symmetric changes occur with each edge~$(v,w)$ incident to~$w$.
\end{observation}

Like Jaccard similarity, we can also calculate update affordability for cosine similarity. However, here for each edge $(u,v)\in E$ we divide the calculation of update affordability into two cases;
\begin{itemize}
  \item $d_{\min}(u,v)\geq 0.81\eps^2\cdot d_{\max}(u,v)$, and
  \item $d_{\min}(u,v) < 0.81\eps^2\cdot d_{\max}(u,v)$
\end{itemize}
The update affordability for each edge $(u,v)$ is computed by the following lemmas.

\begin{lemma}\label{lemma:cosine-dis-sim}
  If an edge~$(u,v)$ is labelled as dissimilar by the $\strategy$, then with probability at least~$1-\delta$,~$(u,v)$ can afford at least~$k=\lf 0.45\rho\eps^2\cdot d_{\max}(u,v)\rf$ affecting updates if $\dmin\geq 0.81\eps^2\cdot \dmax$ or $k = \lf 0.19\eps^2\cdot\dmax \rf$ affecting updates if $\dmin < 0.81\eps^2\cdot \dmax$ before its label flips from dissimilar to similar.
\end{lemma}
\begin{proof}
  First consider the case where $d_{\min}(u,v)\geq 0.81\cdot d_{\max}(u,v)$.
  Initially, let~$a=|N[u]\cap N[v]|\geq 2$. Since~$(u,v)$ is labelled as dissimilar by $\strategy$, we have~$\tilde{\sigma}_c(u,v)<\eps$. Then with probability at least~$1-\delta$,~$\sigma_c(u,v)<(1+\half\rho)\eps$. Without loss of generality, we first consider an update incident to~$u$. If~$\du=a$, then~$\forall w\in N[u]$,~$w\in N[v]$. Thus deleting edge~$(u,w)$ will only decrease~$\suvc$. That is to say, in order to increase~$\suvc$, we need to consider insertions of edge~$(u,w)$. Otherwise if~$\du > a$, since~$\du$ and~$a$ are both integers, we must have~$\du\geq a+1$. Therefore,
  \begin{equation*}
    \du\geq a+1>\frac{2a^2+2a+1}{2a+1}.
  \end{equation*}
  It can be easily verified that
  \begin{equation*}
    \frac{a+1}{\sqrt{(\du+1)\dv}}>\frac{a}{\sqrt{(\du-1)\dv}}.
  \end{equation*}
  As a result, in considering the minimum number of affecting updates to cause$(u,v)$'s label flips from dissimilar to similar, we only need to focus on the edge insertions that increase~$\suvc$. Let~$k=\lf 0.45\rho\eps^2\cdot d_{\max}(u,v)\rf$ be the total number of such updates, and~$t$ be the number of such affecting updates that involve~$u$, by the first bullet of Case~$1$ in Observation~\ref{obs:cosine}:
  \begin{equation*}
    \begin{split}
      \suvc&=\frac{a+k}{\sqrt{(\du+t)(\dv+k-t)}}<\frac{a+k}{\sqrt{\du\dv}} \\
      &=\frac{a}{\sqrt{\du\dv}}+\frac{k}{\sqrt{\du\dv}} \\
      &\leq (1+\half \rho)\eps + \frac{\lf 0.45\rho\eps^2\cdot d_{\max}(u,v)\rf}{\sqrt{\du\dv}} \\
      &\leq (1+\half \rho)\eps + \frac{\half\cdot 0.9\rho\eps^2\dmax}{\sqrt{\du\dv}} \\
      &\leq (1+\half \rho)\eps + \frac{\half \rho\eps\sqrt{\du\dv}}{\sqrt{\du\dv}} \\
      &=(1+\rho)\eps
    \end{split}
  \end{equation*}
  The last inequality is because
  $$
    \sqrt{\du\dv}=\sqrt{\dmin\cdot \dmax}\geq\sqrt{0.81\eps^2\cdot d^2_{\max}(u,v)}=0.9\eps\cdot\dmax.
  $$
  Therefore, after~$k$ arbitrary affecting updates, the dissimilar label of~$(u,v)$ remains valid with probability at least~$1-\delta$.

  Then consider the case where $d_{\min}(u,v) < 0.81\eps^2\cdot d_{\max}(u,v)$. Note that by Lemma~\ref{lemma:eps2}, since $\dmin < \eps^2 \dmax$, $(u,v)$ will only be labelled as dissimilar. To get the minimum number of affecting updates $(u,v)$ can afford before its label flips, we only need to consider the updates that narrow the gap between $\dmin$ and $\dmax$. Suppose there are $t$ edge deletions incident to the vertex with the maximum degree and there are $k-t$ edge insertion incident to the vertex with the minimum degree, where $k=\lf 0.19\eps^2\cdot \dmax$. Then the degree after $k$ updates in total becomes:
  \begin{equation*}
    \begin{aligned}
       & d'_{\max}(u,v) = d_{\max}(u,v)-t    \\
       & d'_{\min}(u,v) = d_{\min}(u,v)+k-t.
    \end{aligned}
  \end{equation*}
  Current degrees satisfy the following inequality:
  \begin{equation*}
    \begin{aligned}
      \eps^2\cdot d'_{\max}(u,v) & =\eps^2d_{\max}(u,v)-\eps^2t                   \\
                                 & >d_{\min}(u,v)+0.19\eps^2\cdot d_{\max}(u,v)-t \\
                                 & \geq d_{\min}(u,v)+k-t=d'_{\min}(u,v)
    \end{aligned}
  \end{equation*}
  The first inequality is because $0.81\eps^2\cdot \dmax=(\eps^2-0.19\eps^2)\cdot \dmax>\dmin$. Therefore, after $k$ arbitrary affecting updates the dissimilar label of $(u,v)$ remains valid.
\end{proof}

\begin{lemma}\label{lemma:cosine-sim-dis}
  If an edge~$(u,v)$ is labelled as similar by the $\strategy$, then with probability at least~$1-\delta$,~$(u,v)$ can afford at least~$k=\lf \half\rho\eps\cdot d_{\min}(u,v)\rf$ affecting updates before its label flips from similar to dissimilar.
\end{lemma}
\begin{proof}
  Initially, let~$a=|N[u]\cap N[v]|\geq 2$. Since~$(u,v)$ is labelled as similar by $\strategy$, we have~$\tilde{\sigma}_c(u,v)\geq\eps$. Then with probability at least~$1-\delta$,~$\sigma_c(u,v)\geq(1-\half\rho)\eps$.
  Without loss of generality, we consider an update incident to $u$. Since we have $d[u]\geq a$ and $a\geq 2$, we have,
  $$
    d[u]\geq a>\frac{2a^2-2a+1}{2a-1}.
  $$
  Thus it can be easily verified that
  $$
    \frac{a-1}{\sqrt{(\du-1)\dv}} < \frac{a}{\sqrt{(\du+1)\dv}}.
  $$
  As a result in considering the minimum number of affecting updates to cause$(u,v)$'s label flips from similar to dissimilar, we only need to focus on the edge deletions that decrease~$\suvc$. Let~$k=\lf \half\rho\eps\cdot d_{\min}(u,v)\rf$ be the total number of such updates, and~$t$ be the number of such affecting updates that involve~$u$, by the first bullet of Case~$2$ in Observation~\ref{obs:cosine}:
  \begin{equation*}
    \begin{split}
      \suvc&=\frac{a-k}{\sqrt{(\du-t)(\dv-k+t)}}>\frac{a-k}{\sqrt{\du\dv}} \\
      &=\frac{a}{\sqrt{\du\dv}}-\frac{k}{\sqrt{\du\dv}} \\
      &\geq (1-\half \rho)\eps - \frac{\lf \half\rho\eps\cdot d_{\min}(u,v)\rf}{\sqrt{\du\dv}} \\
      &\geq (1-\half \rho)\eps - \frac{\half \rho\eps\sqrt{\du\dv}}{\sqrt{\du\dv}} \\
      &=(1-\rho)\eps
    \end{split}
  \end{equation*}
  Therefore, after~$k$ arbitrary affecting updates, the dissimilar label of~$(u,v)$ remains valid with probability at least~$1-\delta$.
\end{proof}

\subsection{Distributed Tracking on Updates}\label{subsec:cosine-tracking}

For an edge~$(u,v)$ incident to~$u$, we first put it into one of these two categories:
\begin{itemize}
  \item if~$d_{\min}(u,v)\geq 0.81\eps^2 \cdot d_{\max}(u,v)$, we track it by a~$\DT$ instance~$\DT(u,v)$ with tracking threshold set to
        \begin{equation}\label{eq:cosine-tau}
          \tau(u,v)=\lfloor 0.45\rho\eps^2\cdot d_{\max}(u,v) \rfloor + 1;
        \end{equation}
  \item if~$d_{\min}(u,v)<0.81\eps^2 \cdot d_{\max}(u,v)$, we track it by the another~$\DT$ instance~$\DT^*(u,v)$ with tracking threshold set to
        \begin{equation}
          \tau^*(u,v)=\lfloor 0.19\eps^2\cdot d_{\max}(u,v) \rfloor + 1;
        \end{equation}
\end{itemize}

The correctness is proven by Lemma~\ref{lemma:cosine-dis-sim} and~\ref{lemma:cosine-sim-dis}. Note that for the second case, instead of using the update affordability as tracking threshold directly, we set the gap between~$\eps^2 d_{\max}(u,v)$ and~$d_{\min}(u,v)$ as the tracking threshold for~$\DT^*(u,v)$ and track the number of affecting updates with~$\DT^*(u,v)$.

For~$\DT(u,v)$ and~$\DT^*(u,v)$, we organize them together with~$\DtHeap(u)$ on~$u$ as in Section~\ref{sec::prelim-dt}.

\subsection{Algorithm Procedure}
In this subsection, we first consider the~$\dynelm$ {\em under cosine similarity} algorithm.

The running process of~$\dynelm$ under cosine similarity is very like the one in Section~\ref{sec:dynelm} and it is outlined as follows:
\begin{itemize}
  \item \textbf{Step 1.} Initialize the set of label-flipping edges $\F \leftarrow \varnothing$; and increment
        $\cnt_u$ and $\cnt_w$ (by~$1$), respectively.
  \item \textbf{Step 2.} There are two cases:
        \begin{itemize}
          \item \underline{\em Case 1: this update is an insertion.} Insert~$(u, w)$ into~$G$ and
                label it by the $\strategy$. If~$(u,w)$ is labelled as similar, add~$(u,w)$ to~$\F$. Moreover, if~$(u,w)$ satisfies the first bullet in Section~\ref{subsec:cosine-tracking}, create $DT(u,w)$ with $\tau(u,w)$; otherwise create~$DT^*(u,w)$ with~$\tau^*(u,w)$.
          \item \underline{\em Case 2: this update is a deletion.} If $(u,w)$ is labelled as similar, add $(u,w)$ to $\F$. Delete $(u,w)$ from $G$; and delete $DT(u,w)$ or~$\DT^*(u,w)$.
        \end{itemize}
  \item \textbf{Step 3.} While there is a checkpoint-ready entry in $\DtHeap(u)$, pop the entry (from the top). Let $\DT(u,v)$ be the DT instance corresponding to this entry. Instruct~$u$ to inform the
        coordinator~$(u,v)$.
        When~$DT(u,v)$ is mature, relabel~$(u,v)$ by the $\strategy$. If its label flipped, add~$(u,v)$
        to~$\F$. Remove its entry from~$\DtHeap(v)$, and if it satisfies the first bullet in Section~\ref{subsec:cosine-tracking}, create~$\DT(u,v)$ with ~$\tau(u,v)$; otherwise, create~$\DT^*(u,v)$ with~$\tau^*(u,v)$.
        Repeat until there is no checkpoint-ready entry in $\DtHeap(u)$.
  \item \textbf{Step 4.} Perform a symmetric process of Step~3 and~4 for~$w$.
  \item \textbf{Step 5.} Return~$\F$, the set of edges whose labels flipped.
\end{itemize}

\subsubsection{Theoretical Analysis}\label{subsec:cosine-algo}
In this subsection, we analyze the details of the~$\dynelm$ algorithm under cosine similarity.

The implementation is the same as in Section~\ref{sec:basic-analysis}, and thus given a failure probability~$\delta^*$ and the total number of invocations of the $\strategy$~$\kappa$, the cost of each invocation of~$\strategy$ is bounded by~$O(\log n\cdot \log\frac{\kappa}{\delta^*})$. Since for an vertex~$u$, its neighbors are maintained in~$\DtHeap(u)$, the space consumption of each vertex is still bounded by~$O(1+\du)$. Thus the overall space consumption directly follows Lemma~\ref{lmm:space}. And the correctness and failure probability directly follows Lemma~\ref{lmm:probability-bound}.

Now it suffices to bound the amortized update cost. For each~$\DT$ instance, the maturity cost maintains the same as in the original work, which is bounded~$O(\log^2 n +\log n\cdot \log\frac{\kappa}{\delta^*})$, where~$\kappa$ is number of invocations of~$\strategy$. Consider an update~$(u,w)$, the same crucial observation is that the update of $(u,w)$ can {\em only} contribute (via a counter increment) to the maturity of the DT instances of its affected edges,
which exist at the {\em current moment}. Let $\DT(u,v^*_1)$ and~$\DT^*(u,v^*_2)$ be the instance with the {\em smallest} threshold value $\tau(u,v^*_1)$ and~$\tau^*(u,v^*_2)$ among all the affected DT instances at the current moment, respectively.

For~$\DT(u,v^*_1)$, we have~$d_{\min}(u,v^*_1)\geq 0.81\eps^2 \cdot d_{\max}(u,v^*_1)$ at the moment when $\DT(u,v^*_1)$ is created. Let $d'[u]$ be the degree of $u$ when $\DT(u,v^*_1)$ was created. We claim that the degree of $u$, $d[u]$, at the current moment is at most $d'[u] + \tau(u,v^*_1)$ for the same reason as in Section~\ref{sec:basic-analysis}. Since~$d_{\min}(u,v^*_1)\geq 0.81\eps^2 \cdot d_{\max}(u,v^*_1)$, we have:
\begin{equation*}
  \begin{split}
    \tau(u,v^*_1)&=\lf 0.45\rho\eps^2 d_{\max}(u,v^*_1)\rf +1 \geq 0.45 \rho\eps^2 d_{\max}(u,v^*_1) \\
    &\geq 0.45\rho\eps^2 d'[u].
  \end{split}
\end{equation*}
Thus,
\begin{equation}
  \du\leq d'[u]+\tau(u,v^*_1)\leq \frac{\tau(u,v^*_1)}{0.45\rho\eps^2}+\tau(u,v^*_1)=O(1)\cdot \tau(u,v^*_1).
\end{equation}

For~$\DT^*(u,v^*_2)$, we have~$d_{\min}(u,v^*_2)<0.81\eps^2\cdot d_{\max}(u,v^*_2)$ at the moment when $\DT^*(u,v^*_2)$ is created. Let $d^{''}[u]$ be the degree of $u$ when $\DT^*(u,v^*_2)$ was created. We claim that the degree of $u$, $d[u]$, at the current moment is at most $d^{''}[u] + \tau(u,v^*_2)$ for the same reason. Therefore
\begin{equation*}
  \tau^*(u,v^*_2)=\lf 0.19\eps^2\cdot d_{\max}(u,v)\rf +1\geq 0.19\eps^2\cdot d_{\max}(u,v)\geq 0.19\eps^2\cdot d^{''}[u].
\end{equation*}
Then we have
\begin{equation}
  \du\leq d^{''}[u]+\tau^*(u,v^*_2)\leq \frac{\tau^*(u,v^*_2)}{0.19\eps^2}+\tau^*(u,v^*_2)=O(1)\cdot \tau^*(u,v^*_2).
\end{equation}

Furthermore, as each of the affected $\DT(u,v)$ and~$\DT^*(u,v)$ requires at least $\tau(u,v)$ and~$\tau^*(u,v)$ affecting updates to mature, respectively,
the current update of $(u,w)$ is actually accounted for only $\frac{1}{\tau(u,v)}$ of the cost of the $\DT(u,v)$ maturity as well as the following edge re-labeling cost and~$\frac{1}{\tau^*(u,v)}$ of the cost corresponding to~$\DT^*(u,v)$. Summing over all neighbors of~$u$, the amortized cost of an update~$(u,w)$ is bounded by:
\begin{equation}
  \begin{split}
    &\sum_{v\in N[u]:\text{$v$ belongs to~$\DT(u,v)$}}\frac{O(\log^2 n + \log n \cdot \log \frac{\kappa}{\delta^*})}{\tau(u,v)}+ \\
    &\sum_{v\in N[u]:\text{$v$ belongs to~$\DT^*(u,v)$}}\frac{O(\log^2 n + \log n \cdot \log \frac{\kappa}{\delta^*})}{\tau^*(u,v)} \\
    &\leq (\frac{\du}{\tau(u,v^*_1)}+\frac{\du}{\tau^*(u,v^*_2)})\cdot O(\log^2 n + \log n \cdot \log \frac{\kappa}{\delta^*}) \\
    &=O(\log^2 n + \log n \cdot \log \frac{\kappa}{\delta^*})
  \end{split}
\end{equation}
By symmetry, all affected~$\DT$ instances corresponding to~$w$ also charge~$O(\log^2 n + \log n \cdot \log \frac{\kappa}{\delta^*})$ cost to the update of edge~$(u,w)$. Combining with the fact that the number of invocations of~$\strategy$ is bounded by the number of updates, the amortized cost of each update remains~$O(\log^2 n + \log n \cdot \log \frac{M}{\delta^*})$, where~$M$ is the number of updates in a sequence.

For the~$\dynstr$ {\em under cosine similarity} algorithm, it follows the same procedures as~$\dynstr$ algorithm except for the maintenance of edge labelling. From the above analysis, it is easy to be seen that all guarantees in the~$\dynstr$ algorithm are admitted. Thus, Theorem~\ref{thm:cosine-algo} is proven.

}

\section{Experiments}\label{sec:exp}

\noindent
\textbf{Datasets.}
We deploy~$15$ real datasets in the experiments.
Detailed descriptions of all these~$15$ datasets can be found at the Stanford Network Analysis Project (SNAP)\footnote{\url{http://snap.stanford.edu/data/index.html}}.
We pre-process each of these datasets in the following way: (i) treat the graph as undirected; (ii) remove all self-loops and duplicate edges; and (iii) relabel vertex identifiers to be in the set~$\{1,\ldots,n\}$.
The meta information of all processed datasets are listed in Table~\ref{tab:meta-info}.
The first five datasets highlighted in bold are chosen as representatives: their vertex counts increase roughly geometrically (factor two), and each has a reasonable average degree;
and they are used to explore both clustering effectiveness and efficiency of the algorithms, with varying parameter settings.
For easy reference, we rename the five representatives as {\em Slashdot}, {\em Notre}, {\em Google}, {\em Wiki} and {\em LiveJ}, respectively.
In addition, the last dataset in bold, renamed as {\em Twitter}, with 1.2 billion edges, is further used to study the scalability of our proposed algorithms.
The remaining nine datasets are then listed, in ascending order of~$n$.

\begin{table}[]
	\caption{Dataset Meta Information and Memory Footage over the Whole Update Sequence (K = $10^3$, M = $10^6$ and B = $10^9$)}
	\label{tab:meta-info}
	\resizebox{\linewidth}{!}{%
		\begin{tabular}{|l|c|c||c||c|c|c|c|}
			\hline
			\multicolumn{1}{|c|}{\multirow{3}{*}{\textbf{Datasets}}} & \multirow{3}{*}{\textbf{\#Vertices}} & \multirow{3}{*}{\textbf{\#Edges}} & \multirow{3}{*}{\textbf{\#Updates}} & \multicolumn{4}{c|}{\multirow{2}{*}{\textbf{Memory Footage (GigaBytes)}}}                                                                             \\
			\multicolumn{1}{|c|}{}                                   &                                      &                                   &                                     & \multicolumn{4}{c|}{}                                                                                                                                 \\ \cline{5-8}
			\multicolumn{1}{|c|}{}                                   &                                      &                                   &                                     & \textbf{$\bm{\dynelm}$}                                                   & \textbf{$\bm{\dynstr}$} & \textbf{$\bm{\pscan}$} & \textbf{$\bm{\hscan}$} \\ \hline
			\textit{\textbf{soc-Slashdot0811}}                       & 77.3K                                & 469K                              & 4.69M                               & 0.50                                                                      & 0.58                    & 0.47                   & 0.82                   \\ \hline
			\textit{\textbf{web-NotreDame}}                          & 326K                                 & 1.09M                             & 10.9M                               & 1.17                                                                      & 1.87                    & 1.10                   & 1.93                   \\ \hline
			\textit{\textbf{web-Google}}                             & 876K                                 & 4.32M                             & 43.2M                               & 4.51                                                                      & 6.23                    & 3.62                   & 7.54                   \\ \hline
			\textit{\textbf{wiki-topcats}}                           & 1.79M                                & 25.4M                             & 254M                                & 25.79                                                                     & 29.39                   & (26.82)                & (51.66)                \\ \hline
			\textit{\textbf{soc-LiveJournal1}}                       & 4.85M                                & 42.9M                             & 429M                                & 43.51                                                                     & 58.85                   & (44.13)                & (87.70)                \\ \hline
			\textit{email-Eu-core}                                   & 0.99K                                & 16.1K                             & 161K                                & 0.02                                                                      & 0.02                    & 0.02                   & 0.03                   \\ \hline
			\textit{ca-GrQc}                                         & 5.24K                                & 14.5K                             & 145K                                & 0.02                                                                      & 0.03                    & 0.02                   & 0.03                   \\ \hline
			\textit{ca-CondMat}                                      & 23.1K                                & 93.4K                             & 934K                                & 0.10                                                                      & 0.16                    & 0.09                   & 0.17                   \\ \hline
			\textit{soc-Epinions1}                                   & 75.8K                                & 406K                              & 4.06M                               & 0.43                                                                      & 0.54                    & 0.41                   & 0.71                   \\ \hline
			\textit{dblp}                                            & 317K                                 & 1.05M                             & 10.5M                               & 1.11                                                                      & 1.97                    & 1.05                   & 1.86                   \\ \hline
			\textit{amazon0601}                                      & 403K                                 & 2.44M                             & 24.4M                               & 2.54                                                                      & 4.08                    & 2.43                   & 4.27                   \\ \hline
			\textit{soc-Pokec}                                       & 1.63M                                & 22.3M                             & 223M                                & 22.44                                                                     & 24.90                   & (18.26)                & (30.85)                \\ \hline
			\textit{as-skitter}                                      & 1.70M                                & 11.1M                             & 111M                                & 11.71                                                                     & 14.27                   & (8.47)                 & (42.32)                \\ \hline
			\textit{wiki-Talk}                                       & 2.39M                                & 4.66M                             & 46.6M                               & 5.48                                                                      & 7.29                    & (4.71)                 & (24.06)                \\ \hline \hline
			\textit{\textbf{twitter-2010}}                           & 41.65M                               & 1.20B                             & 1.32B                               & 204.09                                                                    & 257.46                  & (135.11)               & (273.32)               \\ \hline
		\end{tabular}
	}
\end{table}

\begin{table*}[]
	\caption{Approximate Clustering Quality {\newblue under Jaccard Similarity}: Mis-Labelled Rate, Overall Clustering Quality (ARI) and Individual Cluster Quality}
	\label{tab:quality}
	\resizebox{\linewidth}{!}{
		\begin{tabular}{|c|c|c|c|c||c|c|c|c||c|c|c|c||c|c|c|c||c|c|c|c||c|c|c|c|}
			\hline
			\multirow{2}{*}{\textbf{}}                                                         & \multicolumn{4}{c||}{\textit{\textbf{Slashdot}} $(\eps = 0.15)$} & \multicolumn{4}{c||}{\textit{\textbf{Notre}} $(\eps = 0.19)$} & \multicolumn{4}{c||}{\textit{\textbf{Google}} $(\eps = 0.15)$} & \multicolumn{4}{c||}{\textit{\textbf{Wiki}} $(\eps = 0.19)$} & \multicolumn{4}{c||}{\textit{\textbf{LiveJ}} $(\eps = 0.6)$} & \multicolumn{4}{c|}{\textit{\textbf{Twitter}} $(\eps = 0.2)$}                                                                                                                                                                                                                                                                                                                                                                                                                                                                                \\ \cline{2-25}
			                                                                                   & \multicolumn{2}{c|}{$\mathbf{\rho = 0.01}$}                      & \multicolumn{2}{c||}{$\mathbf{\rho = 0.5}$}                   & \multicolumn{2}{c|}{$\mathbf{\rho = 0.01}$}                    & \multicolumn{2}{c||}{$\mathbf{\rho = 0.5}$}                  & \multicolumn{2}{c|}{$\mathbf{\rho = 0.01}$}                  & \multicolumn{2}{c||}{$\mathbf{\rho = 0.5}$}                   & \multicolumn{2}{c|}{$\mathbf{\rho = 0.01}$} & \multicolumn{2}{c||}{$\mathbf{\rho = 0.5}$} & \multicolumn{2}{c|}{$\mathbf{\rho = 0.01}$} & \multicolumn{2}{c||}{$\mathbf{\rho = 0.5}$} & \multicolumn{2}{c|}{$\mathbf{\rho = 0.01}$} & \multicolumn{2}{c|}{$\mathbf{\rho = 0.5}$}                                                                                                                                                                                             \\ \hline \hline
			\multicolumn{1}{|l|}{\textbf{\%mis-labelled}}                                      & \multicolumn{2}{c|}{0.02\%}                                      & \multicolumn{2}{c||}{2.37\%}                                  & \multicolumn{2}{c|}{0.10\%}                                    & \multicolumn{2}{c||}{5.86\%}                                 & \multicolumn{2}{c|}{0.16\%}                                  & \multicolumn{2}{c||}{8.74\%}                                  & \multicolumn{2}{c|}{0.04\%}                 & \multicolumn{2}{c||}{1.82\%}                & \multicolumn{2}{c|}{0.14\%}                 & \multicolumn{2}{c||}{6.33\%}                & \multicolumn{2}{c|}{0.01\%}                 & \multicolumn{2}{c|}{0.07\%}                                                                                                                                                                                                            \\ \hline \hline
			\textbf{ARI}                                                                       & \multicolumn{2}{c|}{.996386}                                     & \multicolumn{2}{c||}{.971871}                                 & \multicolumn{2}{c|}{.999748}                                   & \multicolumn{2}{c||}{.962753}                                & \multicolumn{2}{c|}{.998872}                                 & \multicolumn{2}{c||}{.970845}                                 & \multicolumn{2}{c|}{.999933}                & \multicolumn{2}{c||}{.989068}               & \multicolumn{2}{c|}{.999767}                & \multicolumn{2}{c||}{.999470}               & \multicolumn{2}{c|}{.994647}                & \multicolumn{2}{c|}{.976576}                                                                                                                                                                                                           \\ \hline \hline
			\multirow{2}{*}{\textbf{\begin{tabular}[c]{@{}c@{}}Top-k\\ Clusters\end{tabular}}} & \multicolumn{4}{c||}{\textbf{Indv. Cluster Quality}}             & \multicolumn{4}{c||}{\textbf{Indv. Cluster Quality}}          & \multicolumn{4}{c||}{\textbf{Indv. Cluster Quality}}           & \multicolumn{4}{c||}{\textbf{Indv. Cluster Quality}}         & \multicolumn{4}{c||}{\textbf{Indv. Cluster Quality}}         & \multicolumn{4}{c|}{\textbf{Indv. Cluster Quality}}                                                                                                                                                                                                                                                                                                                                                                                                                                                                                          \\ \cline{2-25}
			                                                                                   & \textbf{min}                                                     & \textbf{avg}                                                  & \textbf{min}                                                   & \textbf{avg}                                                 & \textbf{min}                                                 & \textbf{avg}                                                  & \textbf{min}                                & \textbf{avg}                                & \textbf{min}                                & \textbf{avg}                                & \textbf{min}                                & \textbf{avg}                               & \textbf{min} & \textbf{avg} & \textbf{min} & \textbf{avg} & \textbf{min} & \textbf{avg} & \textbf{min} & \textbf{avg} & \textbf{min}     & \textbf{avg} & \textbf{min} & \textbf{avg}     \\ \hline
			1                                                                                  & .987                                                             & \underline{.987}                                              & .961                                                           & .961                                                         & 1.00                                                         & 1.00                                                          & 1.00                                        & 1.00                                        & 1.00                                        & 1.00                                        & .999                                        & .999                                       & 1.00         & 1.00         & .999         & .999         & .997         & .997         & .989         & .989         & .990             & .990         & .952         & \underline{.952} \\ \hline
			5                                                                                  & .987                                                             & .996                                                          & .961                                                           & .988                                                         & 1.00                                                         & 1.00                                                          & 1.00                                        & 1.00                                        & .998                                        & .999                                        & .975                                        & .987                                       & .998         & .999         & .953         & .990         & .996         & .997         & .989         & .994         & .990             & .995         & .911         & .963             \\ \hline
			10                                                                                 & .987                                                             & .998                                                          & .961                                                           & .994                                                         & 1.00                                                         & 1.00                                                          & 1.00                                        & 1.00                                        & .987                                        & .998                                        & .880                                        & .980                                       & .992         & .998         & .953         & .990         & .995         & .997         & .983         & .992         & .990             & .997         & .907         & .970             \\ \hline
			20                                                                                 & .987                                                             & .999                                                          & .961                                                           & .997                                                         & 1.00                                                         & 1.00                                                          & .997                                        & .999                                        & .987                                        & .998                                        & .880                                        & .987                                       & .987         & .998         & .650         & .965         & .988         & .998         & .929         & .989         & .990             & .997         & .907         & .979             \\ \hline
			50                                                                                 & .987                                                             & .999                                                          & .961                                                           & .999                                                         & 0.977                                                        & .999                                                          & .761                                        & .989                                        & .987                                        & .998                                        & .880                                        & .989                                       & .985         & .998         & {\bf .345}   & .967         & .952         & .997         & .929         & .990         & {.853}           & .994         & .826         & .979             \\ \hline
			100                                                                                & .987                                                             & .999                                                          & .961                                                           & .999                                                         & 0.877                                                        & .998                                                          & .761                                        & .995                                        & .940                                        & .998                                        & \underline{\bf .116}                        & .982                                       & .983         & .999         & {.345}       & .979         & .909         & .997         & .900         & .993         & \underline{.853} & .996         & .826         & .984             \\ \hline
		\end{tabular}
	}
\end{table*}

\begin{table*}[]
	\caption{\newblue Approximate Clustering Quality under Cosine Similarity: Mis-Labelled Rate, Overall Clustering Quality (ARI) and Individual Cluster Quality}
	\label{tab:cosinequality}
	\resizebox{\linewidth}{!}{
		\begin{tabular}{|c|c|c|c|c|c|c|c|c|c|c|c|c|c|c|c|c|c|c|c|c|}
			\hline
			\multirow{2}{*}{}                                                                   & \multicolumn{4}{c|}{\textit{\textbf{Slashdot}} ($\eps=0.3$)} & \multicolumn{4}{c|}{\textit{\textbf{Notre}} ($\eps=0.36$)} & \multicolumn{4}{c|}{\textit{\textbf{Google}} ($\eps=0.3$)} & \multicolumn{4}{c|}{\textit{\textbf{Wiki}} ($\eps=0.34$)} & \multicolumn{4}{c|}{\textit{\textbf{LiveJ}} ($\eps=0.67$)}                                                                                                                                                                                                                                                                                                                                                                                \\ \cline{2-21}
			                                                                                    & \multicolumn{2}{c|}{$\mathbf{\rho=0.01}$}                    & \multicolumn{2}{c|}{$\mathbf{\rho=0.1}$}                   & \multicolumn{2}{c|}{$\mathbf{\rho=0.01}$}                  & \multicolumn{2}{c|}{$\mathbf{\rho=0.1}$}                  & \multicolumn{2}{c|}{$\mathbf{\rho=0.01}$}                  & \multicolumn{2}{c|}{$\mathbf{\rho=0.1}$} & \multicolumn{2}{c|}{$\mathbf{\rho=0.01}$} & \multicolumn{2}{c|}{$\mathbf{\rho=0.1}$} & \multicolumn{2}{c|}{$\mathbf{\rho=0.01}$} & \multicolumn{2}{c|}{$\mathbf{\rho=0.1}$}                                                                                                                                                       \\ \hline
			\textbf{\%mis-labelled}                                                             & \multicolumn{2}{c|}{0.11\%}                                  & \multicolumn{2}{c|}{0.83\%}                                & \multicolumn{2}{c|}{0.19\%}                                & \multicolumn{2}{c|}{2.91\%}                               & \multicolumn{2}{c|}{0.33\%}                                & \multicolumn{2}{c|}{2.61\%}              & \multicolumn{2}{c|}{0.08\%}               & \multicolumn{2}{c|}{0.61\%}              & \multicolumn{2}{c|}{0.19\%}               & \multicolumn{2}{c|}{1.40\%}                                                                                                                                                                    \\ \hline
			\textbf{ARI}                                                                        & \multicolumn{2}{c|}{.989941}                                 & \multicolumn{2}{c|}{.978194}                               & \multicolumn{2}{c|}{.984992}                               & \multicolumn{2}{c|}{.943016}                              & \multicolumn{2}{c|}{.969585}                               & \multicolumn{2}{c|}{.709121}             & \multicolumn{2}{c|}{.975939}              & \multicolumn{2}{c|}{.794518}             & \multicolumn{2}{c|}{.973321}              & \multicolumn{2}{c|}{.924053}                                                                                                                                                                   \\ \hline
			\multirow{2}{*}{\textbf{\begin{tabular}[c]{@{}c@{}}Top-K \\ Clusters\end{tabular}}} & \multicolumn{4}{c|}{\textbf{Indv. Cluster Quality}}          & \multicolumn{4}{c|}{\textbf{Indv. Cluster Quality}}        & \multicolumn{4}{c|}{\textbf{Indv. Cluster Quality}}        & \multicolumn{4}{c|}{\textbf{Indv. Cluster Quality}}       & \multicolumn{4}{c|}{\textbf{Indv. Cluster Quality}}                                                                                                                                                                                                                                                                                                                                                                                       \\ \cline{2-21}
			                                                                                    & \textbf{min}                                                 & \textbf{avg}                                               & \textbf{min}                                               & \textbf{avg}                                              & \textbf{min}                                               & \textbf{avg}                             & \textbf{min}                              & \textbf{avg}                             & \textbf{min}                              & \textbf{avg}                             & \textbf{min} & \textbf{avg} & \textbf{min} & \textbf{avg} & \textbf{min} & \textbf{avg} & \textbf{min} & \textbf{avg} & \textbf{min} & \textbf{avg} \\ \hline
			1                                                                                   & .981                                                         & .981                                                       & .949                                                       & .949                                                      & .971                                                       & .971                                     & .867                                      & .867                                     & .992                                      & .992                                     & .716         & .716         & .995         & .995         & .833         & .833         & .993         & .993         & .937         & .937         \\ \hline
			5                                                                                   & .958                                                         & .979                                                       & .853                                                       & .916                                                      & .890                                                       & .966                                     & .710                                      & .855                                     & .959                                      & .983                                     & .716         & .754         & .900         & .952         & .550         & .742         & .817         & .925         & .437         & .860         \\ \hline
			10                                                                                  & .958                                                         & .989                                                       & .435                                                       & .846                                                      & .890                                                       & .980                                     & .710                                      & .902                                     & .958                                      & .980                                     & .423         & .696         & .900         & .969         & .550         & .773         & .817         & .961         & .371         & .852         \\ \hline
			20                                                                                  & .857                                                         & .985                                                       & .000                                                       & .787                                                      & .890                                                       & .988                                     & .710                                      & .929                                     & .808                                      & .961                                     & .423         & .681         & .557         & .949         & .288         & .709         & .817         & .978         & .371         & .877         \\ \hline
			50                                                                                  & .545                                                         & .980                                                       & .000                                                       & .790                                                      & .782                                                       & .989                                     & .380                                      & .891                                     & .767                                      & .957                                     & .177         & .656         & .557         & .946         & .276         & .704         & .182         & .966         & .177         & .862         \\ \hline
			100                                                                                 & .545                                                         & .986                                                       & .000                                                       & .817                                                      & .782                                                       & .992                                     & .197                                      & .879                                     & .509                                      & .956                                     & .026         & .607         & .439         & .954         & .057         & .678         & .182         & .978         & .177         & .888         \\ \hline
		\end{tabular}
	}
\end{table*}


\begin{figure*}
	\begin{tabular}{cc}
		\includegraphics[width=0.4\textwidth]{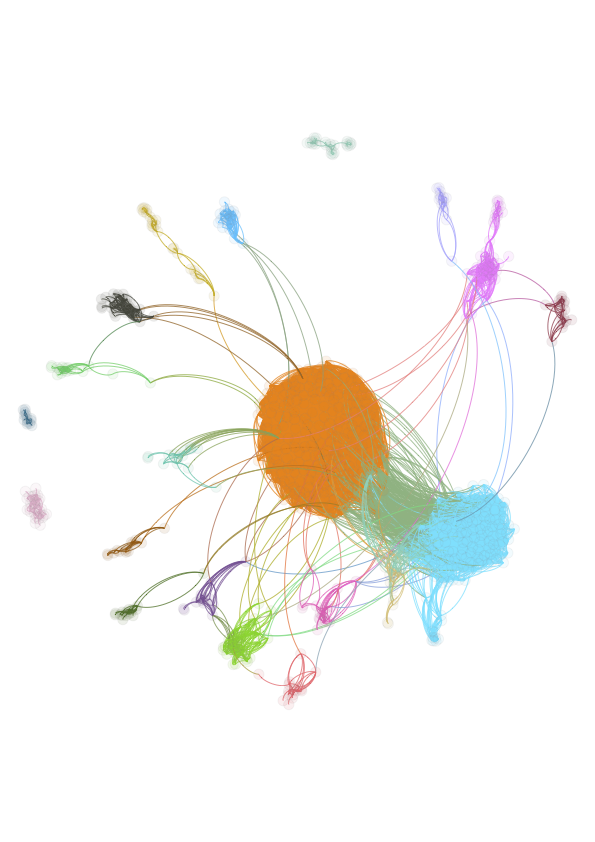} &
		\includegraphics[width=0.4\textwidth]{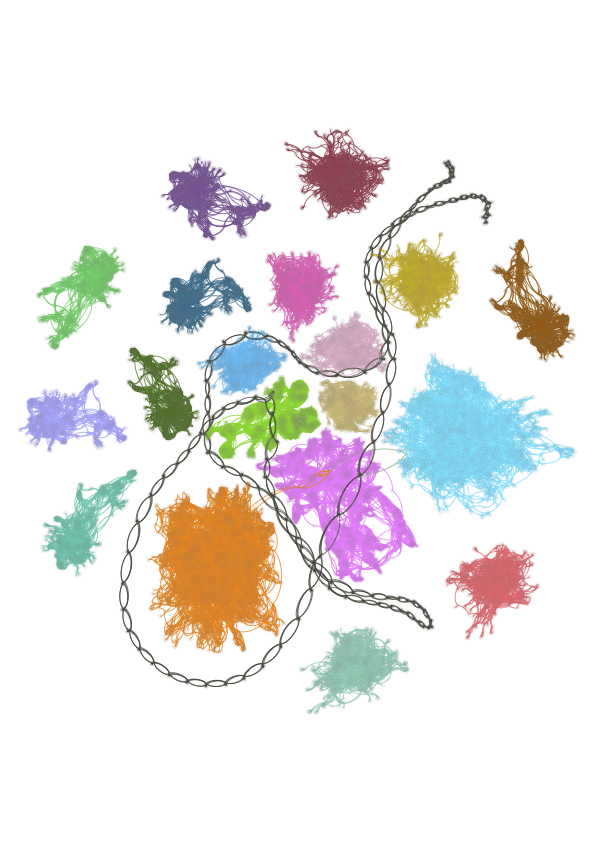}                                       \\
		(a) \textit{Slashdot} ($\eps=0.15$)                                         & (b) \textit{Notre} ($\eps=0.19$) \\
		\includegraphics[width=0.4\textwidth]{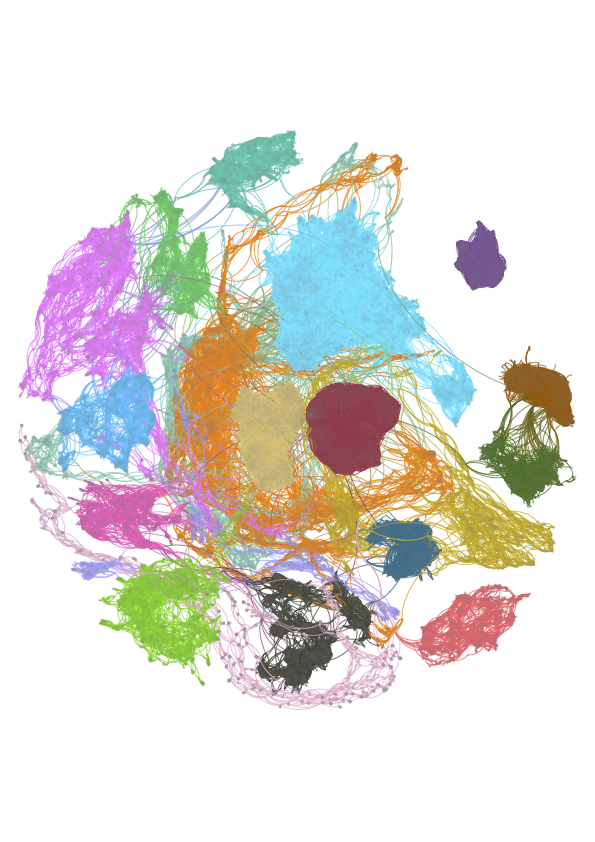}     &
		\includegraphics[width=0.4\textwidth]{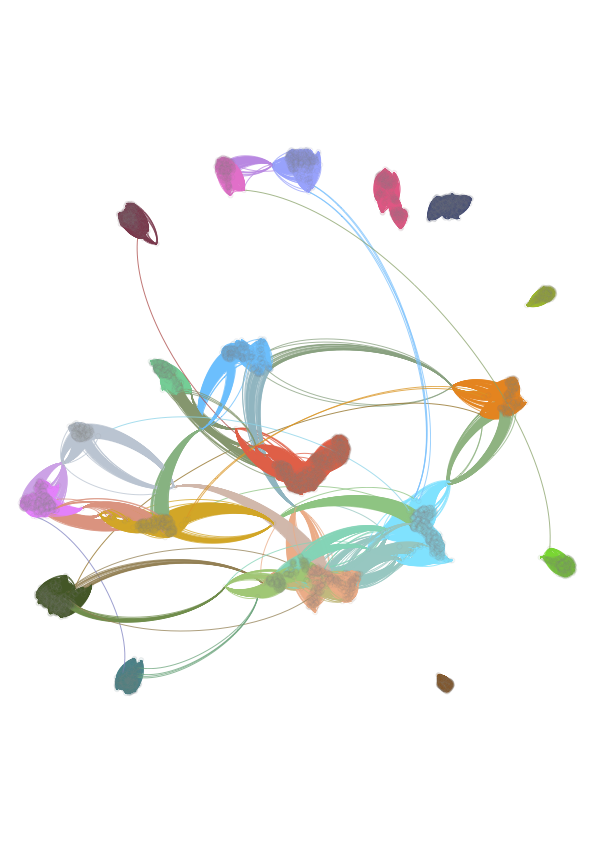}                                       \\
		(c) \textit{Wiki} ($\eps=0.19$)                                             & (d) \textit{LiveJ} ($\eps=0.6$)
	\end{tabular}
	\caption{\newblue Clustering Visualisations under Jaccard Similarity: Top-20 Clusters with~$\mu=5$}
	\label{fig:visualisation}
\end{figure*}

\begin{figure*}
	\centering
	\begin{tabular}{cc}
		\includegraphics[width=0.4\textwidth]{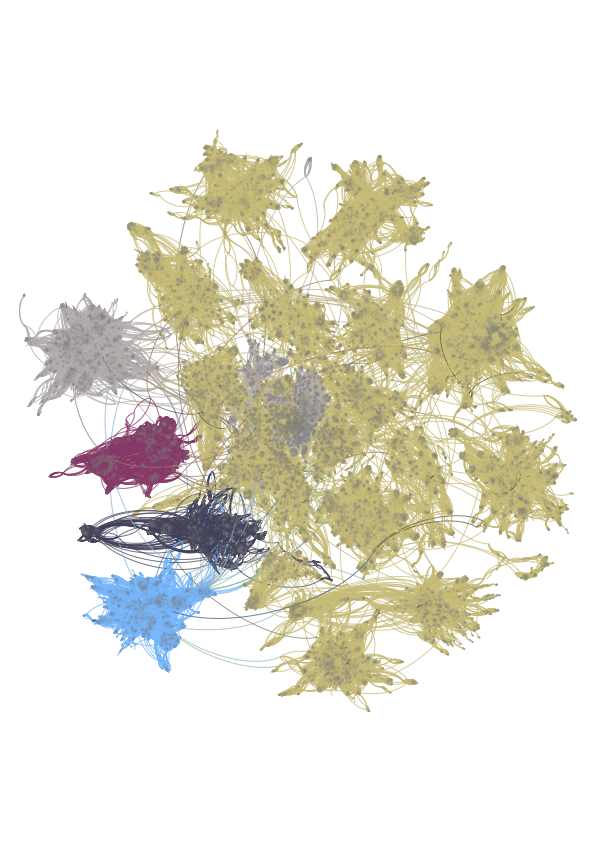} &
		\includegraphics[width=0.4\textwidth]{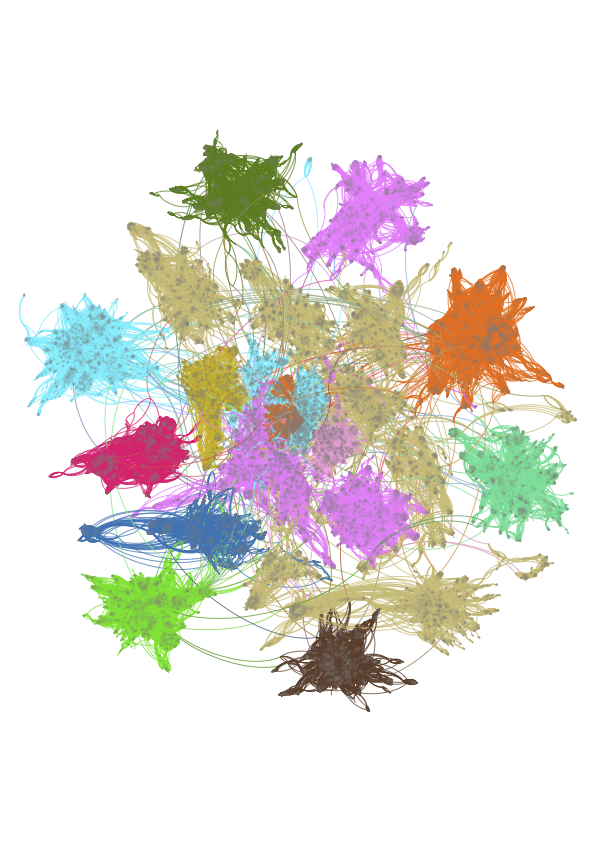}                                     \\
		(a) \textit{Google} ($\eps=0.13$)                                         & (b) \textit{Google} ($\eps=0.135$) \\
		\includegraphics[width=0.4\textwidth]{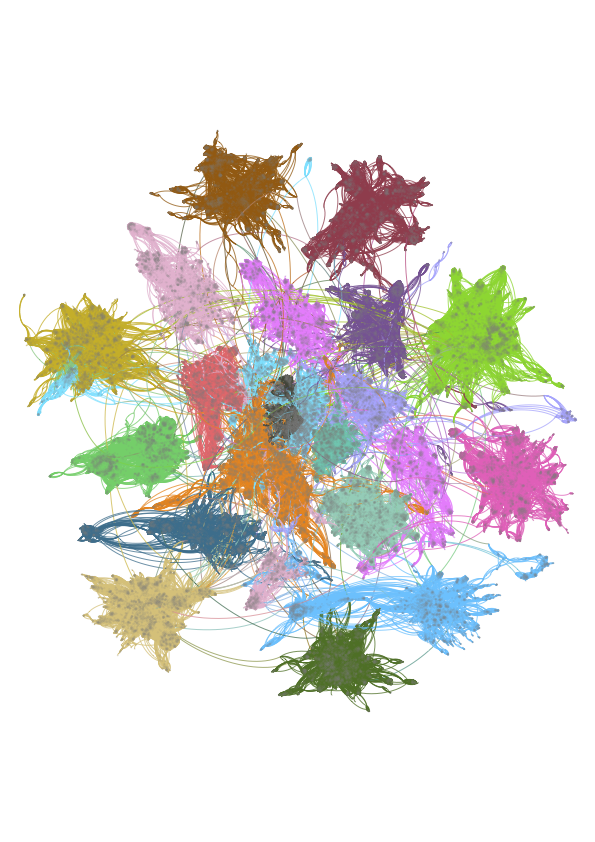} &
		\includegraphics[width=0.4\textwidth]{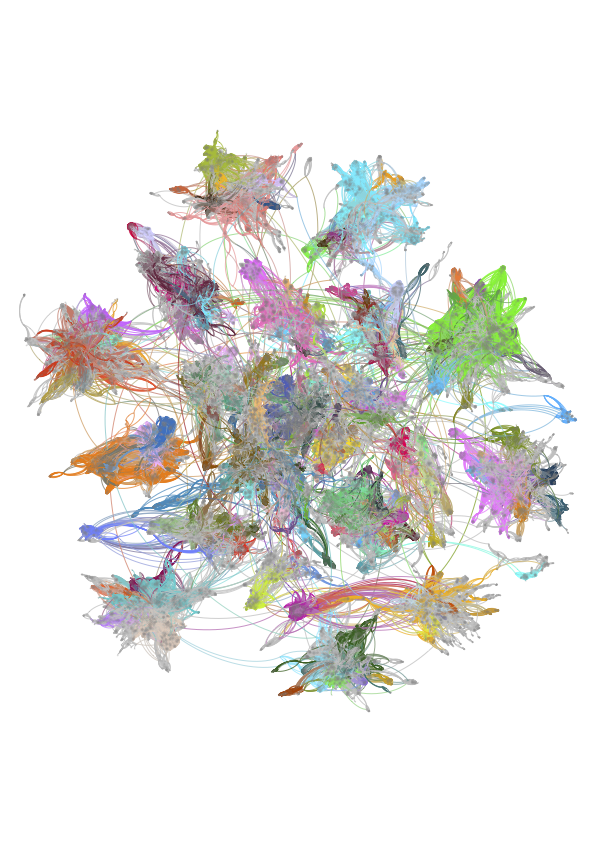}                                      \\
		(c) \textit{Google} ($\eps=0.15$)                                         & (d) \textit{Google} ($\eps=0.2$)
	\end{tabular}
	\caption{\newblue Evolutions of the Top-20 Clusters under Jaccard Similarity on \textit{Google} with varying~$\eps$ and~$\mu=5$}
	\label{fig:google}
\end{figure*}

\begin{figure*}
	\centering
	\begin{tabular}{cc}
		\includegraphics[width=0.3\linewidth]{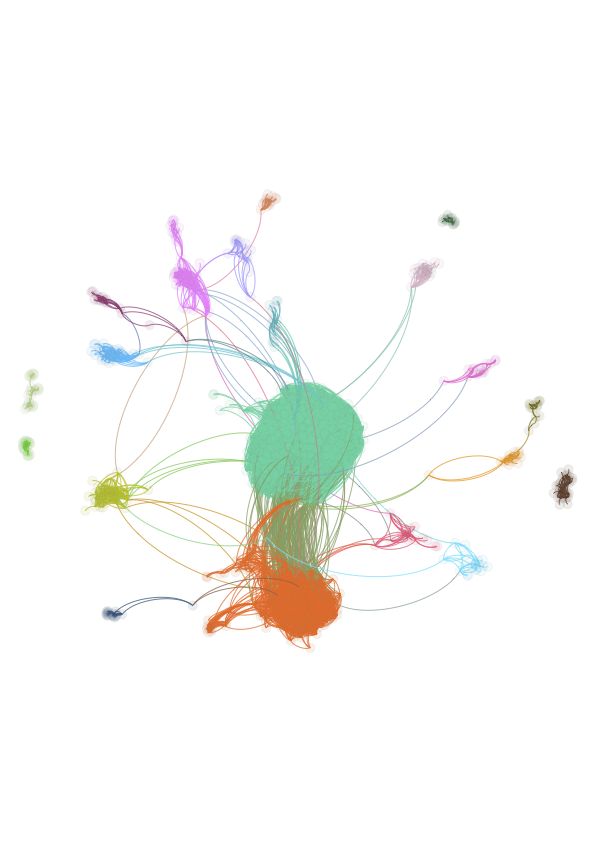} &
		\includegraphics[width=0.3\linewidth]{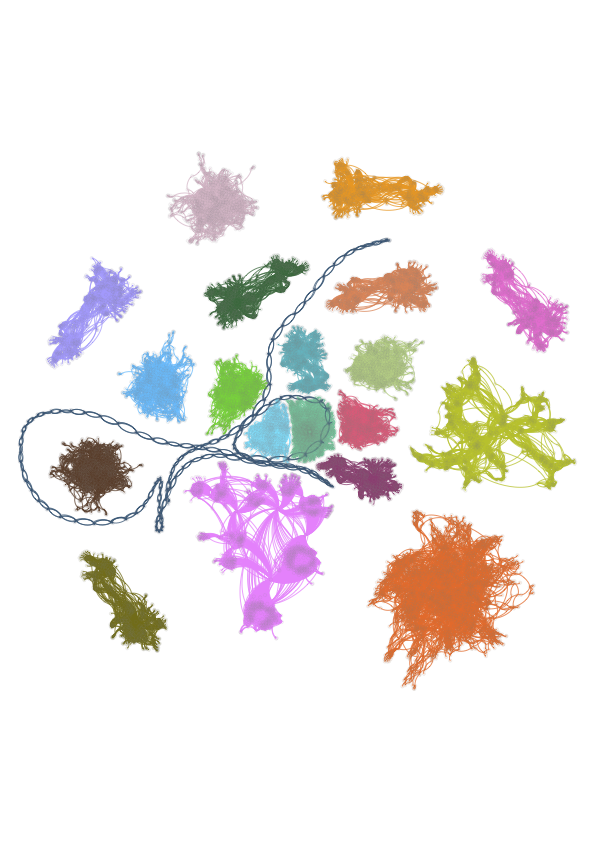}                                    \\
		(a) {\em Slashdot} ($\eps=0.3$)                                              & (b) {\em Notre} ($\eps=0.36$)
	\end{tabular} \\
	\begin{tabular}{ccc}
		\includegraphics[width=0.3\linewidth]{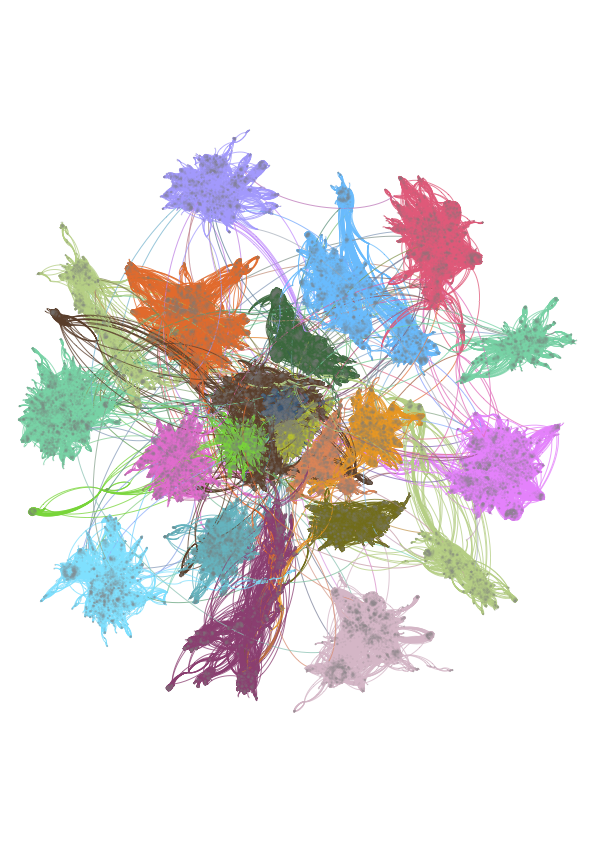} &
		\includegraphics[width=0.3\linewidth]{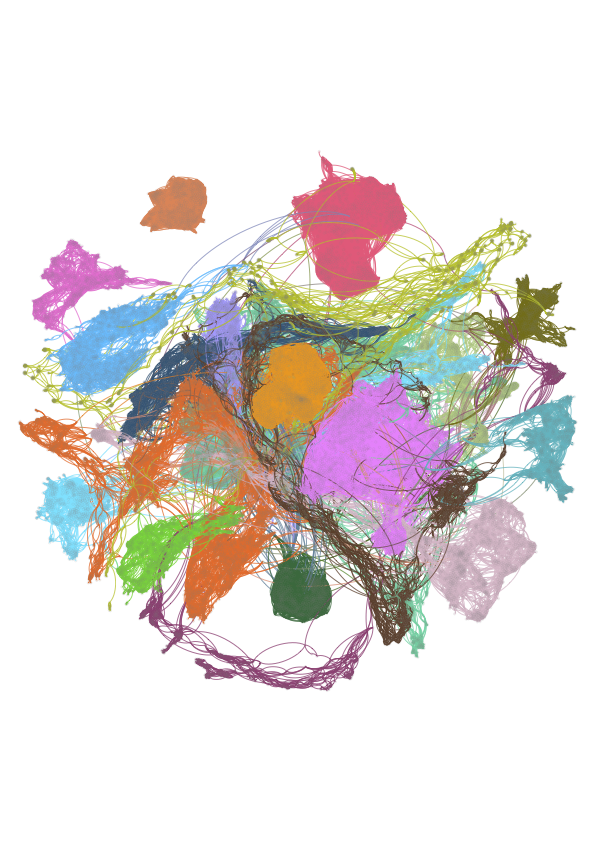}   &
		\includegraphics[width=0.3\linewidth]{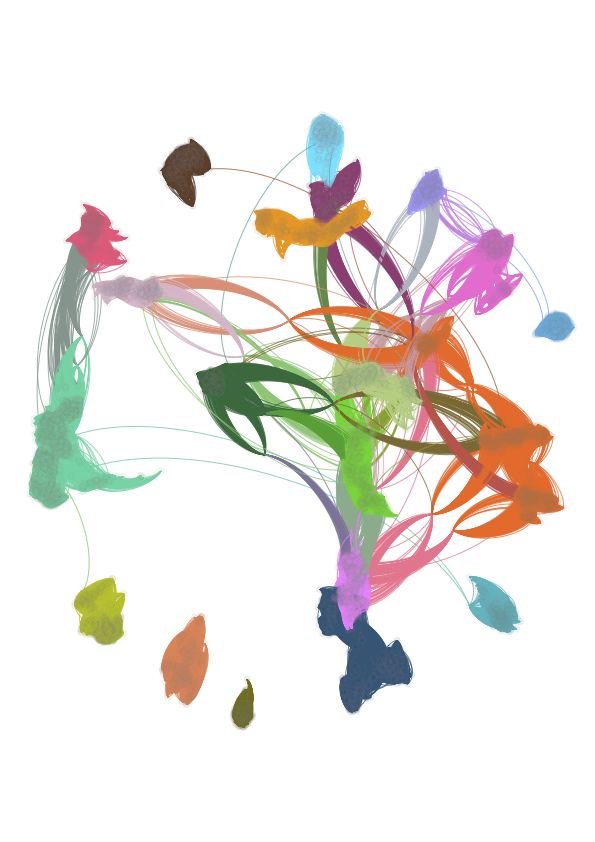}                                                                 \\
		(c) {\em Google} ($\eps=0.3$)                                              & (d) {\em Wiki} ($\eps=0.34$) & (e) {\em LiveJ} ($\eps=0.67$)
	\end{tabular}
	\caption{\newblue Clustering Visualisations under Cosine Similarity: Top-20 Clusters with~$\mu=5$}
	\label{fig:cosine_visualisation}
\end{figure*}

\subsection{Clustering Visualisations}\label{subsec:visual}
{\newblue
	We start with the visualisation results of our 5 representative datasets with $\mu$ set to $5$ under both similarities. Since the sizes of these graphs are large, we only show the clustering results of the top-20 clusters w.r.t. cluster size (i.e., the number of vertices contained in a cluster). Different clusters are shown in different colours. For a hub $u$ that belongs to multiple clusters, we only assign it to the cluster containing $u$'s ``smallest'' similar core neighbour (i.e., the similar core neighbour vertex with the smallest id). Furthermore, we omit the noises in the visualisations.

	In choosing proper $\eps$ for each dataset, our target is that in the clustering results the sizes of each cluster do not vary too much. We take {\em Google} as an example to show the effect of $\eps$ on $\strcluresult$ in Figure~\ref{fig:google}. The vertices shown are those in the top-20 clusters when $\eps=0.15$ as shown in Figure~\ref{fig:google}(c). $\eps=0.15$ is also the proper $\eps$ we choose for {\em Google}. If we increase $\eps$ to $0.2$ in Figure~\ref{fig:google}(d), the clusters are separated into more clusters whose sizes are smaller comparing to $\eps=0.15$. The reason is that when $\eps$ is increased, some edges which are originally labelled as similar under $\eps=0.15$ will become dissimilar. Thus some core vertices will become non-core vertices and some similar edges linking two core vertices will be ``broken''. As a result, more clusters with smaller sizes are produced. On the contrary, if we decrease $\eps$ to $0.135$ and $0.13$ in Figure~\ref{fig:google}(b) and Figure~\ref{fig:google}(a), the clusters begin to merge with each other to form larger clusters. It is because under smaller $\eps$, some edges originally labelled as dissimilar under $\eps=0.15$ become similar. Thus some non-core vertices become core vertices and more dissimilar edges become similar core edges linking core vertices that originally belong to different clusters. The evaluation of $\strcluresult$ on {\em Google} with varying $\eps$ under Jaccard similarity is shown in Figure~\ref{fig:google}. For other datasets, the chosen $\eps$ value and the visualisation results of the top-20 clusters are shown in Figure~\ref{fig:visualisation}.

	For cosine similarity, the visualisation results are shown in Figure~\ref{fig:cosine_visualisation} with different colours representing different clusters. The value of $\eps$ is picked such that the visualisation results are similar to the results shown in Figure~\ref{fig:visualisation} and Figure~\ref{fig:google}(c). From these visualisation results, it can be confirmed that the intra-cluster edges are much denser than the inter-cluster edges, which indicates the quality of structural clustering results is good and the results are meaningful for human to understand.

	On the other hand, comparing Figure~\ref{fig:visualisation}, Figure~\ref{fig:google}(c) and Figure~\ref{fig:cosine_visualisation}, an observation is that while the clustering results is similar, the values of $\eps$ under cosine similarity are generally larger than the values of $\eps$ under Jaccard similarity. The reason is that by the definition of these two similarities, the cosine similarity between two vertices is always no smaller than the Jaccard similarity between them. To see this, consider an edge $(u,v)\in E$ and without loss of generality, suppose $d[u]\geq d[v]$:
	\begin{equation*}
		|N[u]\cup N[v]|\geq d[u]\geq\sqrt{d[u]\cdot d[v]}.
	\end{equation*}
	Therefore,
	\begin{equation*}
		\frac{|N[u]\cap N[v]|}{|N[u]\cup N[v]|}\leq \frac{|N[u]\cap N[v]|}{\sqrt{d[u]\cdot d[v]}}.
	\end{equation*}
}

\subsection{Approximate Clustering Quality}\label{subsec:effectiveness}
{\newblue Next}, we evaluate the quality of the $\rho$-approximate $\strcluresult$ computed by $\dynelm$ (equivalently, by $\dynstr$) on the five representative datasets plus {\em Twitter} {\newblue under Jaccard similarity and on the five representative datasets under cosine similarity.}
Our evaluation adopts three measurements: (i) {\em mis-labelled rate}, (ii) {\em overall clustering quality},
and (iii) {\em individual cluster quality}; the details of these measurements will be introduced shortly.
Furthermore, the approximate clustering results we considered are obtained with $\rho = 0.01$ and $\rho =0.5$, respectively,
under $\mu = 5$ and customized $\eps$'s (shown beside the dataset names in Table~\ref{tab:quality}).
These customized $\eps$ values are chosen based on the visualisations of the clustering results {\newblue shown in Figure~\ref{fig:visualisation}, Figure~\ref{fig:google}(c) and Figure~\ref{fig:cosine_visualisation}}
Under these $\eps$ values, the obtained clustering results are natural to human sensibility: the intra-cluster edges are much denser than the inter-cluster edges.

\vspace{1mm}
\noindent
\textbf{Mis-Labelled Rate.}
Recall that the $\strcluresult$ is uniquely determined by the edge labelling.
If the $\rho$-approximate edge labelling, $\tL$, is highly similar to its exact counterpart, $\L_{\eps}(G)$,
their corresponding clustering results should be highly similar.
To measure the similarity between $\L_{\eps}(G)$ and $\tL$,
we consider the {\em mis-labelled rate}, which is defined as the {\em percentage} of the edges that have different labels in these two labelling's.
As shown in Table~\ref{tab:quality}, when $\rho = 0.5$, the mis-labelled rates across the six datasets are $2.37\%, 5.86\%, 8.74\%, 1.82\%, 6.33\%$ and $0.07\%$ {\newblue under Jaccard similarity}, respectively.
While these mis-labelled rates are
small already,
these numbers can be even significantly smaller when $\rho =0.01$.
They are just: $0.02\%, 0.1\%, 0.16\%, 0.04\%, 0.14\%$ and $0.01\%$, respectively:
none of them is more than $16/10000$.
In other words, when $\rho = 0.01$, all the approximate edge labellings are almost identical to the exact labellings on these datasets.
Therefore, one can expect that by setting $\rho = 0.01$,
the approximate results computed by $\dynelm$ are ``{\em merely identical}'' to the exact ones {\newblue under Jaccard similarity}.
To quantify this intuition, we study the {\em overall clustering quality} and the {\em individual cluster quality}.

\vspace{1mm}
\noindent
\textbf{Overall Clustering Quality.}
We quantify the overall clustering quality with the {\em Adjusted Rand Index} (ARI)~\cite{ha-classification-1985}, which is adopted by some of the authors of SCAN in their later work~\cite{yxs-hicss-2008} to measure the quality of structural clustering results.
In fact, the ARI is a widely adopted~\cite{ha-classification-1985, mc-mbr-1986, bp-tsmc-1998,mtmg-ml-2003,kv-tpam-2006,kht-icif-2006,veb-jmlr-2010} similarity measurement between two {\em partitions} of a same set.
The ARI values are
between $0$ and $1$, where {\em the closer the ARI value is to $1$, the more similar the two partitions are}.
However, in general, ARI is not applicable to $\strcluresult$'s,
because the $\strcluresult$ is not necessarily a partition of the vertices (some non-core vertices may belong to none or multiple clusters).
To address this subtlety, we assign each non-core vertex $u$ {\em only} to the cluster which contains $u$'s ``smallest'' similar core neighbour (in terms of the identifier value), and ignore all the noise vertices.
In Table~\ref{tab:quality}, when $\rho = 0.5$, the ARI scores (between the approximate and the exact clusterings) across all the six datasets are at least $0.96$.
Even better, when $\rho = 0.01$, these scores are at least $0.994$ (this worst value occurs on {\em Twitter}), where, impressively, the score is up to $0.9999$ on {\em Wiki}.
These high ARI scores indicate that the {overall} $\rho$-approximate clusterings are of very high quality {\newblue under Jaccard similarity}.

\vspace{1mm}
\noindent
\textbf{Individual Cluster Quality.}
Since the high overall clustering quality may not necessarily reflect the high quality of each individual cluster,
we thus look into the individual cluster quality of the approximate results.
Specifically,
consider a cluster $C$ in an $\rho$-approximate clustering result;
let $S\subseteq C$ be the set of all the vertices in $C$ that are {\em core} under
the {\em exact} edge labelling.
Moreover, let $\C^*$ be the set of {\em exact clusters} that contain at least one core vertices in $S$.
The {\em individual cluster quality} of the approximate cluster $C$ is defined as the {\em largest} Jaccard similarity between $C$ and each cluster $C' \in \C^*$, i.e.,
$\max_{C' \in \C^*} \frac{|C \cap C'|}{|C \cup C'|}\,.$
The closer this value is to $1$, the more similar of the approximate cluster $C$ is to an exact cluster $C'$, and hence, the higher quality of $C$.
Table~\ref{tab:quality} shows the {\em minimum} and {\em average} individual cluster qualities among the top-$k$ largest (in terms of size) $\rho$-approximate clusters on the six datasets, where
$k \in \{1, 5, 10, 20, 50, 100\}$.
The average value is the average quality of the top-$k$ clusters; and the minimum value shows how bad the cluster with the {\em least} quality is.
Furthermore, the ``worst'' minimum and average values across all cases are underlined in the table.

For $\rho = 0.5$, the average individual qualities are at least $0.952$ across all cases.
However, as highlighted in bold in Table~\ref{tab:quality}, we do see two big drops in the minimum quality among the top-$100$ and top-$50$ clusters respectively on {\em Google} and {\em Wiki}.
We looked into these two cases and eventually found out the cause behind these drops.
When $\rho = 0.5$, some similar edges
have been mis-labelled as dissimilar due to the does-not-matter case.
As a result, the corresponding exact clusters happened to split into two smaller clusters in the approximate clusterings, resulting in the low individual cluster qualities.
In contrast, for $\rho = 0.01$, these cases did not happen and both the average and minimum individual cluster qualities are consistently good across all cases;
they are $0.987$ and $0.853$, respectively, where the latter indicates that even for the ``worst'' cluster, it still has $0.853$ Jaccard similarity with its corresponding exact cluster.
Let alone that in most of other cases, the minimum individual quality is at least $0.95$.

In summary,
the quality of the approximate results obtained with $\rho = 0.01$ are consistently high in terms of all the three measurements {\newblue under Jaccard similarity}.
We thus set $\rho  =0.01$ as our default value in the subsequent experiments.
As we will see shortly,
with such a tiny sacrifice in the clustering quality,
our algorithms can gain up to {\em three-orders-of-magnitude} improvements in efficiency.

\subsection{\newblue Comparison between Jaccard Similarity and Cosine Similarity in Approximate Quality}
{\newblue
	For cosine similarity, the results of the above three measures are shown in Table~\ref{tab:cosinequality}.

	In Table~\ref{tab:quality}, if we list the five representative datasets in ascending orders with respect to the mis-labelled rate, the order will be: \textit{Slashdot}, \textit{Wiki}, \textit{Notre}, \textit{LiveJ} and \textit{Google} with $\rho=0.01$; and \textit{Wiki}, \textit{Slashdot}, \textit{Notre}, \textit{LiveJ} and \textit{Google} with $\rho=0.1$. Comparing this order to the mis-labelled rate presented in Table~\ref{tab:cosinequality}, these two orders above match the orders obtained from Table~\ref{tab:cosinequality} under $\rho=0.01$ and $\rho=0.5$, respectively. Moreover, when $\rho$ is both set to $0.01$, the mis-labelled rates for all $5$ datasets under cosine similarity are all higher than those under Jaccard similarity. As discussed above, when producing similar clustering results, the value of $\eps$ under cosine similarity is in general larger than under Jaccard similarity. Therefore, the value of $\rho\eps$ is larger under cosine similarity so the range of ``don't care case'' is larger for cosine similarity. As a result, an edge where the similarity between its endpoints is close to $\eps$ is more likely to be mis-labelled.Thus the mis-labelled rate is higher under cosine similarity. The value of ARI and the result of individual cluster quality also confirm that the approximation quality for cosine similarity is worse than that for Jaccard similarity when $\rho$ is both set to $0.01$.

	When $\rho$ is set to $0.1$, however, the approximation quality for cosine similarity drops significantly, which is shown by both ARI and individual cluster quality. Although for most datasets, the worst average individual cluster quality is larger than $0.6$, the worst minimum individual cluster similarities all go smaller than $0.2$. One case worth to mention is \textit{Slashdot}. The 23-rd largest cluster in the approximation clustering is consist of some vertices that are not core vertices in the exact clustering result. Therefore, the Jaccard similarity between this cluster and any other clusters in the exact clustering result is all 0, making the minimum individual cluster similarity to be 0. Note that from the visualisation of \textit{Slashdot}, it can be seen that there are only two big clusters in this graph, and the sizes of remaining clusters are far smaller than the big ones. That's why the ARI of \textit{Slashdot} under $\rho=0.1$ is still very large.

	From the above analysis, we can confirm that when the target is to produce similar clustering results with $\rho$-approximate notion, Jaccard similarity would be more favourable since its $\strcluresult$ better reflects the exact clustering result.
}

\subsection{Efficiency Experiment Setup}\label{subsec:efficency_setup}

\noindent
\textbf{Update Simulations.}
In addition to the original edges, we generate a sequence of edge insertions and deletions
to simulate the {\em update process} for each dataset.
Specifically, for a fixed value of~$\eta$,
we generate an update {independently} with probability
$\frac{1}{1 + \eta}$ to insert a {\em new} edge, and probability $\frac{\eta}{1 + \eta}$ to delete an {\em existing} edge.
In this way, on
average, the frequency of deletions is roughly~$\eta$ proportion of the frequency of the insertions.
For a deletion, the edge to be removed is picked {\em independently and uniformly at random} from the current edges.
For an insertion, we generate all of them {\em consistently}, with one of the following three strategies:
\begin{itemize}[leftmargin = *]
	\item {\em Random-Random} ($\mathcal{RR}$): Uniformly-at-random pick an edge that is {\em not} in the current graph.
	\item {\em Degree-Random} ($\mathcal{DR}$): First, choose a vertex~$u$ with probability $\frac{d[u]}{2m}$, where~$m$ is the current number of edges in the graph. {If the degree of~$u$ is~$n -1$, repeat this step.}
	      Second, uniformly-at-random pick a vertex~$v$ from those vertices not currently adjacent to~$u$.
	\item {\em Degree-Degree} ($\mathcal{DD}$): Similar to $\mathcal{DR}$, but the second vertex~$v$ is also chosen (independently) with probability~$\frac{d[v]}{2m}$.
	      If~$(u,v)$ is already in the graph {or a self loop}, repeat.
\end{itemize}
Denote the number of {\em original} edges in the graph (after our pre-processing) by $m_0$, shown in the meta information.
Except {\em Twitter}, for each of the 14 datasets,
we generate~$9 \cdot m_0$ updates, including both deletions (if~$\eta > 0$) and insertions.
The {\em update process} is simulated as follows. Starting from an empty graph, insert each of the~$m_0$ original edges one by one, and then perform each of the~$9\cdot m_0$ generated updates.
Therefore, in the update process, in total~$10\cdot m_0$ updates are performed on each graph.
As for {\em Twitter}, we only further generate $0.1 \cdot m_0$ updates, because $m_0 = 1.2 \times 10^9$ is already large enough.
The total numbers of updates can be found in Table~\ref{tab:meta-info}.


\begin{figure*}[t]
	\centering
	\includegraphics[width=0.4\linewidth]{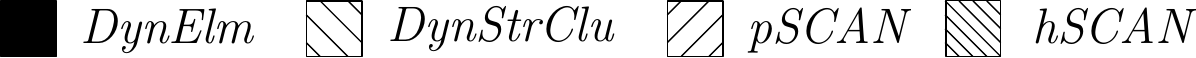} \\
	\includegraphics[width=0.8\linewidth]{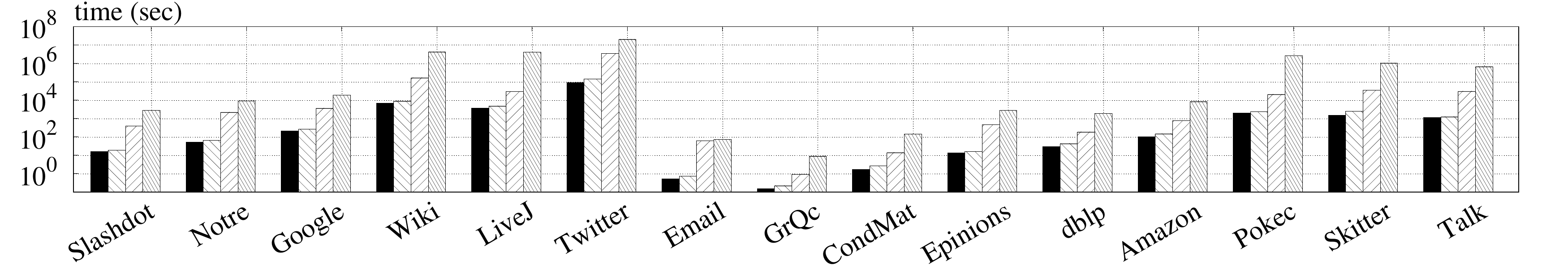}
	\caption{Overall running time (logarithmic scale) under default settings for all four algorithms on all~$15$ datasets.
	}
	\label{fig:time_vs_algo}
\end{figure*}

\begin{figure*}
	\includegraphics[width=0.4\linewidth]{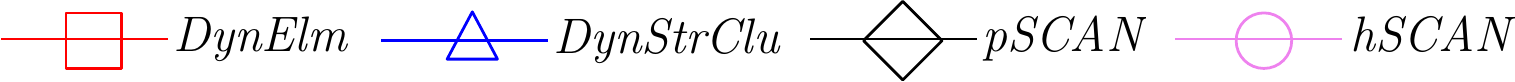} \\
	\resizebox{\linewidth}{!}{%
		\begin{tabular}{ccc}
			\hspace{-4mm}
			\includegraphics[width=0.3\linewidth]{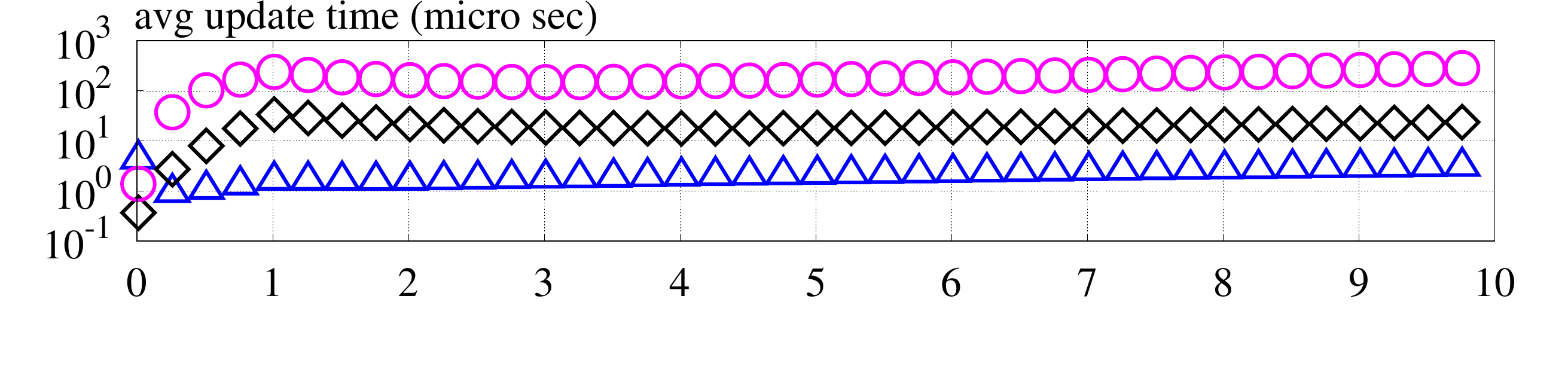} &
			\hspace{4mm}
			\includegraphics[width=0.3\linewidth]{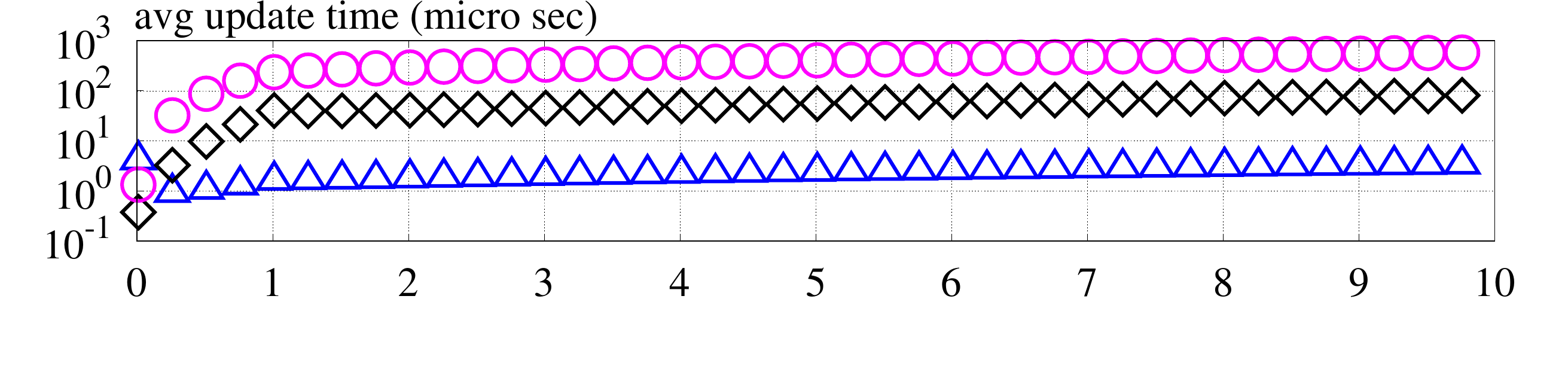} &
			\hspace{4mm}
			\includegraphics[width=0.3\linewidth]{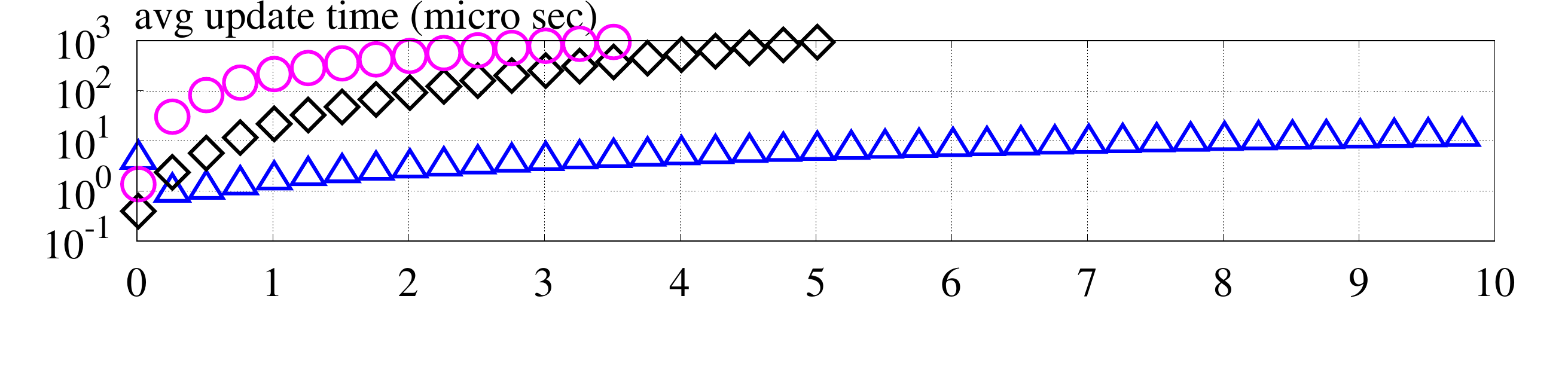}                                                                                                       \\
			\hspace{-6mm} (a) {\em Slashdot} ($\mathcal{RR}$)                                      & \hspace{4mm} (b) {\em Slashdot} ($\mathcal{DR}$) & \hspace{4mm} (c) {\em Slashdot} ($\mathcal{DD}$) \\

			\hspace{-4mm}
			\includegraphics[width=0.3\linewidth]{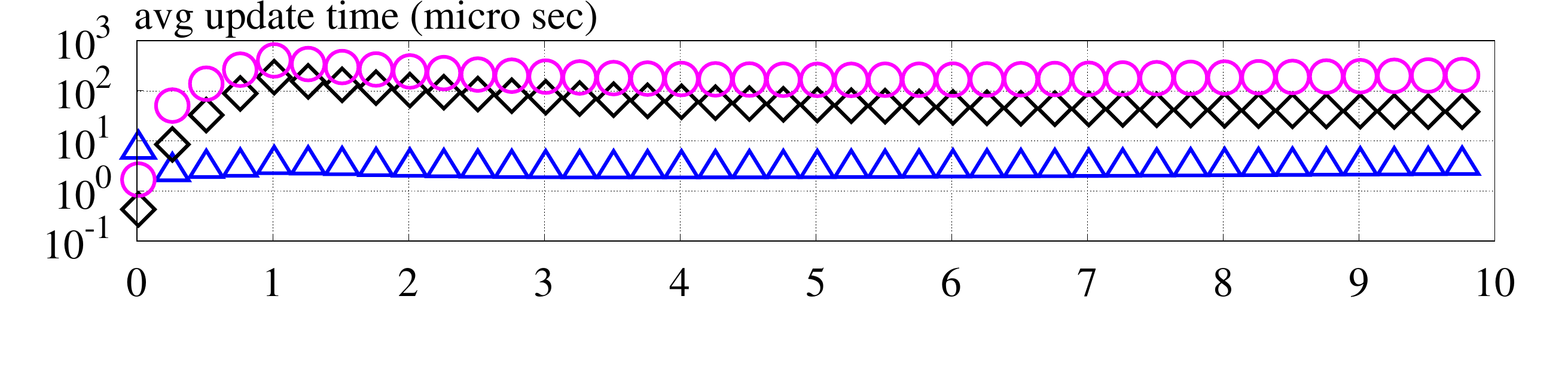}    &
			\hspace{4mm}
			\includegraphics[width=0.3\linewidth]{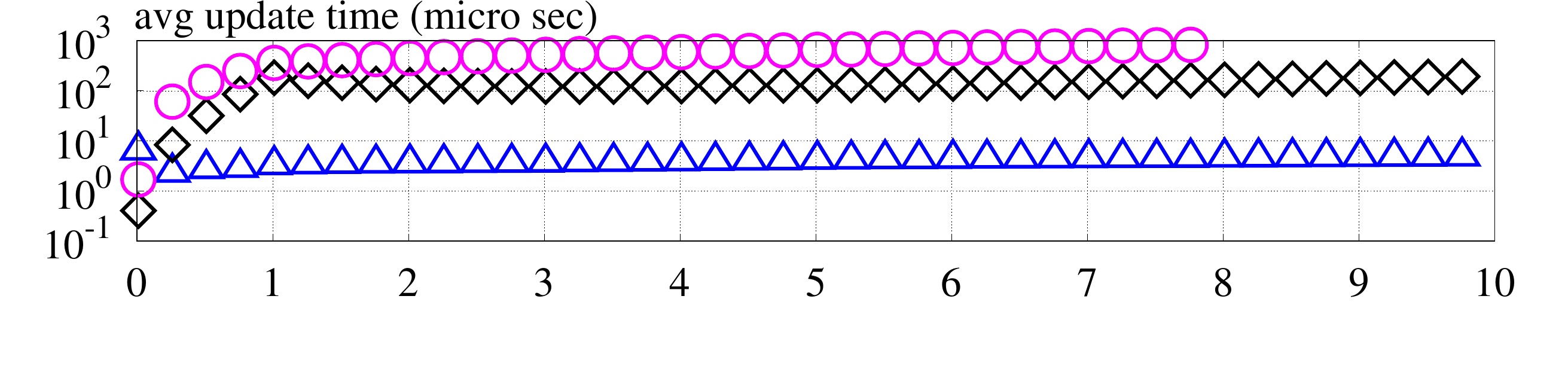}    &
			\hspace{4mm}
			\includegraphics[width=0.3\linewidth]{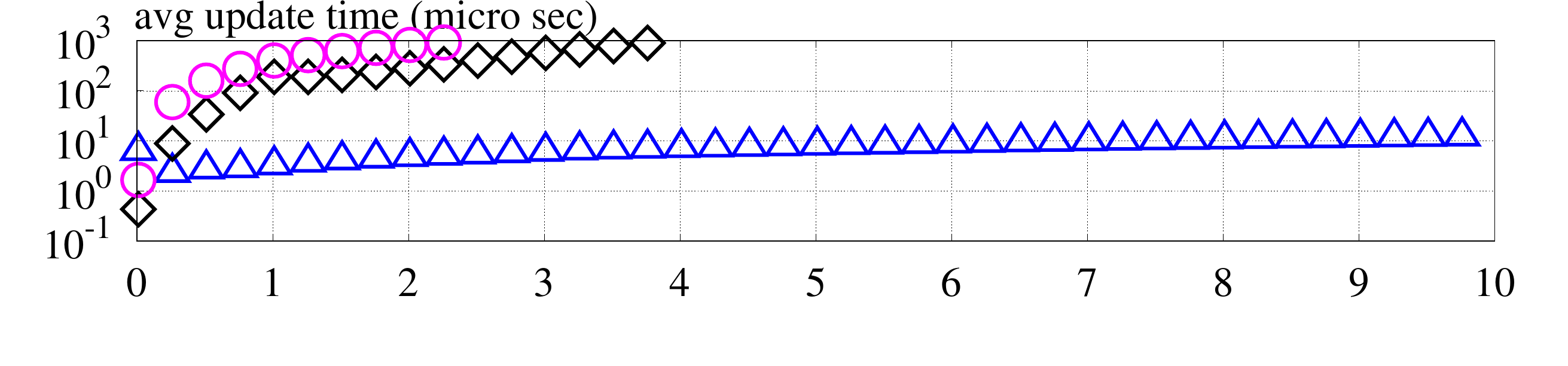}                                                                                                          \\
			\hspace{-10mm} (a) {\em Notre} ($\mathcal{RR}$)                                        & \hspace{1mm} (b) {\em Notre} ($\mathcal{DR}$)    & \hspace{1mm} (c) {\em Notre} ($\mathcal{DD}$)    \\

			\hspace{-4mm}
			\includegraphics[width=0.3\linewidth]{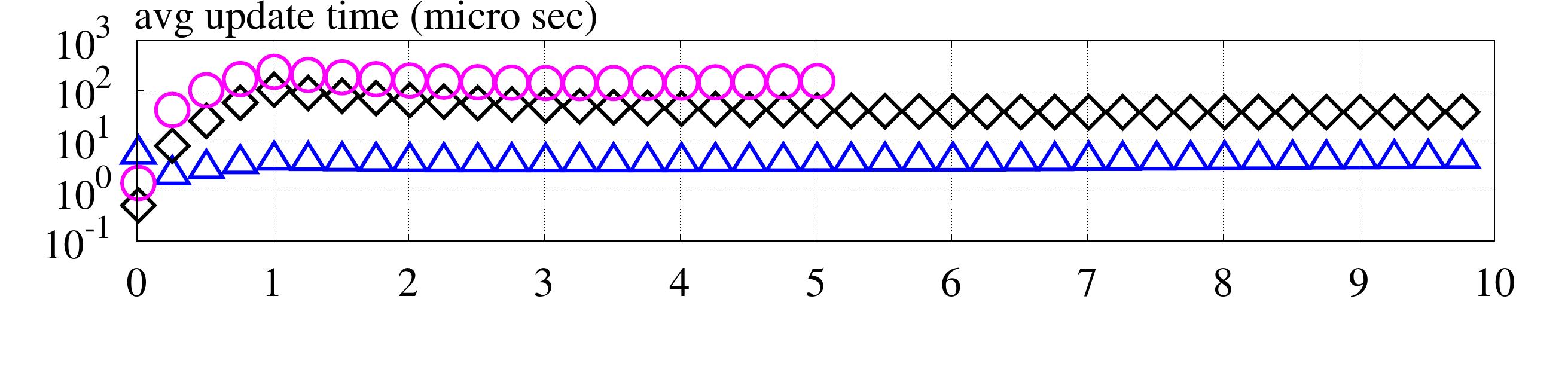}   &
			\hspace{4mm}
			\includegraphics[width=0.3\linewidth]{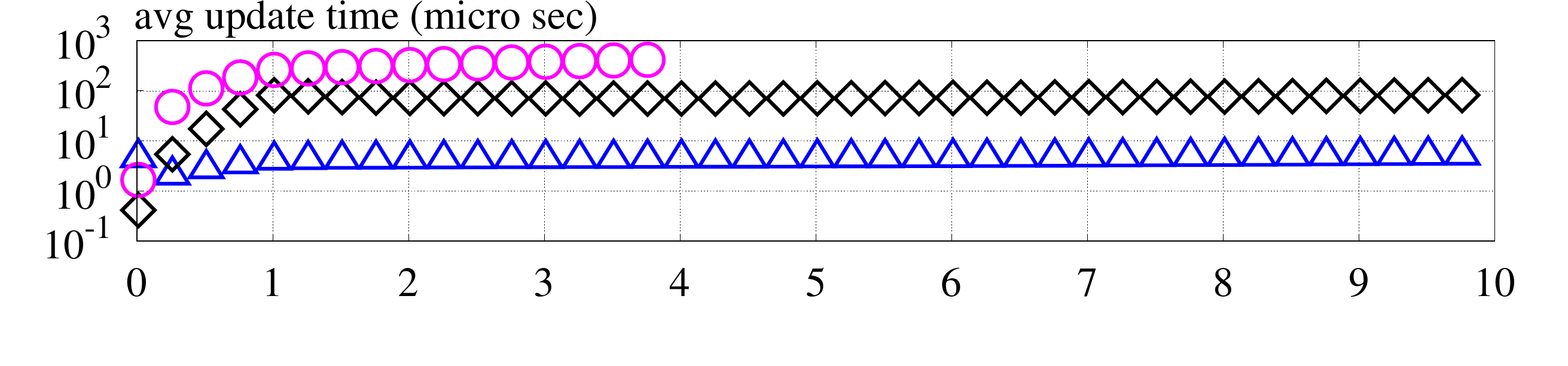}   &
			\hspace{4mm}
			\includegraphics[width=0.3\linewidth]{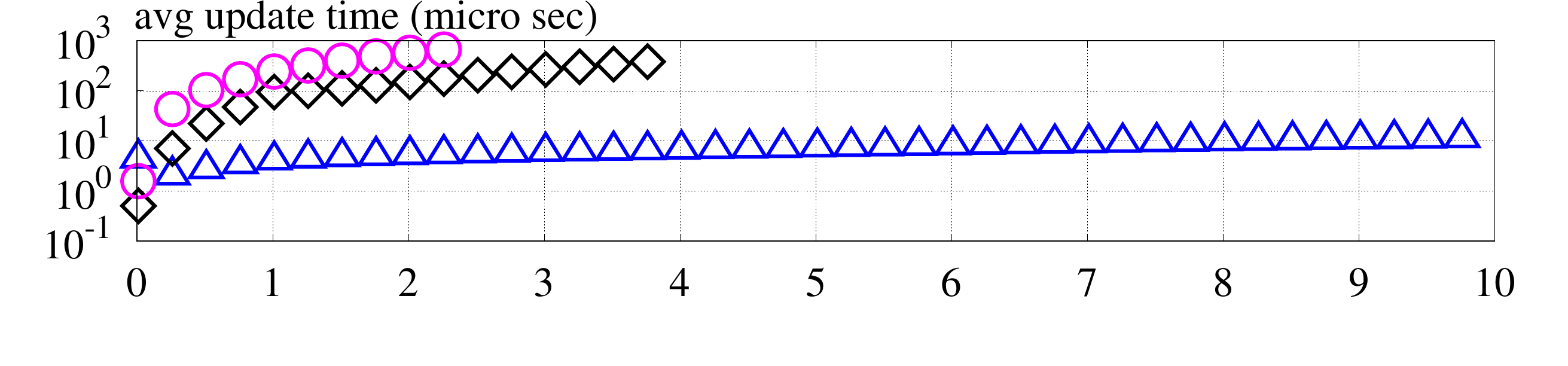}                                                                                                         \\
			\hspace{-8mm} (a) {\em Google} ($\mathcal{RR}$)                                        & \hspace{3mm} (b) {\em Google} ($\mathcal{DR}$)   & \hspace{3mm} (c) {\em Google} ($\mathcal{DD}$)   \\

			\hspace{-4mm}
			\includegraphics[width=0.3\linewidth]{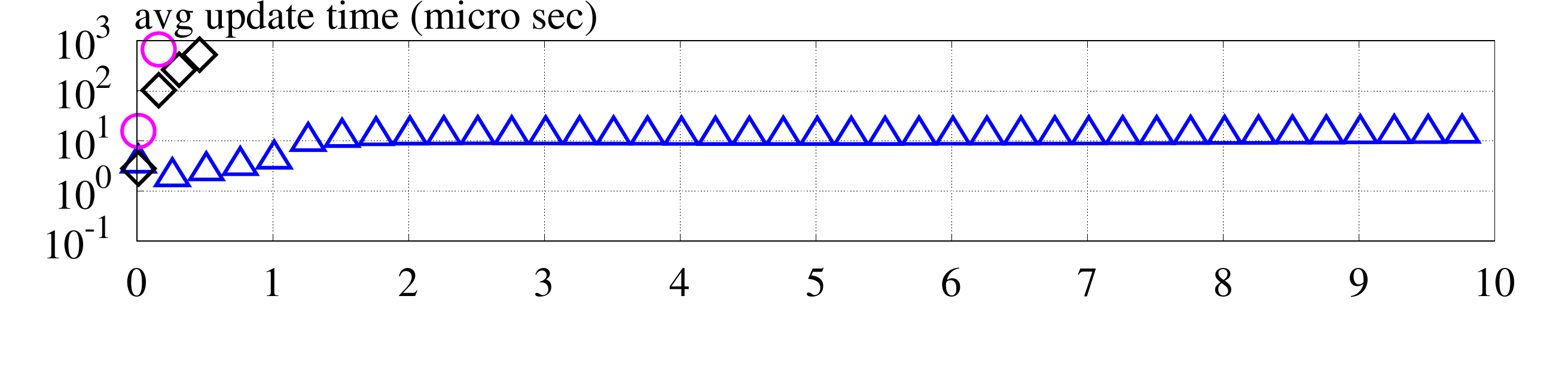}     &
			\hspace{4mm}
			\includegraphics[width=0.3\linewidth]{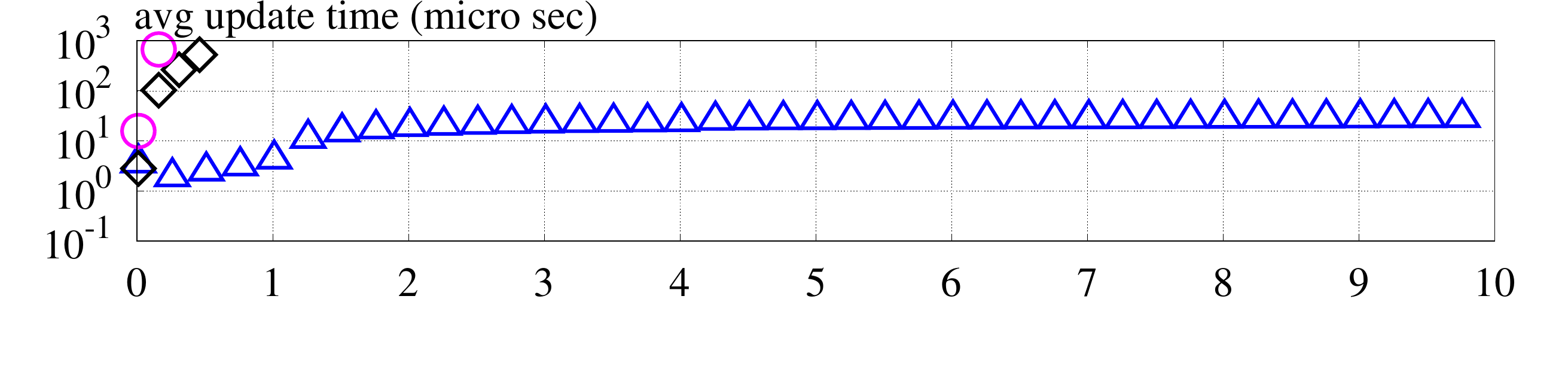}     &
			\hspace{4mm}
			\includegraphics[width=0.3\linewidth]{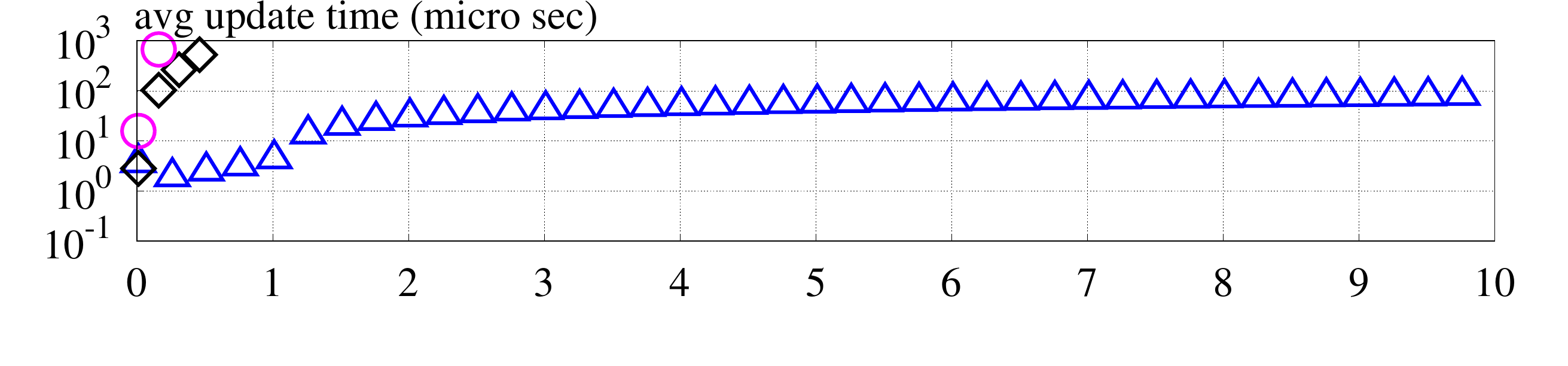}                                                                                                           \\
			\hspace{-10mm} (a) {\em Wiki} ($\mathcal{RR}$)                                         & \hspace{1mm} (b) {\em Wiki} ($\mathcal{DR}$)     & \hspace{1mm} (c) {\em Wiki} ($\mathcal{DD}$)     \\

			\hspace{-4mm}
			\includegraphics[width=0.3\linewidth]{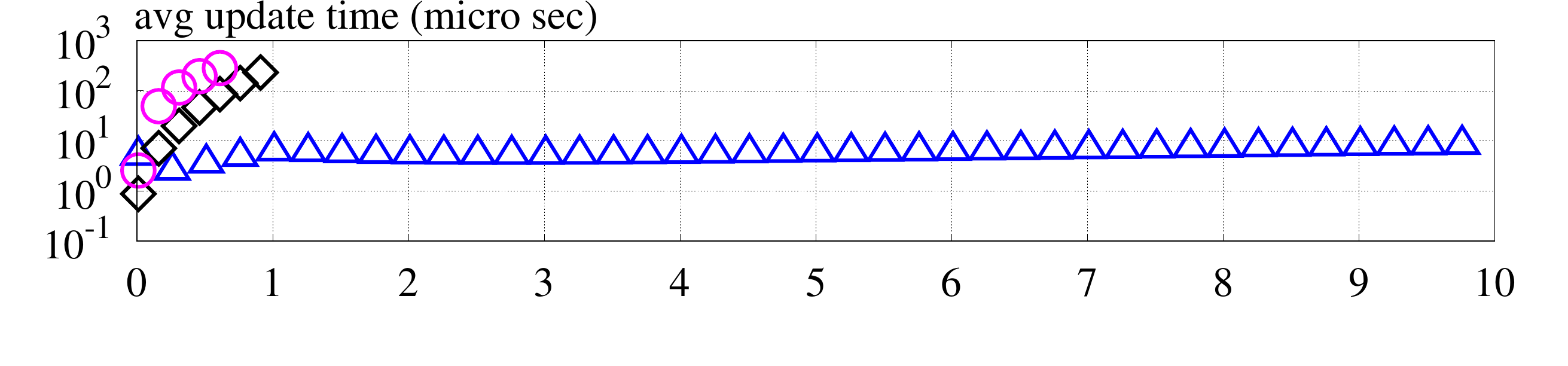}    &
			\hspace{4mm}
			\includegraphics[width=0.3\linewidth]{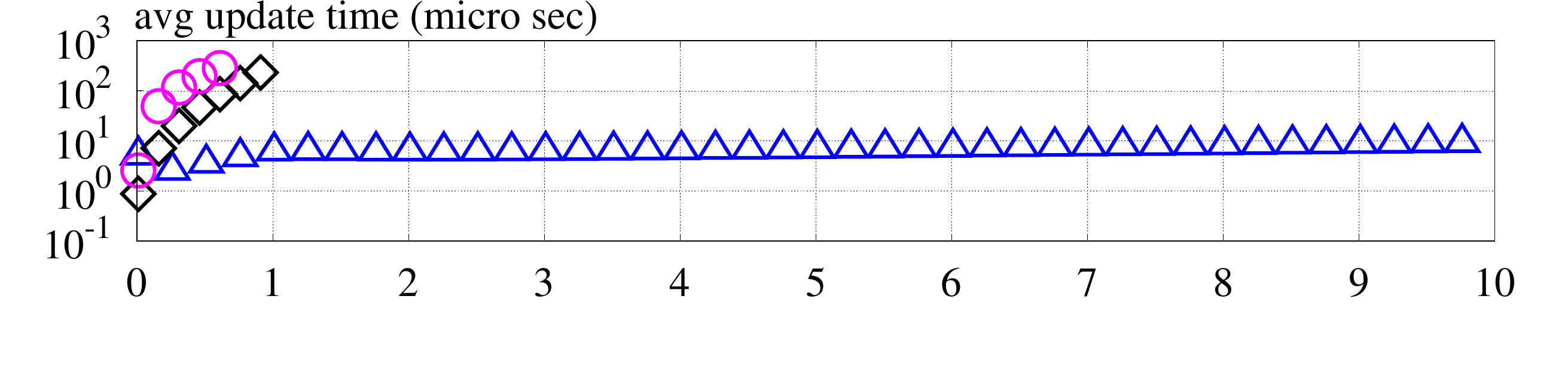}    &
			\hspace{4mm}
			\includegraphics[width=0.3\linewidth]{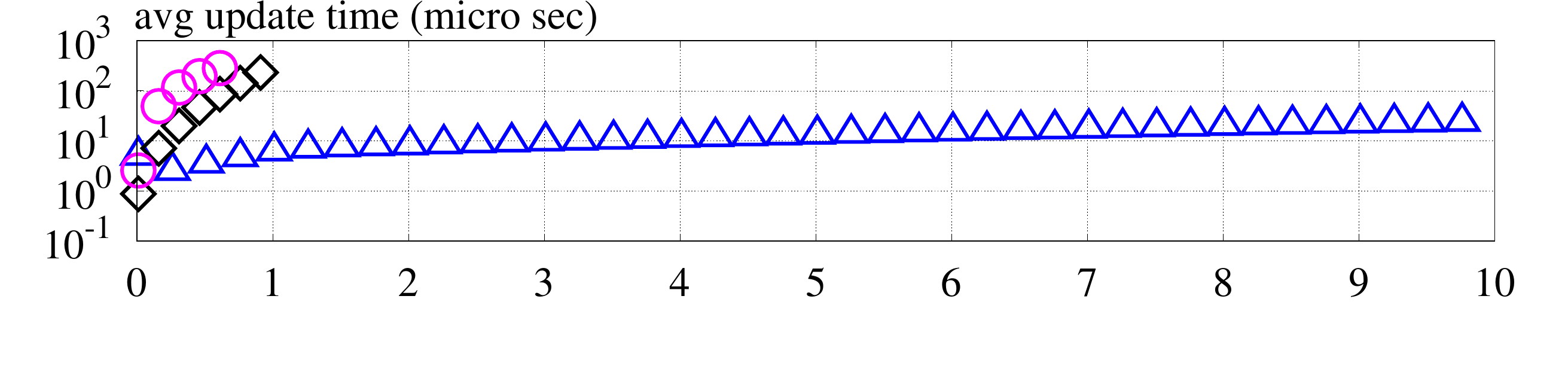}                                                                                                          \\
			\hspace{-10mm} (a) {\em LiveJ} ($\mathcal{RR}$)                                        & \hspace{1mm} (b) {\em LiveJ} ($\mathcal{DR}$)    & \hspace{2mm} (c) {\em LiveJ} ($\mathcal{DD}$)    \\
			\hspace{-4mm}
			\includegraphics[width=0.3\linewidth]{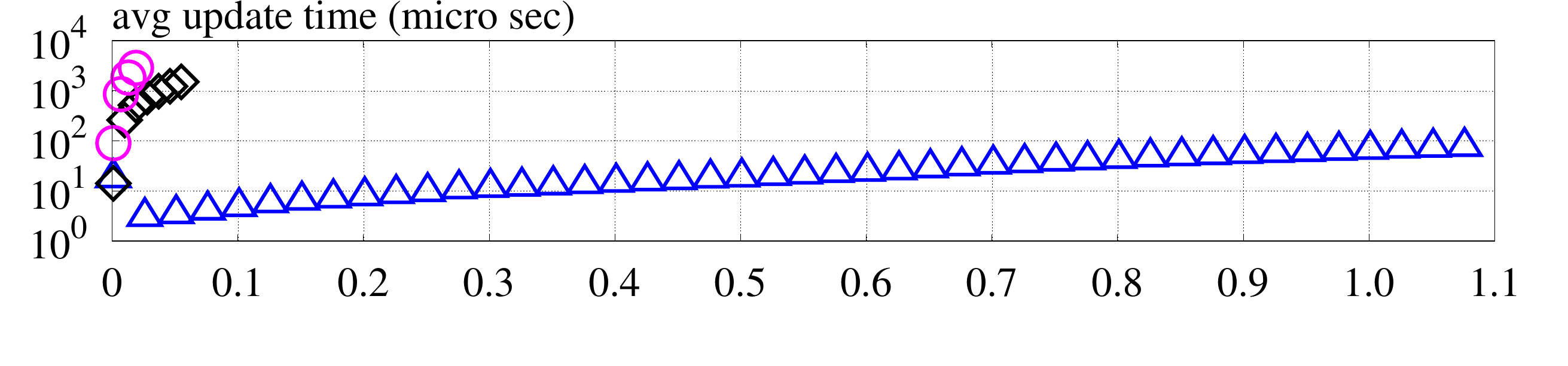}  &
			\hspace{4mm}
			\includegraphics[width=0.3\linewidth]{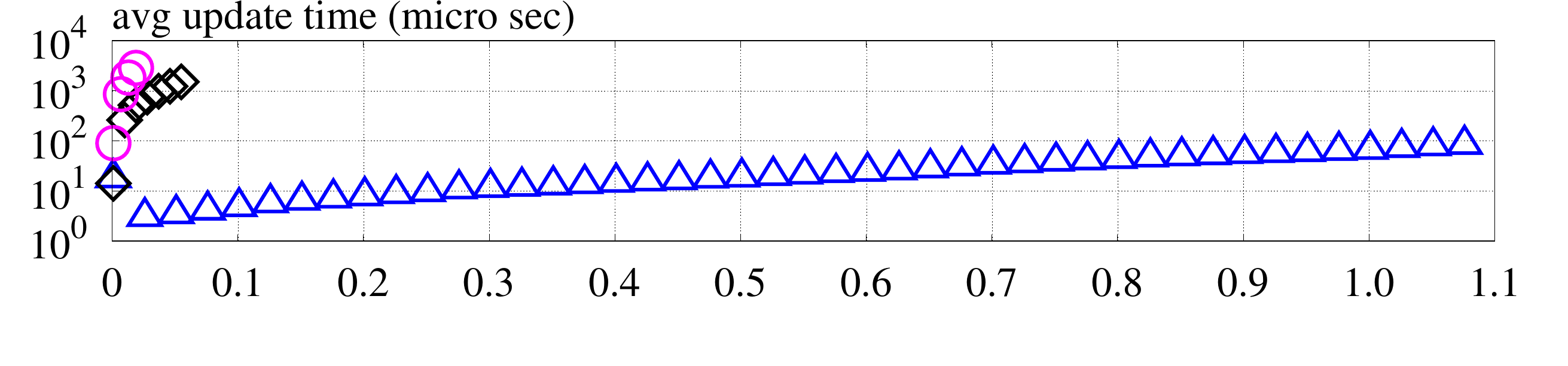}  &
			\hspace{4mm}
			\includegraphics[width=0.3\linewidth]{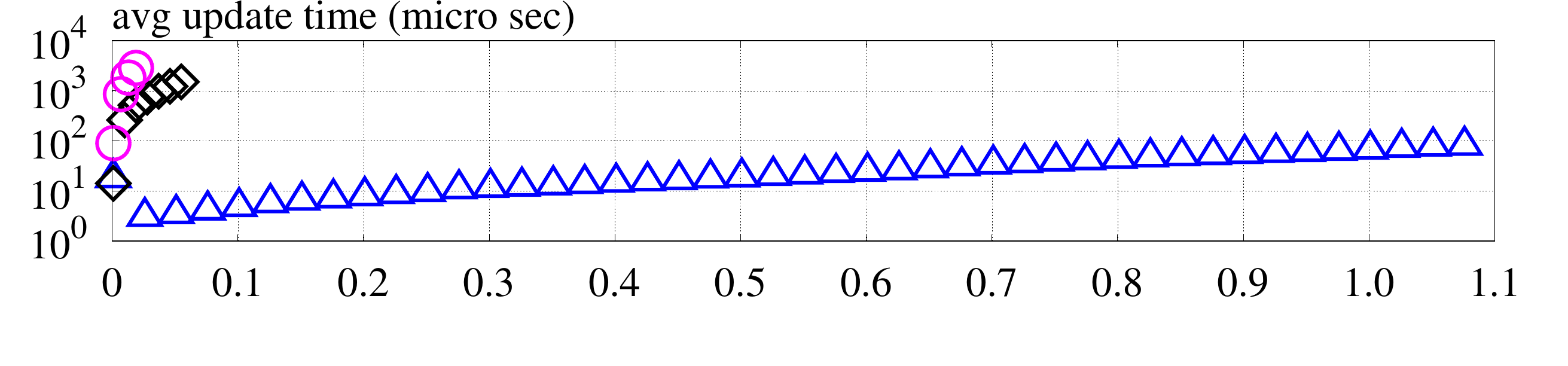}                                                                                                        \\
			\hspace{-8mm} (a) {\em Twitter} ($\mathcal{RR}$)                                       & \hspace{3mm} (b) {\em Twitter} ($\mathcal{DR}$)  & \hspace{4mm} (c) {\em Twitter} ($\mathcal{DD}$)
		\end{tabular}
	}
	\figcapup
	\caption{Average update cost in micro-seconds (i.e., $10^{-6}$ seconds) v.s. update timestamp ($\times~m_0$) {\newblue under Jaccard similarity}}
	\label{fig:avg_vs_update}
	\figcapdown
\end{figure*}

\begin{figure*} [!t]
	\centering
	\vspace{1mm}
	\resizebox{\linewidth}{!}{%
		\begin{tabular}{ccccc}
			\hspace{-6mm} \includegraphics[width=0.2\linewidth]{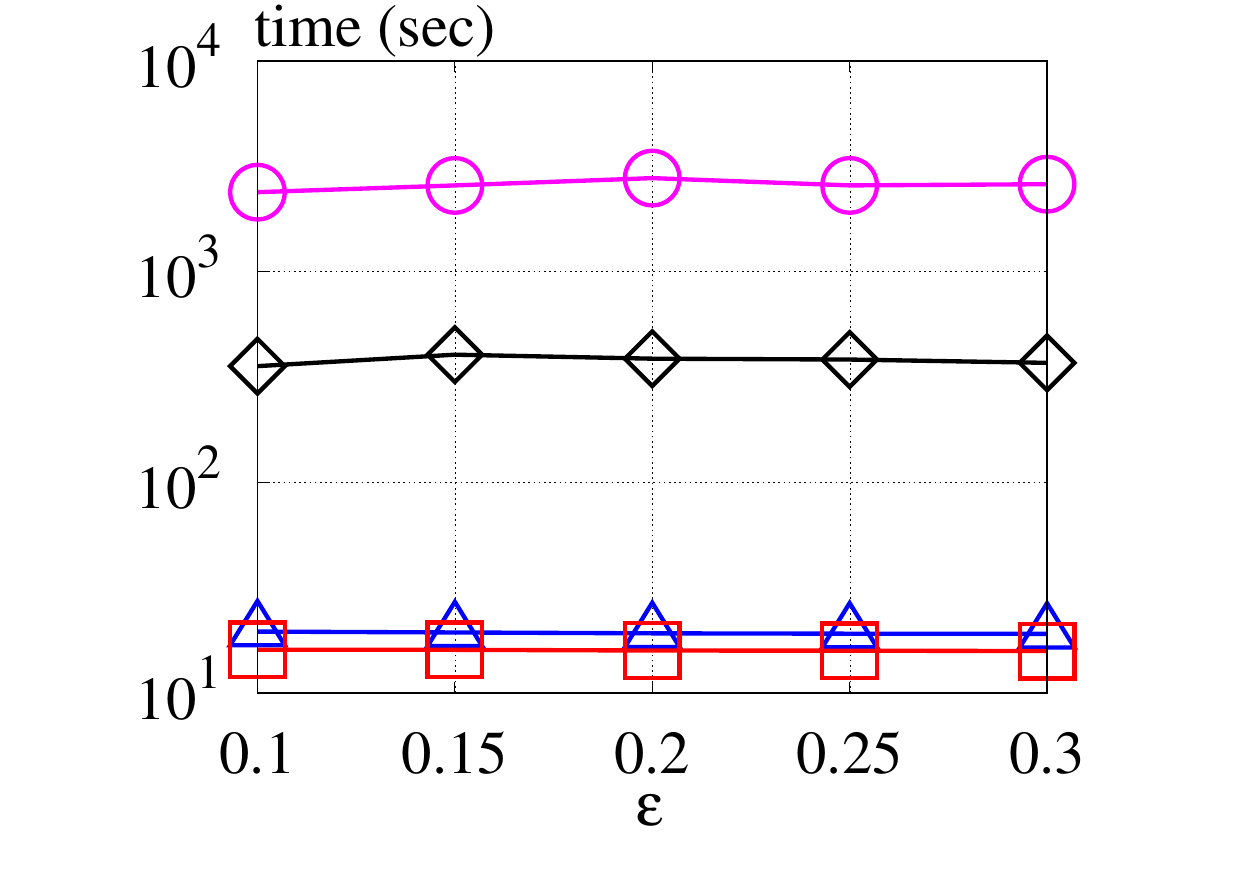} &
			\hspace{-2mm} \includegraphics[width=0.2\linewidth]{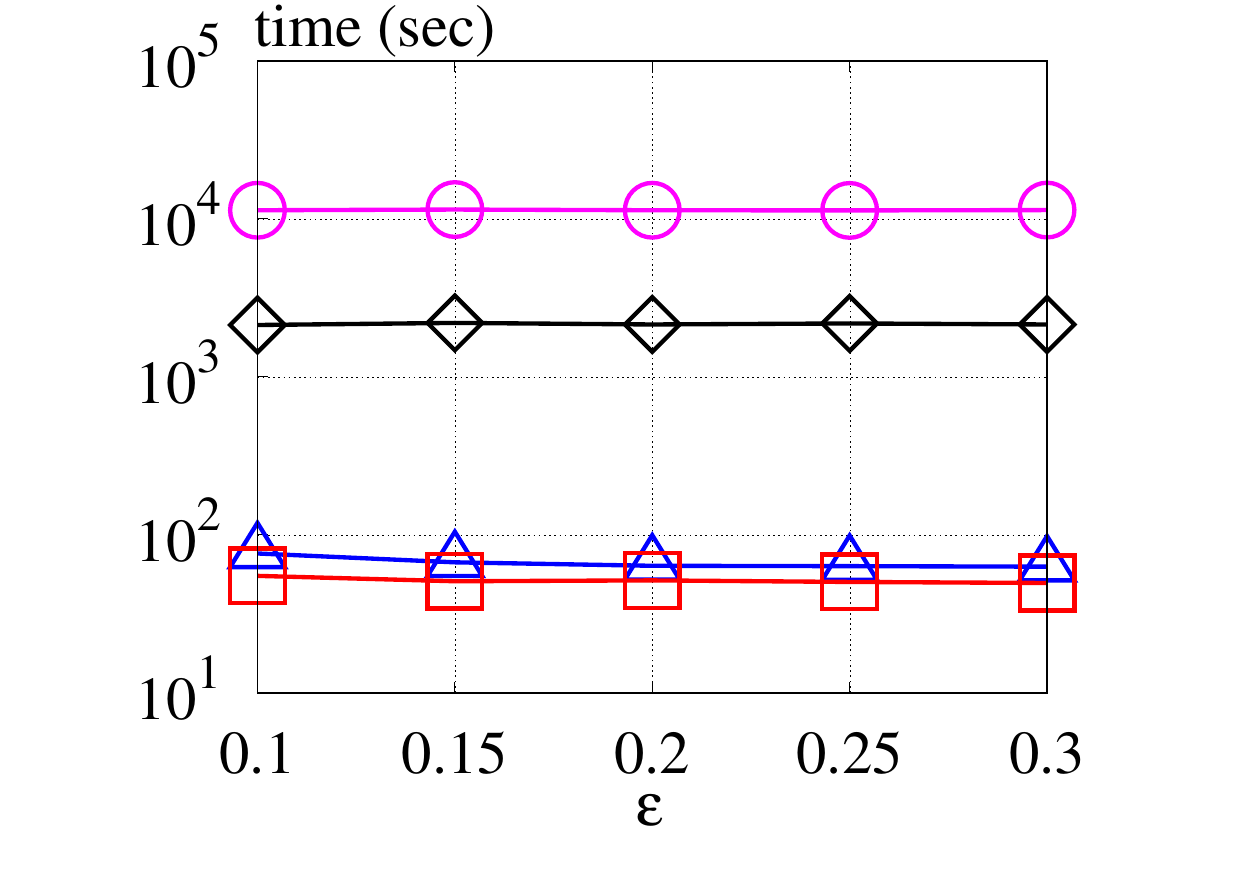}    &
			\hspace{-2mm} \includegraphics[width=0.2\linewidth]{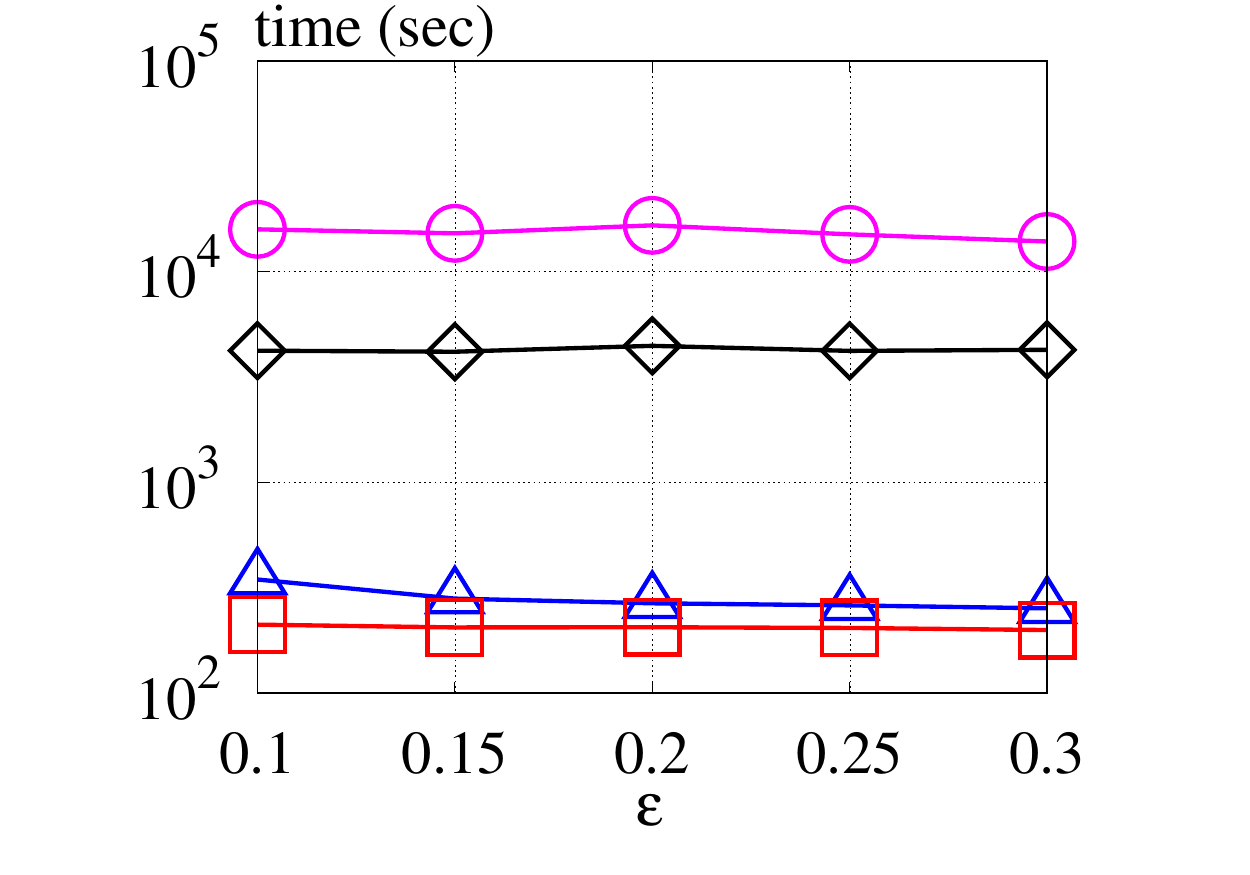}   &
			\hspace{-2mm} \includegraphics[width=0.2\linewidth]{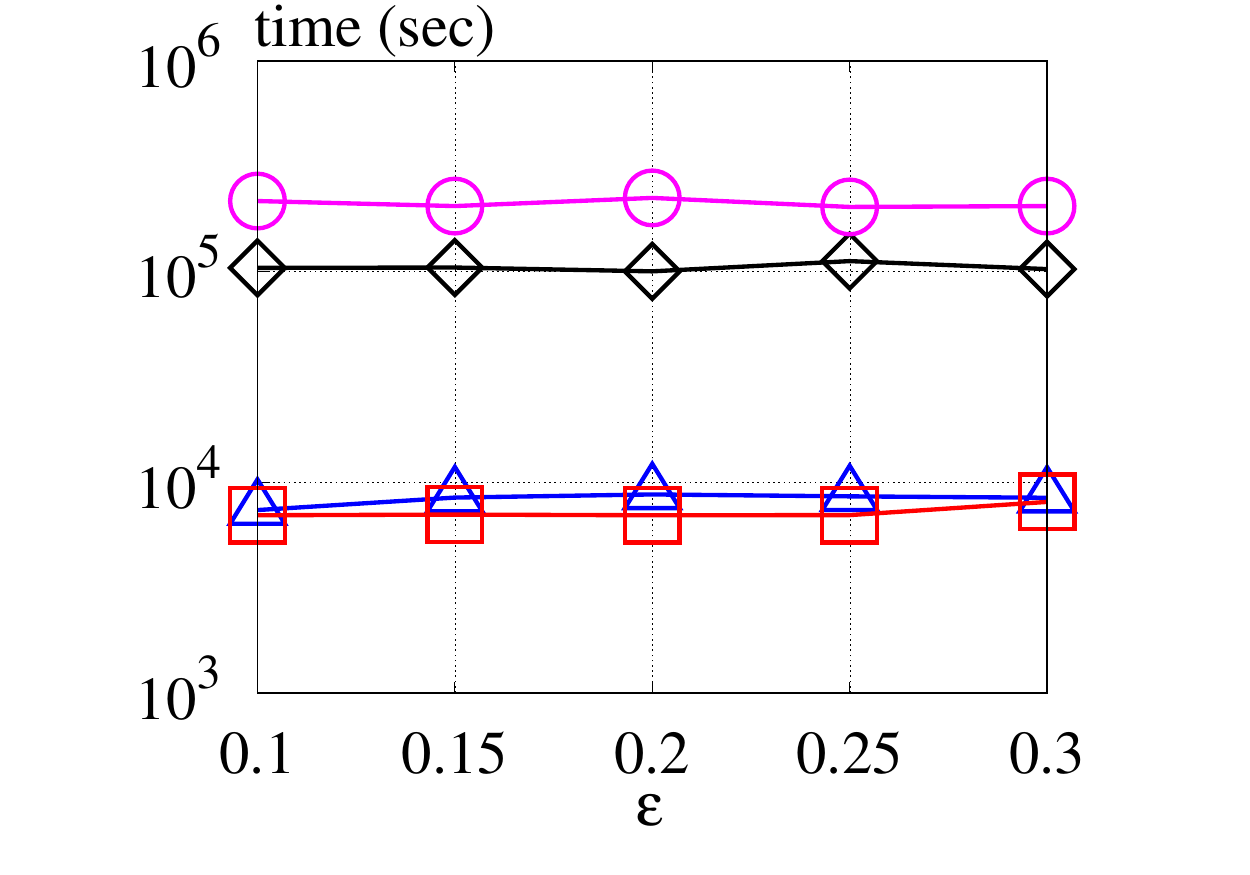}     &
			\hspace{-2mm} \includegraphics[width=0.2\linewidth]{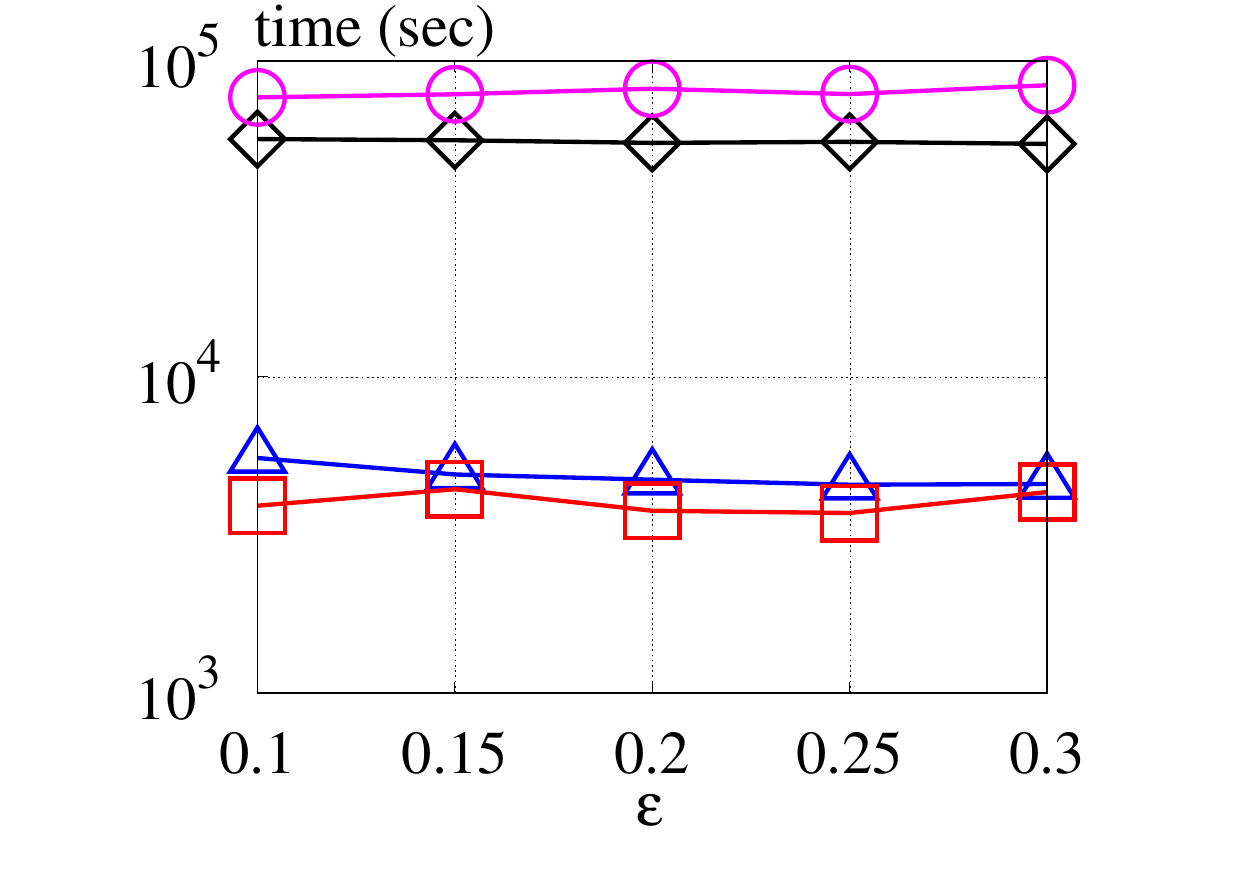}      \\
			\hspace{-6mm} (a) {\em Slashdot}                                                     &
			\hspace{-1mm} (b) {\em Notre}                                                        &
			\hspace{-0mm} (c) {\em Google}                                                       &
			\hspace{-2mm} (d) {\em Wiki}                                                         &
			\hspace{-1mm} (e) {\em LiveJ}
		\end{tabular}
	}
	\figcapup
	\caption{Overall running time v.s. $\eps$ on default settings ($\mu = 5$ and $\rho = 0.01$) {\newblue under Jaccard similarity}}
	\label{fig:time_vs_eps}
	\figcapdown
\end{figure*}

\begin{figure*} [!t]
	\centering
	\resizebox{\linewidth}{!}{%
		\begin{tabular}{ccccc}
			\hspace{-6mm} \includegraphics[width=0.2\linewidth]{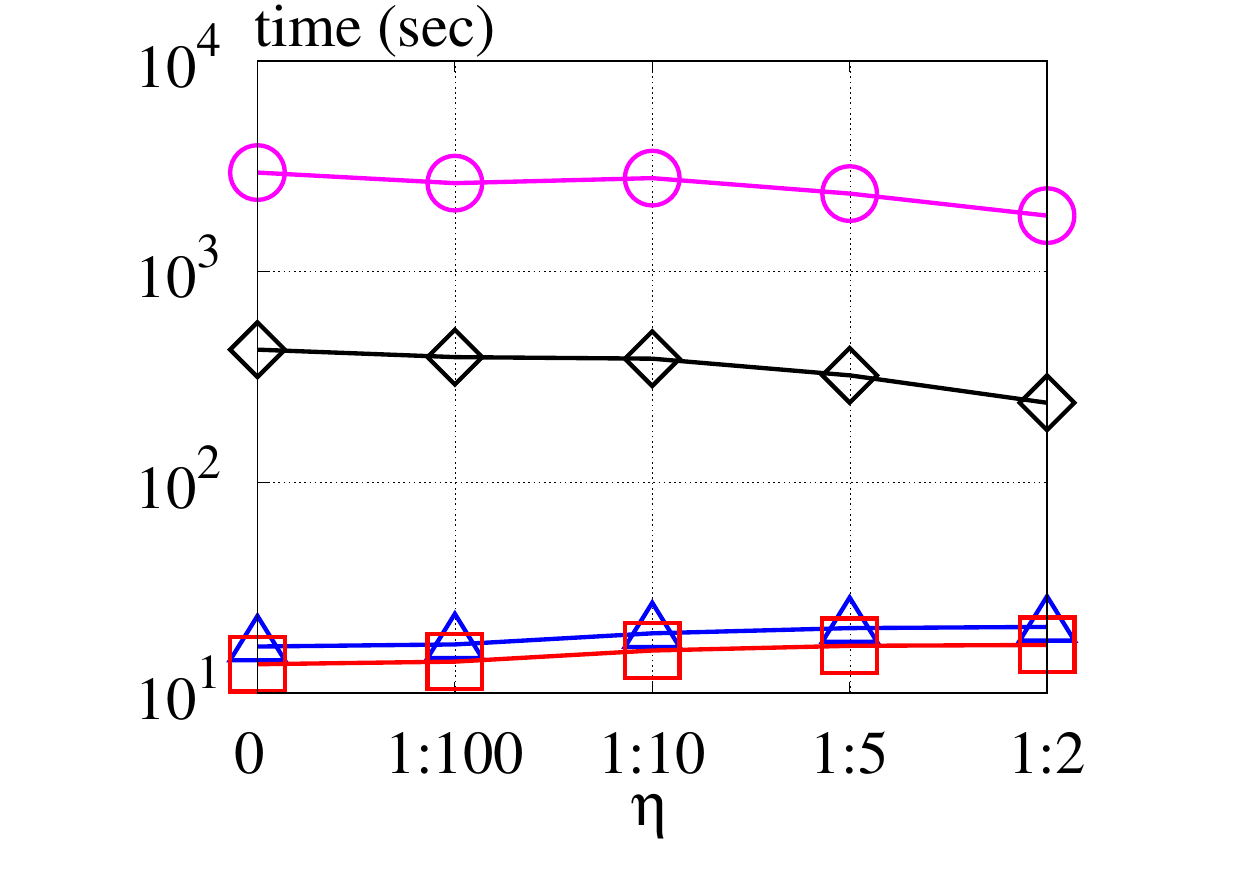} &
			\hspace{-2mm} \includegraphics[width=0.2\linewidth]{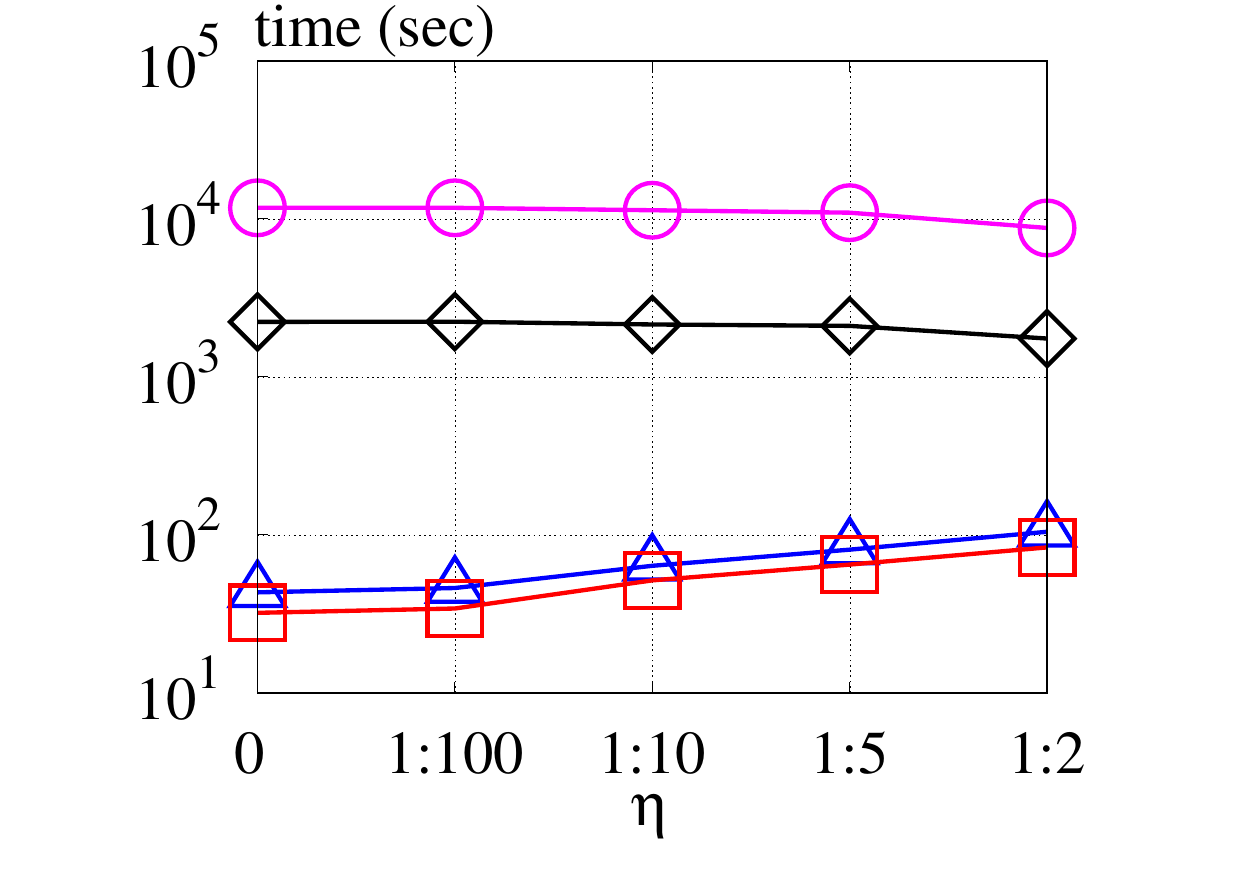}    &
			\hspace{-2mm} \includegraphics[width=0.2\linewidth]{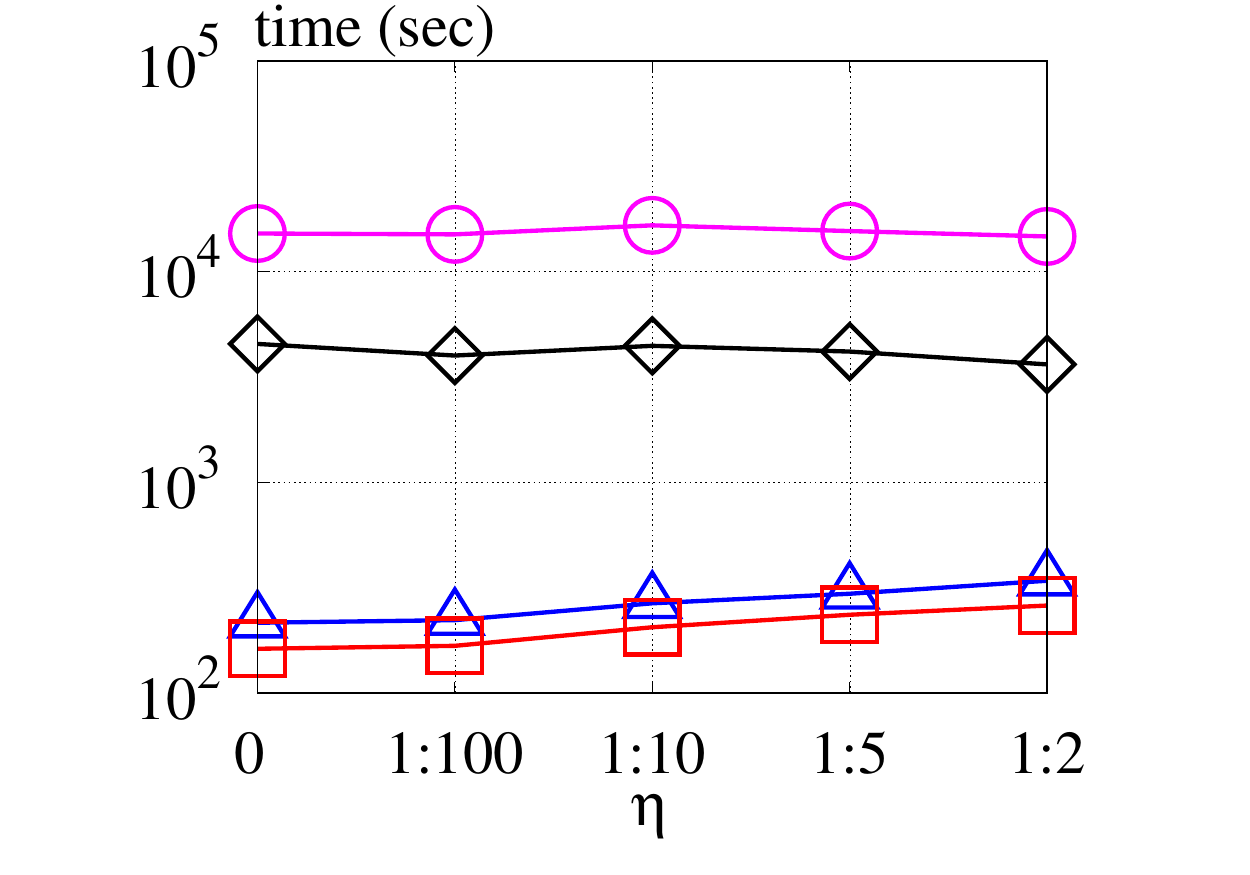}   &
			\hspace{-2mm} \includegraphics[width=0.2\linewidth]{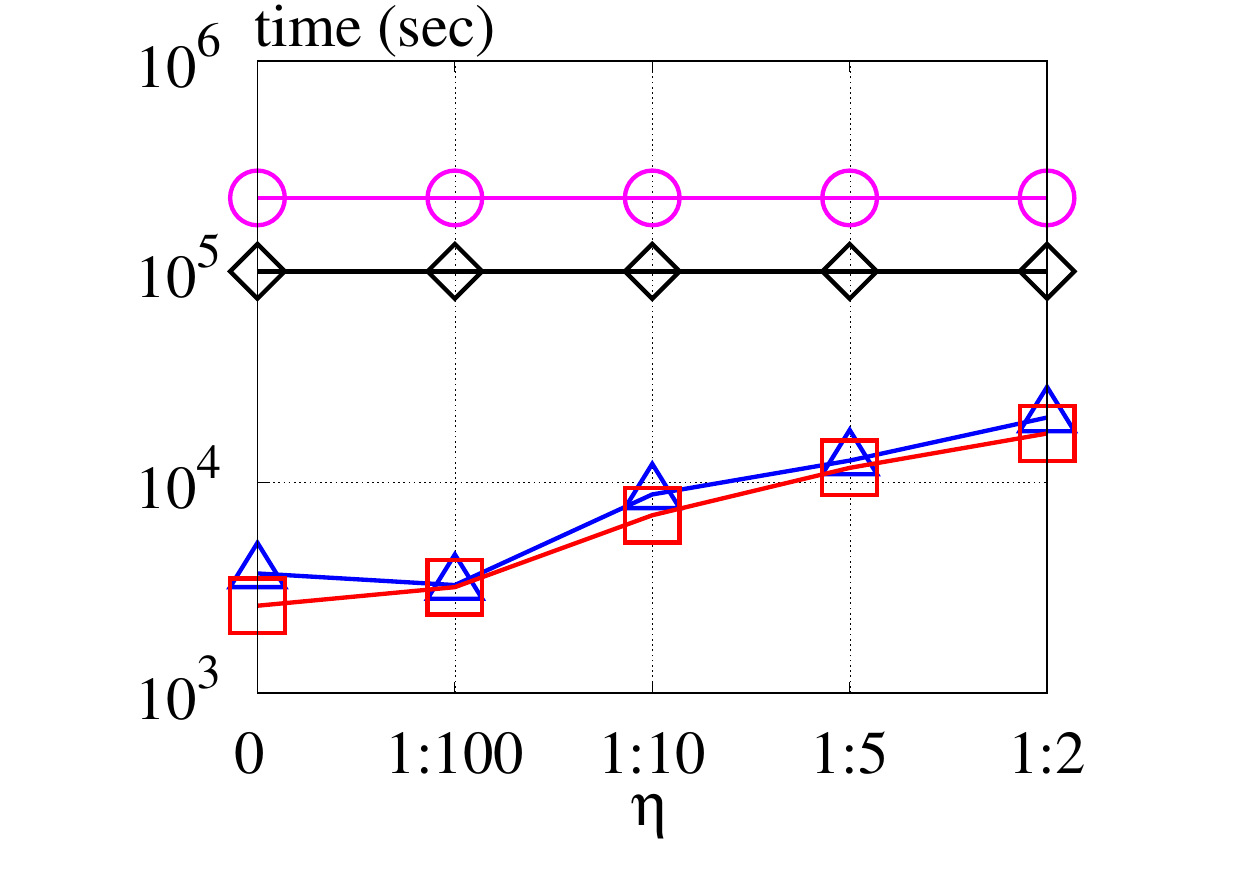}     &
			\hspace{-2mm} \includegraphics[width=0.2\linewidth]{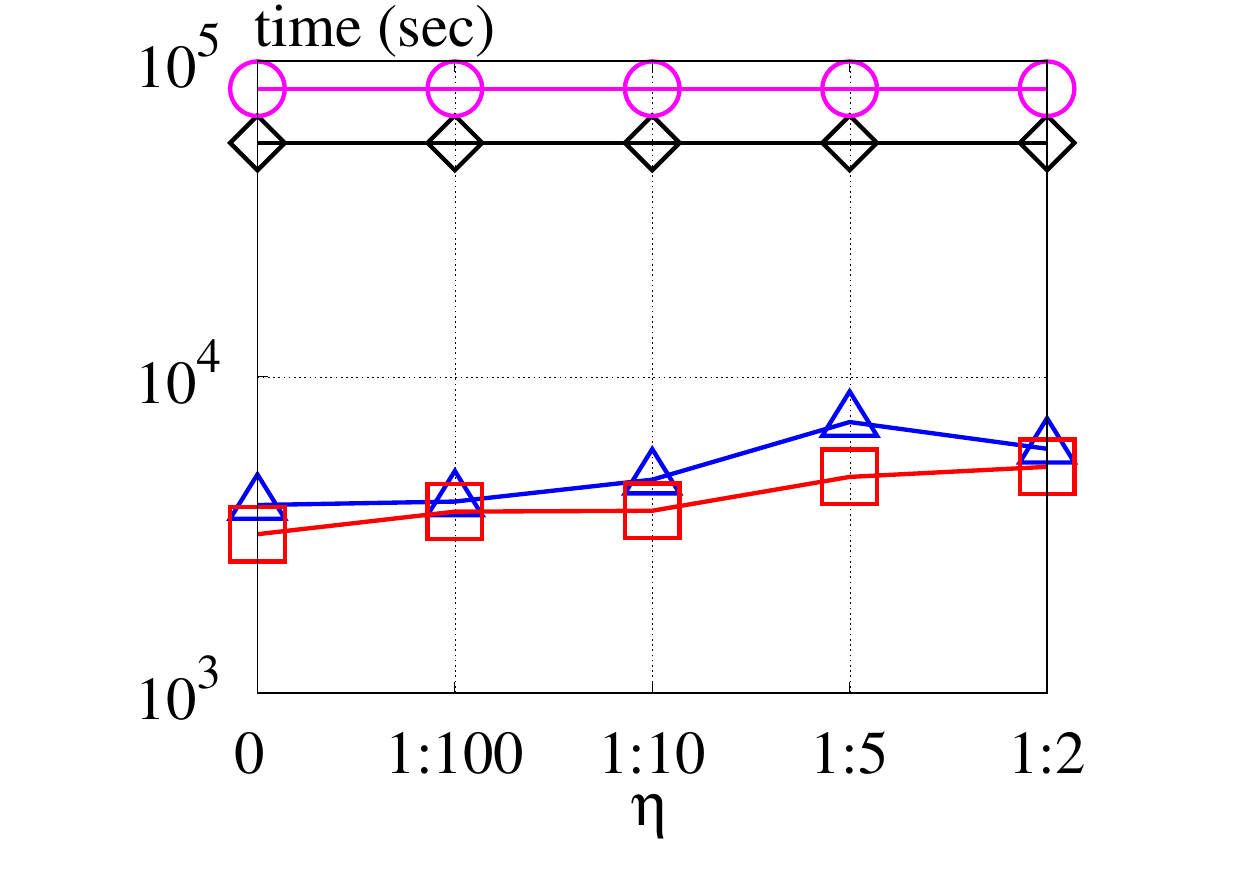}      \\
			\hspace{-6mm} (a) {\em Slashdot}                                                     &
			\hspace{-1mm} (b) {\em Notre}                                                        &
			\hspace{-0mm} (c) {\em Google}                                                       &
			\hspace{-2mm} (d) {\em Wiki}                                                         &
			\hspace{-1mm} (e) {\em LiveJ}
		\end{tabular}
	}
	\figcapup
	\caption{Overall running time v.s. $\eta$ on default settings ($\eps =0.2$, $\mu = 5$ and $\rho = 0.01$) {\newblue under Jaccard similarity}}
	\label{fig:time_vs_eta}
	\vspace{-4mm}
\end{figure*}

\begin{figure*}[!t]
	\centering
	\includegraphics[width=.5\linewidth]{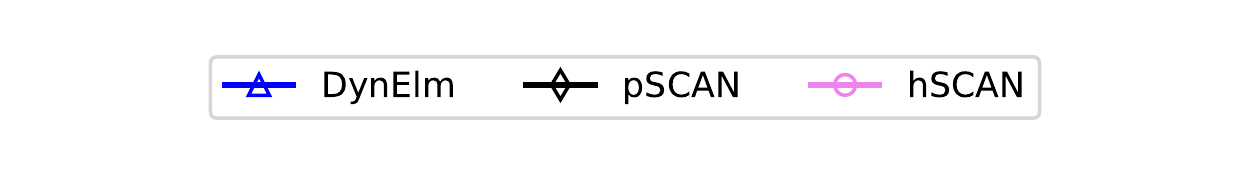}
	\begin{tabular}{cc}
		\includegraphics[width=0.45\linewidth]{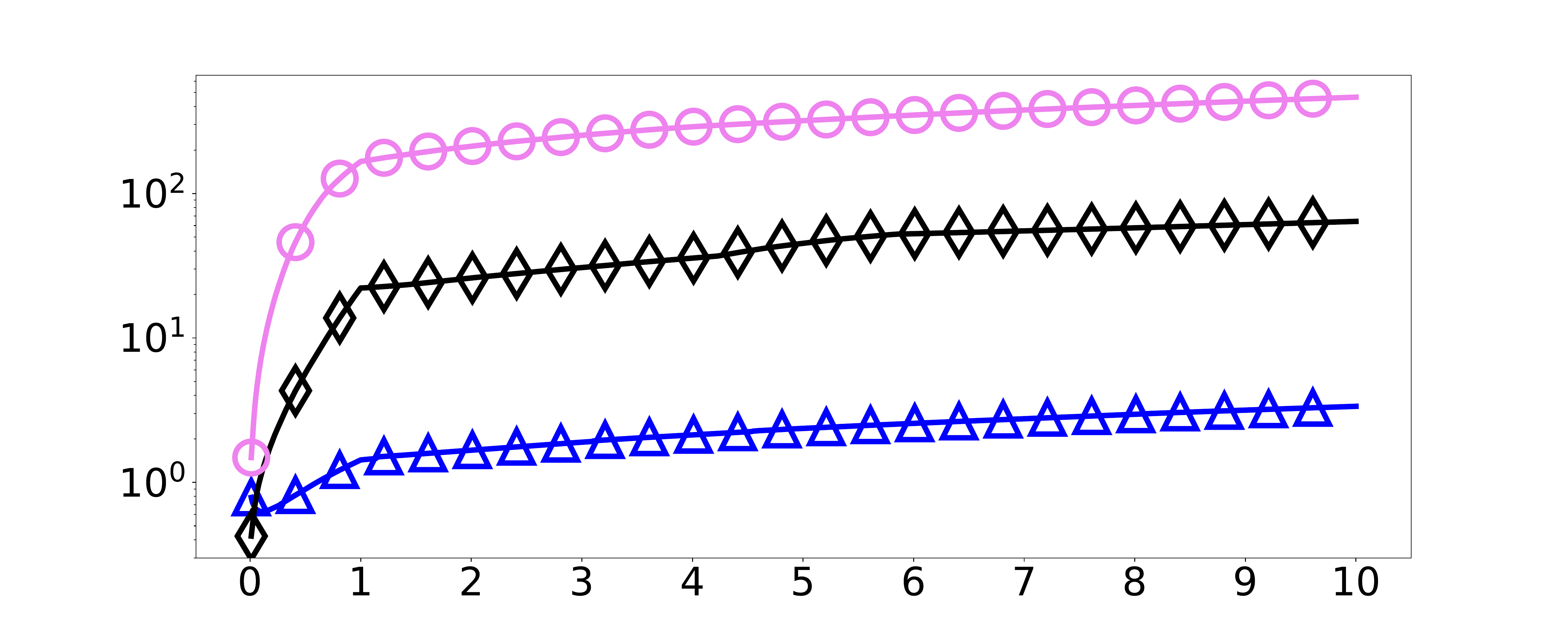} &
		\includegraphics[width=0.45\linewidth]{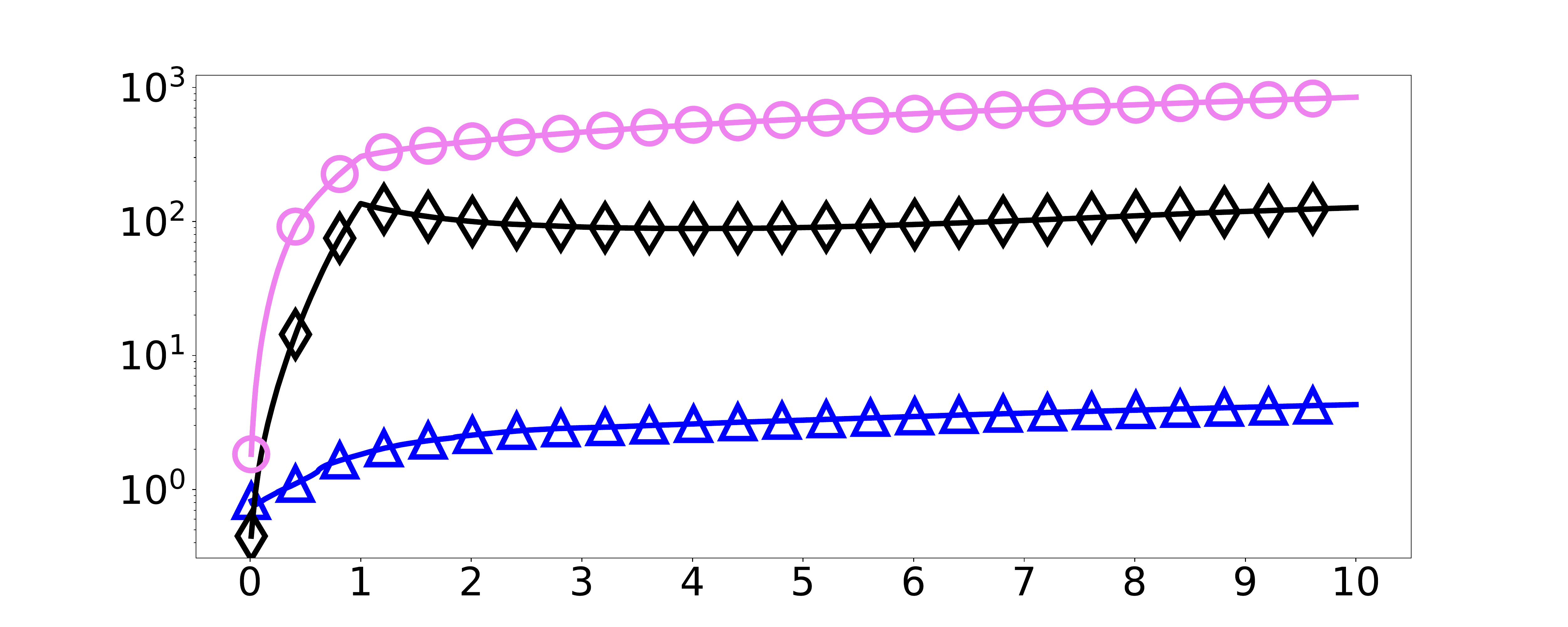}                                 \\
		(a) \textit{Slashdot}                                                      & (b) \textit{Notre} \\
		\includegraphics[width=0.45\linewidth]{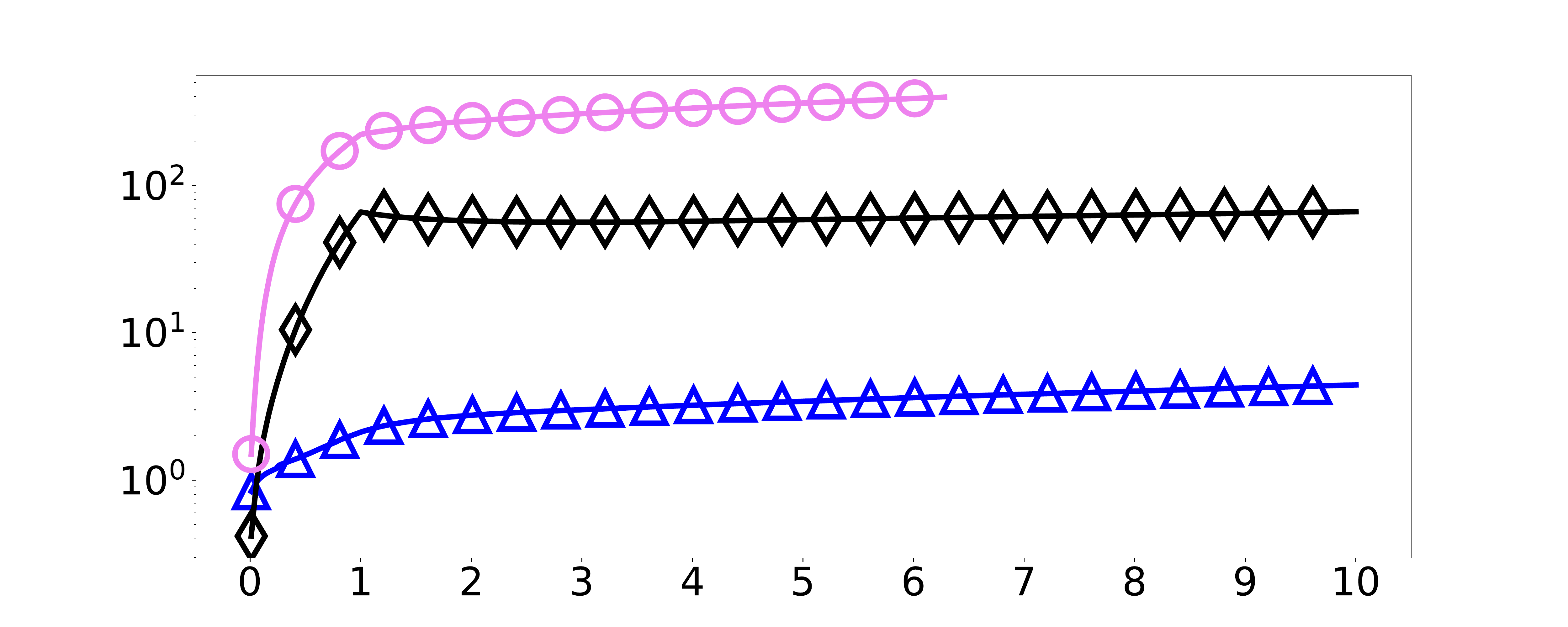}           &
		\includegraphics[width=0.45\linewidth]{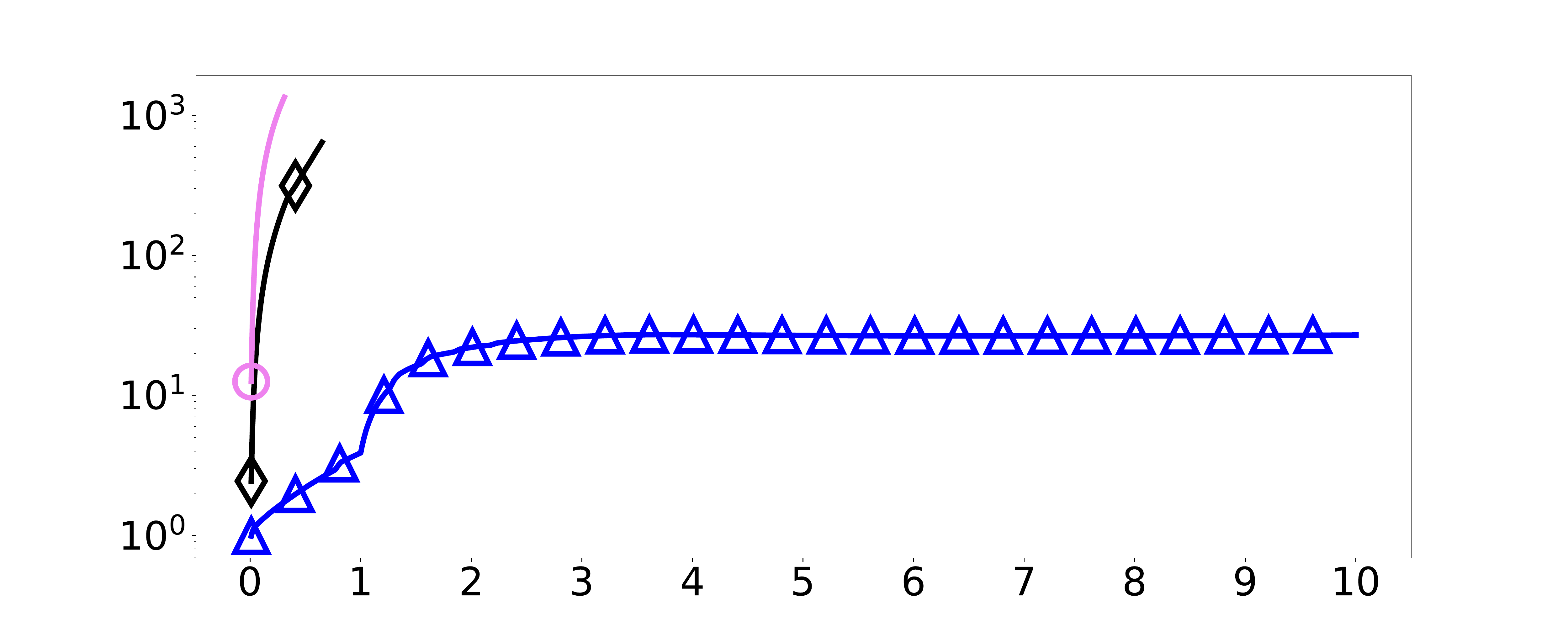}                          \\
		(c) \textit{Google}                                                        & (d) \textit{Wiki}
	\end{tabular} \\
	\begin{tabular}{c}
		\includegraphics[width=0.45\linewidth]{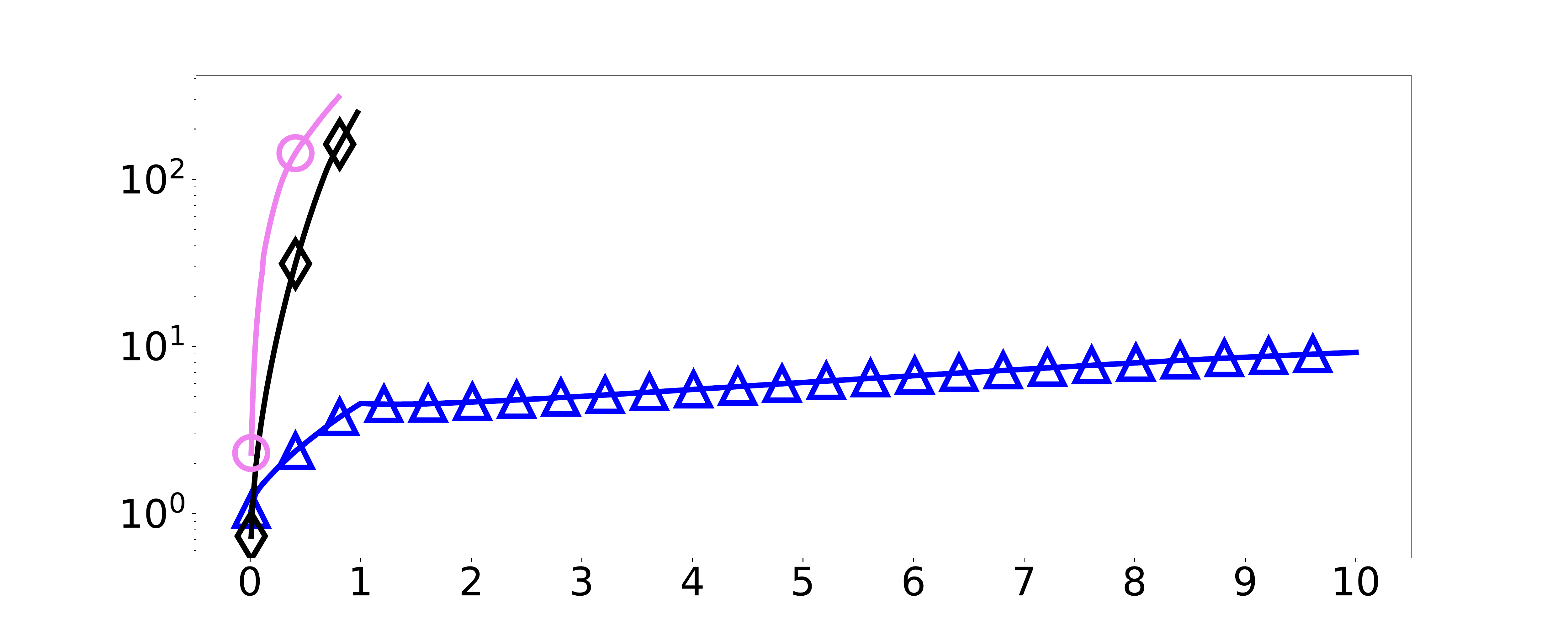} \\
		(e) \textit{LiveJ}
	\end{tabular}
	\caption{\newblue Average update cost in micro-seconds v.s. update timestamp under cosine similarity}
	\label{fig:cosine_efficiency}
\end{figure*}

\vspace{1mm}
\noindent
\textbf{Parameters.} The experiments are conducted with the following parameter settings {\newblue under Jaccard similarity}, where the default value of each parameter is highlighted in
bold and underlined.
\begin{itemize}
	\item $\mu = \underline{\bm{5}}$ and $\delta^* = \underline{\bm{1/n}}$;
	\item $\eps \in \{ 0.1, 0.15, \underline{\bm{0.2}}, 0.25, 0.3\}$;
	\item $\rho \in \{\underline{\bm{0.01}}, 0.05, 0.1, 0.2, 0.5\}$;
	\item $\eta \in \{ 0, 1/100, \underline{\bm{1/10}}, 1/5, 1/2\}$;
	\item insertion strategies: $\{\mathcal{RR}, \underline{\bm{\mathcal{DR}}}, \mathcal{DD}\}$.
\end{itemize}
Unless stated otherwise, when a particular parameter is varied, all the other parameters are set to their default values.

	{\newblue For cosine similarity, we set all parameters to the default setting above except for $\eps=0.6$, which is also the default value of $\eps$ in state-of-the-art works~\cite{cllqz-icde-2016,clqzy-tkde-2017,wqzcl-pvldb-2017}. }

\vspace{1mm}
\noindent
\textbf{Competitors.}
As the superiority of $\pscan$ and $\hscan$ over other existing methods has been shown in their seminal papers~\cite{clqzy-tkde-2017, wqzcl-vldbj-2019}, in our experiments, we focus on the
comparisons between $\pscan$, $\hscan$, and our methods: $\dynelm$ and $\dynstr$.
All these four methods are implemented in C++ (compiled by gcc 9.2.0 with -O3).
The implementations of $\pscan$ and $\hscan$ are provided by their respective authors. Following the instructions in their seminal papers, we adapted them to work with Jaccard similarity. {\newblue We use their source code directly for the experiments under cosine similarity.}

\vspace{1mm}
\noindent
\textbf{Machine and OS.} All the experiments are run on a machine equipped with an Intel(R) Xeon(R) CPU (E7-4830 v2 @ 2.20GHz) and 1TB memory running on Linux (CentOS 7.2).

\begin{figure*}[t]
	\centering
	\resizebox{\linewidth}{!}{%
		\begin{tabular}{cc}
			\includegraphics[width=0.45\linewidth]{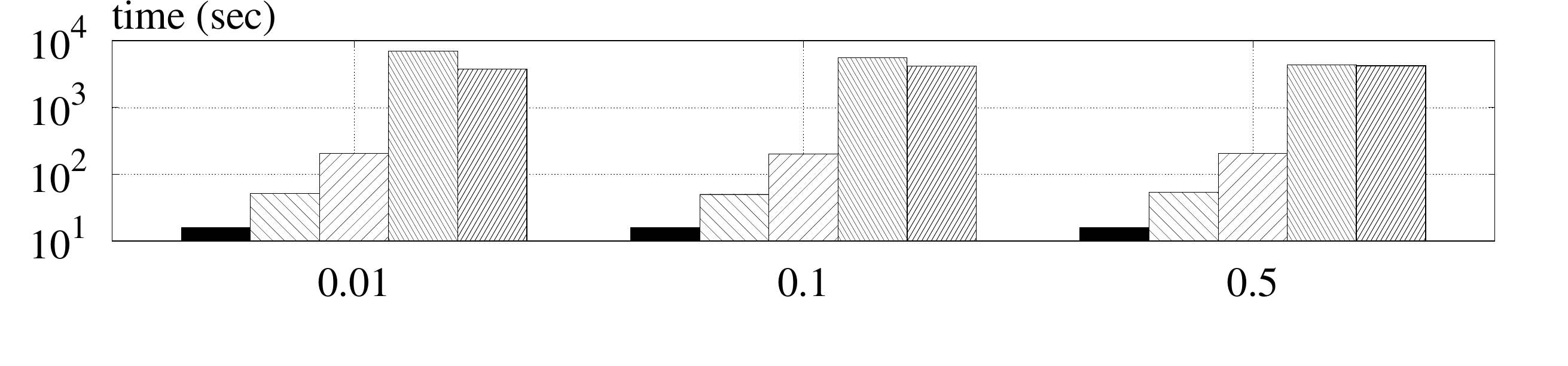}
			                                                        &
			\hfill \includegraphics[width=0.45\linewidth]{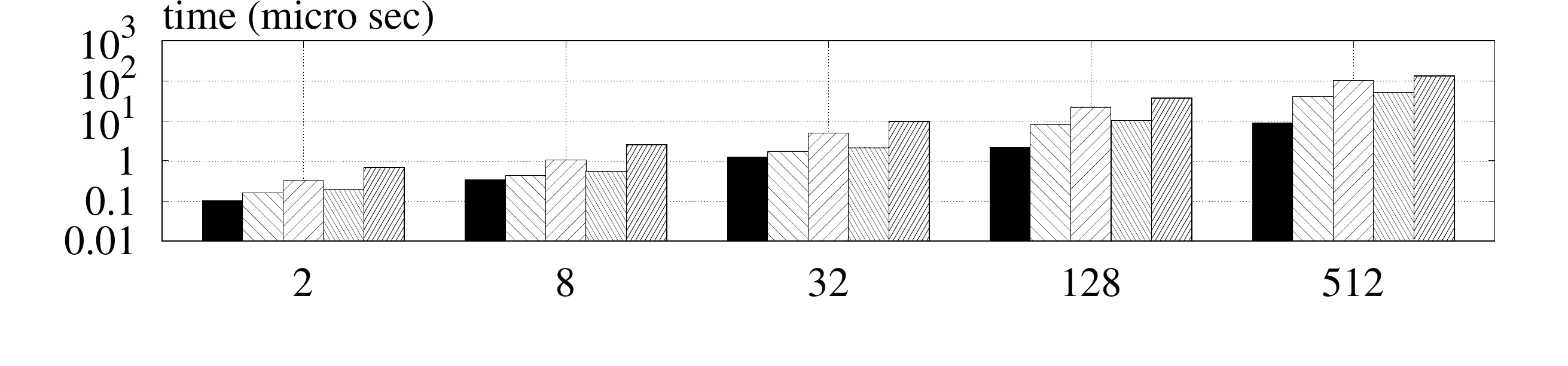} \\
			\hspace{2mm} (a) {\em Overall running time v.s. $\rho$} &
			\hspace{2mm} (b) {\em Query time v.s. query size}
		\end{tabular}
	}
	\figcapup
	\caption{The bars in each group respectively represent in order: {\em Slashdot}, {\em Notre}, {\em Google}, {\em Wiki}, and {\em LiveJ}.}
	\figcapdown
\end{figure*}

\subsection{Overall Performance on All Datasets}\label{sec::all-ds}
\noindent
\textbf{Overall Efficiency on All Datasets.}
We next compare the time efficiency under default settings {\newblue under Jaccard similarity}.
The results are shown in Figure~\ref{fig:time_vs_algo}.
$\dynelm$ is always the most efficient, followed by the slightly slower $\dynstr$. The reason for this is that although $\dynstr$ needs to maintain some extra data structures, the major computational cost still lies in maintaining edge labels.
Except {\em Twitter}, for each dataset, both of our methods can process the whole update sequence within three hours (and within $30$ hours on {\em Twitter}).
However, $\pscan$ and $\hscan$ cannot make that bound.
Worse still,
on {\em Twitter}, for $30$ hours, both $\pscan$ and $\hscan$ can only complete $0.057 \cdot m_0$ and $0.023 \cdot m_0$ updates, respectively;
on the other five large datasets: {\em Wiki}, {\em LiveJ}, {\em Pokec}, {\em Skitter}, and {\em Talk}, they also failed to finish the first~$m_0$ insertions of the corresponding original edges within three hours.
We therefore {\em underestimate} their running times.
For {\em Twitter}, we simply scale the $30$ hours according to their progresses to the end of the whole update process, i.e., $1.1\cdot m_0$,
as the update cost increases with the growth of the graph.
For the other five large datasets, the estimated time is computed as follows:
(i) we run the two algorithms on the graph with all original edges, as if in the static case;
(ii) based on  the state after first step, let the two algorithms process the~$9\cdot m_0$ generated updates for one hour;
(iii) according to the percentage of updates processed within the hour, we estimate the total update time for the~$9\cdot m_0$ updates by scaling;
and (iv) add the three hours for the first~$m_0$ insertions to this estimated time.

From Figure~\ref{fig:time_vs_algo}, except for {\em GrQc}, {\em CondMat} and {\em dblp}, both of our methods are
$10\times$ faster than $\pscan$ on every dataset.
In particular, on the aforementioned large datasets,
our methods outperform $\pscan$ by up to two orders of magnitude.
This is because on these datasets, the edge in an update is more likely to incident on a large-degree vertex~$u$, for which $\pscan$ needs~$O(d[u])$ time, while our methods guarantee~$O(\log^2 n)$ amortized cost (according to our current parameter setting).
Finally, $\hscan$ is slower than $\pscan$ on every dataset, as its update cost complexity is $O(n\log n)$, having a factor of $\log n$ over the update cost of $\pscan$.
Particularly on {\em Wiki}, {\em LiveJ} and {\em Pokec}, $\dynelm$ is almost~$1000\times$ faster than $\hscan$.

\vspace{1mm}
\noindent
\textbf{Memory Consumption.}
The last four columns in Table~\ref{tab:meta-info} show the {peak} memory usage of the four competing methods
over the whole update process {\newblue under Jaccard similarity},
where the numbers in parentheses are estimates, for the corresponding methods fail to complete the update process within a reasonable time.
From the table, we can see that the space consumptions of $\dynelm$ and $\pscan$ are similar, and they are the most space efficient.
$\dynstr$ consumes about~$10\%$ to~$20\%$ more space than $\dynelm$, because it needs to maintain the CC structures. Nonetheless,  the space
consumptions of all are linear in the graph size, consistent with the theoretical guarantees.

\subsection{Efficiency on the Representative Datasets}\label{sec::efficiency}
\noindent
\textbf{Average Cost v.s. Updates.}
Next, we study the average update cost over the simulated update process, with varying insertion generation strategies {\newblue under Jaccard similarity}.
That is for each of $\mathcal{RR}$, $\mathcal{DR}$ and $\mathcal{DD}$, we assess the running time on the five representative datasets
plus {\em Twitter} with all other parameters set to default values.
During the update process, the average update cost at {\em timestamp}~$t$ is calculated as the average running time over all the first~$t$ updates.
The results are shown in Figure~\ref{fig:avg_vs_update}.
As explained earlier, $\dynelm$ is slightly faster than $\dynstr$, leading to the curves of their average update costs over updates are very close to each other in a log-scale y-axis.
For clarity, we omit the curve of $\dynelm$ from Figure~\ref{fig:avg_vs_update}.

For all the strategies on all the datasets, as expected, $\dynstr$ is the most efficient, followed by $\pscan$, and then $\hscan$.
For a fixed dataset, the average update costs of all the three algorithms increase with the strategies in the order of $\mathcal{RR}$, $\mathcal{DR}$ and $\mathcal{DD}$.
This is because, the chance of a generated update incident on a large-degree vertex increases with the strategy changing from $\mathcal{RR}$, to $\mathcal{DR}$ and then to $\mathcal{DD}$.
Since both $\pscan$ and $\hscan$ need to scan the neighbourhoods of the vertices of the edge in an update, they have to pay more costs on those updates generated with $\mathcal{DD}$.
This trend can be seen from Figure~\ref{fig:avg_vs_update}, and it also explains why $\pscan$ fails to complete the update sequence within three hours
on all the five datasets under $\mathcal{DD}$.
In contrast, as our methods guarantee a~$O(\log^2 n)$ amortized update cost; and are somewhat immune to the insertion strategy change.
Moreover, $\pscan$ fails to complete the first~$m_0$ insertions within three hours on {\em Wiki} and {\em LiveJ}, and within 30  hours on {\em Twitter}.
However, according to the first few
data points plotted in the figures, $\dynstr$ already outperforms $\pscan$ by up to two-to-three orders of magnitude on {\em LiveJ}, {\em Wiki} and {\em Twitter}, respectively, with all three strategies.
Not to mention $\hscan$,  which is even slower than $\pscan$.
Moreover, it is worth mentioning that even on the dataset with 1.2 billion edges, {\em Twitter}, our algorithm can still
perform an update, in average, less than $100$ micro-seconds.
This clearly shows the scalability of our methods.

	{\newblue
		For cosine similarity, since $\dynelm$ and $\dynstr$ performs similarly, we choose $\dynelm$, $\pscan$ and $\hscan$ as three competing methods.The results are shown in Figure~\ref{fig:cosine_efficiency}. The curves are very like the curves shown in Figure~\ref{fig:avg_vs_update}. Among all three algorithms, $\dynelm$ is always the most efficient one on every representative datasets, which matches our expectation. Under cosine similarity, both $\pscan$ and $\hscan$ still need to scan the neighbourhoods of the effected vertices to process one single update.
		On \textit{Wiki} and \textit{LiveJ}, our algorithm has already outperformed $\pscan$ by two-to-three orders of magnitude after 3 hours. Since $\hscan$ takes more time in processing one update than $\pscan$, the gap between it and our algorithm will be even huger.

		Although from the theoretical analysis, the amortized cost under cosine similarity suffer from a roughly $O(\frac{1}{\eps})$ factor loss, the real performances of our algorithm under these two similarities are nearly identical. The reason is that, as mentioned above, the value of $\eps$ under cosine similarity is generally larger. Therefore, the theoretical performance of our algorithms under these two similarities won't vary much. This is an evidence that our algorithms are highly capable under different forms of structural similarity.

	}

\vspace{1mm}
\noindent
\textbf{Varying Parameters under Default Setting.}
Figure~\ref{fig:time_vs_eps} and Figure~\ref{fig:time_vs_eta} show the overall running time on the five representatives, varying~$\eps$ and~$\eta$, respectively.
Again, in both of these experiments, our methods are up to $1000\times$ faster than the two competitors.
With~$\eps$ increasing, the overall running times of our algorithms drop slightly, while with~$\eta$ increasing, i.e., more deletions in the update process, both $\dynelm$ and $\dynstr$ have the running time increased.
However, in this case, the running times of $\pscan$ and $\hscan$ both decrease slightly: with more deletions, the degrees of the vertices tend to decrease.
Finally, the last two bar charts show the experimental results on overall running time of $\dynelm$ v.s. $\rho$ and cluster-group-by query time of $\dynstr$ v.s. query size, both on the five representative datasets.
As for the former, $\dynelm$ is not as sensitive to~$\rho$, as the theoretical complexity suggested, while
for the latter, the query time of $\dynstr$ increases roughly linearly with the query size, consistent
with the query time complexity.

\section{Conclusion}\label{sec:conclusion}
In this paper, we study structural clustering ($\strclu$) graphs subject to edge insertions and deletions with the Jaccard similarity {\newblue and cosine similarity}.
Given a failure probability $\delta^*$, for every sequence of $M$ updates,
our algorithm $\dynelm$ can process each update in $O(\log^2 n + \log n \cdot \log \frac{M}{\delta^*})$ amortized time -- a significant improvement on the state-of-the-art~$O(n)$ bound {\newblue under both similarities}.
Furthermore, its space consumption is linear in the current size of the graph, i.e., $O(n + m)$, at all times.
Meanwhile, it guarantees that the $\strcluresult$ can be retrieved in $O(n + m)$ time upon request and the clustering result is correct under the $\rho$-approximate notion with probability at least $1 - \delta^*$.
Based on $\dynelm$, our ultimate algorithm $\dynstr$ not only achieves all the above guarantees of $\dynelm$, but also answers any cluster-group-by query of $Q\subseteq V$ in $O(|Q|\cdot \log n)$ time.
We conducted extensive experiments on 15 real datasets including a billion-edge dataset {\em Twitter}.
The experimental results confirm that both of our methods are up to $1000 \times$ faster than the state-of-the-art competitors in handling updates, while still produce quality clustering. {\newblue We also study the difference between these two similarities in terms of approximation quality. We find that the quality of produced approximate clustering results are better under Jaccard similarity while the performances are similar. This may serve as a guidance when choosing proper similarity definition for real applications.}

\section{Acknowledgement}
In this work, Junhao Gan is supported by Australian Research Council (ARC)
Discovery Early Career Researcher Award (DECRA) DE190101118, and Anthony Wirth is supported in part by ARC Discovery Project (DP) DP190102078.

\bibliographystyle{ACM-Reference-Format}
\bibliography{ref}


\begin{thebibliography}{42}


\ifx \showCODEN    \undefined \def \showCODEN     #1{\unskip}     \fi
\ifx \showDOI      \undefined \def \showDOI       #1{#1}\fi
\ifx \showISBNx    \undefined \def \showISBNx     #1{\unskip}     \fi
\ifx \showISBNxiii \undefined \def \showISBNxiii  #1{\unskip}     \fi
\ifx \showISSN     \undefined \def \showISSN      #1{\unskip}     \fi
\ifx \showLCCN     \undefined \def \showLCCN      #1{\unskip}     \fi
\ifx \shownote     \undefined \def \shownote      #1{#1}          \fi
\ifx \showarticletitle \undefined \def \showarticletitle #1{#1}   \fi
\ifx \showURL      \undefined \def \showURL       {\relax}        \fi
\providecommand\bibfield[2]{#2}
\providecommand\bibinfo[2]{#2}
\providecommand\natexlab[1]{#1}
\providecommand\showeprint[2][]{arXiv:#2}

\bibitem[\protect\citeauthoryear{Bansal, Blum, and Chawla}{Bansal
  et~al\mbox{.}}{2004}]%
        {bbc-ml-2004}
\bibfield{author}{\bibinfo{person}{Nikhil Bansal}, \bibinfo{person}{Avrim
  Blum}, {and} \bibinfo{person}{Shuchi Chawla}.}
  \bibinfo{year}{2004}\natexlab{}.
\newblock \showarticletitle{Correlation Clustering}.
\newblock \bibinfo{journal}{\emph{Machine Learning}} \bibinfo{volume}{56},
  \bibinfo{number}{1-3} (\bibinfo{year}{2004}), \bibinfo{pages}{89--113}.
\newblock


\bibitem[\protect\citeauthoryear{Bastian, Heymann, and Jacomy}{Bastian
  et~al\mbox{.}}{2009}]%
        {ICWSM09154}
\bibfield{author}{\bibinfo{person}{Mathieu Bastian}, \bibinfo{person}{Sebastien
  Heymann}, {and} \bibinfo{person}{Mathieu Jacomy}.}
  \bibinfo{year}{2009}\natexlab{}.
\newblock \bibinfo{title}{Gephi: An Open Source Software for Exploring and
  Manipulating Networks}.
\newblock
\newblock
\urldef\tempurl%
\url{http://www.aaai.org/ocs/index.php/ICWSM/09/paper/view/154}
\showURL{%
\tempurl}


\bibitem[\protect\citeauthoryear{Bezdek and Pal}{Bezdek and Pal}{1998}]%
        {bp-tsmc-1998}
\bibfield{author}{\bibinfo{person}{James~C Bezdek} {and}
  \bibinfo{person}{Nikhil~R Pal}.} \bibinfo{year}{1998}\natexlab{}.
\newblock \showarticletitle{Some new indexes of cluster validity}.
\newblock \bibinfo{journal}{\emph{IEEE Transactions on Systems, Man, and
  Cybernetics, Part B (Cybernetics)}} \bibinfo{volume}{28}, \bibinfo{number}{3}
  (\bibinfo{year}{1998}), \bibinfo{pages}{301--315}.
\newblock


\bibitem[\protect\citeauthoryear{Broder}{Broder}{1997}]%
        {b-sequences-1997}
\bibfield{author}{\bibinfo{person}{Andrei~Z. Broder}.}
  \bibinfo{year}{1997}\natexlab{}.
\newblock \showarticletitle{On the resemblance and containment of documents}.
  In \bibinfo{booktitle}{\emph{Compression and Complexity of {SEQUENCES} 1997,
  Positano, Amalfitan Coast, Salerno, Italy, June 11-13, 1997, Proceedings}}.
  \bibinfo{pages}{21--29}.
\newblock


\bibitem[\protect\citeauthoryear{Chang, Li, Lin, Qin, and Zhang}{Chang
  et~al\mbox{.}}{2016}]%
        {cllqz-icde-2016}
\bibfield{author}{\bibinfo{person}{Lijun Chang}, \bibinfo{person}{Wei Li},
  \bibinfo{person}{Xuemin Lin}, \bibinfo{person}{Lu Qin}, {and}
  \bibinfo{person}{Wenjie Zhang}.} \bibinfo{year}{2016}\natexlab{}.
\newblock \showarticletitle{pSCAN: Fast and exact structural graph clustering}.
  In \bibinfo{booktitle}{\emph{32nd {IEEE} International Conference on Data
  Engineering, {ICDE} 2016, Helsinki, Finland, May 16-20, 2016}}.
  \bibinfo{pages}{253--264}.
\newblock


\bibitem[\protect\citeauthoryear{Chang, Li, Qin, Zhang, and Yang}{Chang
  et~al\mbox{.}}{2017}]%
        {clqzy-tkde-2017}
\bibfield{author}{\bibinfo{person}{Lijun Chang}, \bibinfo{person}{Wei Li},
  \bibinfo{person}{Lu Qin}, \bibinfo{person}{Wenjie Zhang}, {and}
  \bibinfo{person}{Shiyu Yang}.} \bibinfo{year}{2017}\natexlab{}.
\newblock \showarticletitle{pSCAN: Fast and Exact Structural Graph Clustering}.
\newblock \bibinfo{journal}{\emph{{IEEE} Trans. Knowl. Data Eng.}}
  \bibinfo{volume}{29}, \bibinfo{number}{2} (\bibinfo{year}{2017}),
  \bibinfo{pages}{387--401}.
\newblock


\bibitem[\protect\citeauthoryear{Chawathe}{Chawathe}{2019}]%
        {c-cmbda-2019}
\bibfield{author}{\bibinfo{person}{Sudarshan~S Chawathe}.}
  \bibinfo{year}{2019}\natexlab{}.
\newblock \showarticletitle{Clustering blockchain data}.
\newblock In \bibinfo{booktitle}{\emph{Clustering Methods for Big Data
  Analytics}}. \bibinfo{publisher}{Springer}, \bibinfo{pages}{43--72}.
\newblock


\bibitem[\protect\citeauthoryear{Cormode, Muthukrishnan, and Yi}{Cormode
  et~al\mbox{.}}{2011}]%
        {gsk-ta-2011}
\bibfield{author}{\bibinfo{person}{Graham Cormode}, \bibinfo{person}{S.
  Muthukrishnan}, {and} \bibinfo{person}{Ke Yi}.}
  \bibinfo{year}{2011}\natexlab{}.
\newblock \showarticletitle{Algorithms for Distributed Functional Monitoring}.
\newblock \bibinfo{journal}{\emph{ACM Trans. Algorithms}} \bibinfo{volume}{7},
  \bibinfo{number}{2}, Article \bibinfo{articleno}{21} (\bibinfo{date}{March}
  \bibinfo{year}{2011}), \bibinfo{numpages}{21:1--21:20}~pages.
\newblock
\showISSN{1549-6325}


\bibitem[\protect\citeauthoryear{Ding, He, Zha, Gu, and Simon}{Ding
  et~al\mbox{.}}{2001}]%
        {dhzgs-icdm-2001}
\bibfield{author}{\bibinfo{person}{Chris H.~Q. Ding}, \bibinfo{person}{Xiaofeng
  He}, \bibinfo{person}{Hongyuan Zha}, \bibinfo{person}{Ming Gu}, {and}
  \bibinfo{person}{Horst~D. Simon}.} \bibinfo{year}{2001}\natexlab{}.
\newblock \showarticletitle{A Min-max Cut Algorithm for Graph Partitioning and
  Data Clustering}. In \bibinfo{booktitle}{\emph{Proceedings of the 2001 {IEEE}
  International Conference on Data Mining, 29 November - 2 December 2001, San
  Jose, California, {USA}}}. \bibinfo{pages}{107--114}.
\newblock


\bibitem[\protect\citeauthoryear{Ding, Chen, Liu, Ding, Ye, Zhang, Kelly, Guo,
  Su, Harris, et~al\mbox{.}}{Ding et~al\mbox{.}}{2012}]%
        {DCL-BMC-2012}
\bibfield{author}{\bibinfo{person}{Yijun Ding}, \bibinfo{person}{Minjun Chen},
  \bibinfo{person}{Zhichao Liu}, \bibinfo{person}{Don Ding},
  \bibinfo{person}{Yanbin Ye}, \bibinfo{person}{Min Zhang},
  \bibinfo{person}{Reagan Kelly}, \bibinfo{person}{Li Guo},
  \bibinfo{person}{Zhenqiang Su}, \bibinfo{person}{Stephen~C Harris},
  {et~al\mbox{.}}} \bibinfo{year}{2012}\natexlab{}.
\newblock \showarticletitle{atBioNet--an integrated network analysis tool for
  genomics and biomarker discovery}.
\newblock \bibinfo{journal}{\emph{BMC genomics}} \bibinfo{volume}{13},
  \bibinfo{number}{1} (\bibinfo{year}{2012}), \bibinfo{pages}{325}.
\newblock


\bibitem[\protect\citeauthoryear{Gan, Gleich, Veldt, Wirth, and Zhang}{Gan
  et~al\mbox{.}}{2019}]%
        {ggvwz-mfcs-2019}
\bibfield{author}{\bibinfo{person}{Junhao Gan}, \bibinfo{person}{David~F.
  Gleich}, \bibinfo{person}{Nate Veldt}, \bibinfo{person}{Anthony Wirth}, {and}
  \bibinfo{person}{Xin Zhang}.} \bibinfo{year}{2019}\natexlab{}.
\newblock \showarticletitle{Graph Clustering in All Parameter Regimes}.
\newblock \bibinfo{journal}{\emph{CoRR}}  \bibinfo{volume}{abs/1910.06435}
  (\bibinfo{year}{2019}).
\newblock
\urldef\tempurl%
\url{http://arxiv.org/abs/1910.06435}
\showURL{%
\tempurl}


\bibitem[\protect\citeauthoryear{Gan and Tao}{Gan and Tao}{2017a}]%
        {gt-sigmod-2017}
\bibfield{author}{\bibinfo{person}{Junhao Gan} {and} \bibinfo{person}{Yufei
  Tao}.} \bibinfo{year}{2017}\natexlab{a}.
\newblock \showarticletitle{Dynamic Density Based Clustering}. In
  \bibinfo{booktitle}{\emph{Proceedings of ACM Management of Data ({SIGMOD})}}.
  \bibinfo{pages}{1493--1507}.
\newblock


\bibitem[\protect\citeauthoryear{Gan and Tao}{Gan and Tao}{2017b}]%
        {gt-tods-2017}
\bibfield{author}{\bibinfo{person}{Junhao Gan} {and} \bibinfo{person}{Yufei
  Tao}.} \bibinfo{year}{2017}\natexlab{b}.
\newblock \showarticletitle{On the Hardness and Approximation of Euclidean
  {DBSCAN}}.
\newblock \bibinfo{journal}{\emph{{ACM} Trans. Database Syst.}}
  \bibinfo{volume}{42}, \bibinfo{number}{3} (\bibinfo{year}{2017}),
  \bibinfo{pages}{14:1--14:45}.
\newblock


\bibitem[\protect\citeauthoryear{Gan and Tao}{Gan and Tao}{2018}]%
        {gt-sigmod-2018}
\bibfield{author}{\bibinfo{person}{Junhao Gan} {and} \bibinfo{person}{Yufei
  Tao}.} \bibinfo{year}{2018}\natexlab{}.
\newblock \showarticletitle{Fast Euclidean {OPTICS} with Bounded Precision in
  Low Dimensional Space}. In \bibinfo{booktitle}{\emph{Proceedings of ACM
  Management of Data ({SIGMOD})}}. \bibinfo{pages}{1067--1082}.
\newblock


\bibitem[\protect\citeauthoryear{Gleich, Veldt, and Wirth}{Gleich
  et~al\mbox{.}}{2018}]%
        {gvw-isaac-2018}
\bibfield{author}{\bibinfo{person}{David~F. Gleich}, \bibinfo{person}{Nate
  Veldt}, {and} \bibinfo{person}{Anthony Wirth}.}
  \bibinfo{year}{2018}\natexlab{}.
\newblock \showarticletitle{Correlation Clustering Generalized}. In
  \bibinfo{booktitle}{\emph{29th International Symposium on Algorithms and
  Computation, {ISAAC} 2018, December 16-19, 2018, Jiaoxi, Yilan, Taiwan}}.
  \bibinfo{pages}{44:1--44:13}.
\newblock


\bibitem[\protect\citeauthoryear{G{\"{u}}nnemann, F{\"{a}}rber, Boden, and
  Seidl}{G{\"{u}}nnemann et~al\mbox{.}}{2014}]%
        {gfbs-kis-2014}
\bibfield{author}{\bibinfo{person}{Stephan G{\"{u}}nnemann},
  \bibinfo{person}{Ines F{\"{a}}rber}, \bibinfo{person}{Brigitte Boden}, {and}
  \bibinfo{person}{Thomas Seidl}.} \bibinfo{year}{2014}\natexlab{}.
\newblock \showarticletitle{GAMer: a synthesis of subspace clustering and dense
  subgraph mining}.
\newblock \bibinfo{journal}{\emph{Knowl. Inf. Syst.}} \bibinfo{volume}{40},
  \bibinfo{number}{2} (\bibinfo{year}{2014}), \bibinfo{pages}{243--278}.
\newblock


\bibitem[\protect\citeauthoryear{Hoeffding}{Hoeffding}{1994}]%
        {hoeffding-1994}
\bibfield{author}{\bibinfo{person}{Wassily Hoeffding}.}
  \bibinfo{year}{1994}\natexlab{}.
\newblock \showarticletitle{Probability inequalities for sums of bounded random
  variables}.
\newblock In \bibinfo{booktitle}{\emph{The Collected Works of Wassily
  Hoeffding}}. \bibinfo{publisher}{Springer}, \bibinfo{pages}{409--426}.
\newblock


\bibitem[\protect\citeauthoryear{Holm, de~Lichtenberg, and Thorup}{Holm
  et~al\mbox{.}}{2001}]%
        {hlt-acm-2001}
\bibfield{author}{\bibinfo{person}{Jacob Holm}, \bibinfo{person}{Kristian de
  Lichtenberg}, {and} \bibinfo{person}{Mikkel Thorup}.}
  \bibinfo{year}{2001}\natexlab{}.
\newblock \showarticletitle{Poly-logarithmic deterministic fully-dynamic
  algorithms for connectivity, minimum spanning tree, 2-edge, and
  biconnectivity}.
\newblock \bibinfo{journal}{\emph{J. {ACM}}} \bibinfo{volume}{48},
  \bibinfo{number}{4} (\bibinfo{year}{2001}), \bibinfo{pages}{723--760}.
\newblock


\bibitem[\protect\citeauthoryear{Huang, Yi, and Zhang}{Huang
  et~al\mbox{.}}{2019}]%
        {hyz-algorithmica-2019}
\bibfield{author}{\bibinfo{person}{Zengfeng Huang}, \bibinfo{person}{Ke Yi},
  {and} \bibinfo{person}{Qin Zhang}.} \bibinfo{year}{2019}\natexlab{}.
\newblock \showarticletitle{Randomized Algorithms for Tracking Distributed
  Count, Frequencies, and Ranks}.
\newblock \bibinfo{journal}{\emph{Algorithmica}} \bibinfo{volume}{81},
  \bibinfo{number}{6} (\bibinfo{year}{2019}), \bibinfo{pages}{2222--2243}.
\newblock


\bibitem[\protect\citeauthoryear{Hubert and Arabie}{Hubert and Arabie}{1985}]%
        {ha-classification-1985}
\bibfield{author}{\bibinfo{person}{Lawrence Hubert} {and}
  \bibinfo{person}{Phipps Arabie}.} \bibinfo{year}{1985}\natexlab{}.
\newblock \showarticletitle{Comparing partitions}.
\newblock \bibinfo{journal}{\emph{Journal of classification}}
  \bibinfo{volume}{2}, \bibinfo{number}{1} (\bibinfo{year}{1985}),
  \bibinfo{pages}{193--218}.
\newblock


\bibitem[\protect\citeauthoryear{Keralapura, Cormode, and
  Ramamirtham}{Keralapura et~al\mbox{.}}{2006}]%
        {kcr-sigmod-2006}
\bibfield{author}{\bibinfo{person}{Ram Keralapura}, \bibinfo{person}{Graham
  Cormode}, {and} \bibinfo{person}{Jeyashankher Ramamirtham}.}
  \bibinfo{year}{2006}\natexlab{}.
\newblock \showarticletitle{Communication-efficient distributed monitoring of
  thresholded counts}. In \bibinfo{booktitle}{\emph{Proceedings of the {ACM}
  {SIGMOD} International Conference on Management of Data, Chicago, Illinois,
  USA, June 27-29, 2006}}. \bibinfo{pages}{289--300}.
\newblock


\bibitem[\protect\citeauthoryear{Kuncheva, Hadjitodorov, and Todorova}{Kuncheva
  et~al\mbox{.}}{2006}]%
        {kht-icif-2006}
\bibfield{author}{\bibinfo{person}{Ludmila~I Kuncheva},
  \bibinfo{person}{Stefan~Todorov Hadjitodorov}, {and}
  \bibinfo{person}{Ludmila~P Todorova}.} \bibinfo{year}{2006}\natexlab{}.
\newblock \showarticletitle{Experimental comparison of cluster ensemble
  methods}. In \bibinfo{booktitle}{\emph{2006 9th International Conference on
  Information Fusion}}. IEEE, \bibinfo{pages}{1--7}.
\newblock


\bibitem[\protect\citeauthoryear{Kuncheva and Vetrov}{Kuncheva and
  Vetrov}{2006}]%
        {kv-tpam-2006}
\bibfield{author}{\bibinfo{person}{Ludmila~I Kuncheva} {and}
  \bibinfo{person}{Dmitry~P Vetrov}.} \bibinfo{year}{2006}\natexlab{}.
\newblock \showarticletitle{Evaluation of stability of k-means cluster
  ensembles with respect to random initialization}.
\newblock \bibinfo{journal}{\emph{IEEE transactions on pattern analysis and
  machine intelligence}} \bibinfo{volume}{28}, \bibinfo{number}{11}
  (\bibinfo{year}{2006}), \bibinfo{pages}{1798--1808}.
\newblock


\bibitem[\protect\citeauthoryear{Lim, Ryu, Kwon, Jung, and Lee}{Lim
  et~al\mbox{.}}{2014}]%
        {lrkjl-icde-2014}
\bibfield{author}{\bibinfo{person}{Sungsu Lim}, \bibinfo{person}{Seungwoo Ryu},
  \bibinfo{person}{Sejeong Kwon}, \bibinfo{person}{Kyomin Jung}, {and}
  \bibinfo{person}{Jae{-}Gil Lee}.} \bibinfo{year}{2014}\natexlab{}.
\newblock \showarticletitle{LinkSCAN*: Overlapping community detection using
  the link-space transformation}. In \bibinfo{booktitle}{\emph{{IEEE} 30th
  International Conference on Data Engineering, Chicago, {ICDE} 2014, IL, USA,
  March 31 - April 4, 2014}}. \bibinfo{pages}{292--303}.
\newblock


\bibitem[\protect\citeauthoryear{Mete, Tang, Xu, and Yuruk}{Mete
  et~al\mbox{.}}{2008}]%
        {MTX-BMC-2008}
\bibfield{author}{\bibinfo{person}{Mutlu Mete}, \bibinfo{person}{Fusheng Tang},
  \bibinfo{person}{Xiaowei Xu}, {and} \bibinfo{person}{Nurcan Yuruk}.}
  \bibinfo{year}{2008}\natexlab{}.
\newblock \showarticletitle{A structural approach for finding functional
  modules from large biological networks}. In \bibinfo{booktitle}{\emph{Bmc
  Bioinformatics}}, Vol.~\bibinfo{volume}{9}. Springer, \bibinfo{pages}{S19}.
\newblock


\bibitem[\protect\citeauthoryear{Milligan and Cooper}{Milligan and
  Cooper}{1986}]%
        {mc-mbr-1986}
\bibfield{author}{\bibinfo{person}{Glenn~W Milligan} {and}
  \bibinfo{person}{Martha~C Cooper}.} \bibinfo{year}{1986}\natexlab{}.
\newblock \showarticletitle{A study of the comparability of external criteria
  for hierarchical cluster analysis}.
\newblock \bibinfo{journal}{\emph{Multivariate behavioral research}}
  \bibinfo{volume}{21}, \bibinfo{number}{4} (\bibinfo{year}{1986}),
  \bibinfo{pages}{441--458}.
\newblock


\bibitem[\protect\citeauthoryear{Monti, Tamayo, Mesirov, and Golub}{Monti
  et~al\mbox{.}}{2003}]%
        {mtmg-ml-2003}
\bibfield{author}{\bibinfo{person}{Stefano Monti}, \bibinfo{person}{Pablo
  Tamayo}, \bibinfo{person}{Jill Mesirov}, {and} \bibinfo{person}{Todd Golub}.}
  \bibinfo{year}{2003}\natexlab{}.
\newblock \showarticletitle{Consensus clustering: a resampling-based method for
  class discovery and visualization of gene expression microarray data}.
\newblock \bibinfo{journal}{\emph{Machine learning}} \bibinfo{volume}{52},
  \bibinfo{number}{1-2} (\bibinfo{year}{2003}), \bibinfo{pages}{91--118}.
\newblock


\bibitem[\protect\citeauthoryear{Moser, Colak, Rafiey, and Ester}{Moser
  et~al\mbox{.}}{2009}]%
        {mcre-sdm-2009}
\bibfield{author}{\bibinfo{person}{Flavia Moser}, \bibinfo{person}{Recep
  Colak}, \bibinfo{person}{Arash Rafiey}, {and} \bibinfo{person}{Martin
  Ester}.} \bibinfo{year}{2009}\natexlab{}.
\newblock \showarticletitle{Mining Cohesive Patterns from Graphs with Feature
  Vectors}. In \bibinfo{booktitle}{\emph{Proceedings of the {SIAM}
  International Conference on Data Mining, {SDM} 2009, April 30 - May 2, 2009,
  Sparks, Nevada, {USA}}}. \bibinfo{pages}{593--604}.
\newblock


\bibitem[\protect\citeauthoryear{Palla, Der{\'e}nyi, Farkas, and Vicsek}{Palla
  et~al\mbox{.}}{2005}]%
        {pdfv-nature-2005}
\bibfield{author}{\bibinfo{person}{Gergely Palla}, \bibinfo{person}{Imre
  Der{\'e}nyi}, \bibinfo{person}{Ill{\'e}s Farkas}, {and}
  \bibinfo{person}{Tam{\'a}s Vicsek}.} \bibinfo{year}{2005}\natexlab{}.
\newblock \showarticletitle{Uncovering the overlapping community structure of
  complex networks in nature and society}.
\newblock \bibinfo{journal}{\emph{nature}} \bibinfo{volume}{435},
  \bibinfo{number}{7043} (\bibinfo{year}{2005}), \bibinfo{pages}{814}.
\newblock


\bibitem[\protect\citeauthoryear{Papadopoulos, Kompatsiaris, Vakali, and
  Spyridonos}{Papadopoulos et~al\mbox{.}}{2012}]%
        {PKV-DMKD-2012}
\bibfield{author}{\bibinfo{person}{Symeon Papadopoulos},
  \bibinfo{person}{Yiannis Kompatsiaris}, \bibinfo{person}{Athena Vakali},
  {and} \bibinfo{person}{Ploutarchos Spyridonos}.}
  \bibinfo{year}{2012}\natexlab{}.
\newblock \showarticletitle{Community detection in social media}.
\newblock \bibinfo{journal}{\emph{Data Mining and Knowledge Discovery}}
  \bibinfo{volume}{24}, \bibinfo{number}{3} (\bibinfo{year}{2012}),
  \bibinfo{pages}{515--554}.
\newblock


\bibitem[\protect\citeauthoryear{Papadopoulos, Zigkolis, Kompatsiaris, and
  Vakali}{Papadopoulos et~al\mbox{.}}{2010}]%
        {PZKV-IEEE-2010}
\bibfield{author}{\bibinfo{person}{S Papadopoulos}, \bibinfo{person}{C
  Zigkolis}, \bibinfo{person}{Y Kompatsiaris}, {and} \bibinfo{person}{A
  Vakali}.} \bibinfo{year}{2010}\natexlab{}.
\newblock \showarticletitle{Cluster-based landmark and event detection on
  tagged photo collections.}
\newblock \bibinfo{journal}{\emph{IEEE}}  \bibinfo{volume}{99}
  (\bibinfo{year}{2010}), \bibinfo{pages}{1--1}.
\newblock


\bibitem[\protect\citeauthoryear{Sankar, Ravindran, and Shivashankar}{Sankar
  et~al\mbox{.}}{2015}]%
        {srs-ijcai-2015}
\bibfield{author}{\bibinfo{person}{Vishnu Sankar}, \bibinfo{person}{Balaraman
  Ravindran}, {and} \bibinfo{person}{S. Shivashankar}.}
  \bibinfo{year}{2015}\natexlab{}.
\newblock \showarticletitle{{CEIL:} {A} Scalable, Resolution Limit Free
  Approach for Detecting Communities in Large Networks}. In
  \bibinfo{booktitle}{\emph{Proceedings of the Twenty-Fourth International
  Joint Conference on Artificial Intelligence, {IJCAI} 2015, Buenos Aires,
  Argentina, July 25-31, 2015}}. \bibinfo{pages}{2097--2103}.
\newblock


\bibitem[\protect\citeauthoryear{Schaeffer}{Schaeffer}{2007}]%
        {s-csr-2007}
\bibfield{author}{\bibinfo{person}{Satu~Elisa Schaeffer}.}
  \bibinfo{year}{2007}\natexlab{}.
\newblock \showarticletitle{Graph clustering}.
\newblock \bibinfo{journal}{\emph{Computer science review}}
  \bibinfo{volume}{1}, \bibinfo{number}{1} (\bibinfo{year}{2007}),
  \bibinfo{pages}{27--64}.
\newblock


\bibitem[\protect\citeauthoryear{Shiokawa, Fujiwara, and Onizuka}{Shiokawa
  et~al\mbox{.}}{2015}]%
        {sfo-pvldb-2015}
\bibfield{author}{\bibinfo{person}{Hiroaki Shiokawa}, \bibinfo{person}{Yasuhiro
  Fujiwara}, {and} \bibinfo{person}{Makoto Onizuka}.}
  \bibinfo{year}{2015}\natexlab{}.
\newblock \showarticletitle{{SCAN++:} Efficient Algorithm for Finding Clusters,
  Hubs and Outliers on Large-scale Graphs}.
\newblock \bibinfo{journal}{\emph{{PVLDB}}} \bibinfo{volume}{8},
  \bibinfo{number}{11} (\bibinfo{year}{2015}), \bibinfo{pages}{1178--1189}.
\newblock


\bibitem[\protect\citeauthoryear{Thorup}{Thorup}{2000}]%
        {t-stoc-2000}
\bibfield{author}{\bibinfo{person}{Mikkel Thorup}.}
  \bibinfo{year}{2000}\natexlab{}.
\newblock \showarticletitle{Near-optimal fully-dynamic graph connectivity}. In
  \bibinfo{booktitle}{\emph{Proceedings of the Thirty-Second Annual {ACM}
  Symposium on Theory of Computing, May 21-23, 2000, Portland, OR, {USA}}}.
  \bibinfo{pages}{343--350}.
\newblock


\bibitem[\protect\citeauthoryear{Vinh, Epps, and Bailey}{Vinh
  et~al\mbox{.}}{2010}]%
        {veb-jmlr-2010}
\bibfield{author}{\bibinfo{person}{Nguyen~Xuan Vinh}, \bibinfo{person}{Julien
  Epps}, {and} \bibinfo{person}{James Bailey}.}
  \bibinfo{year}{2010}\natexlab{}.
\newblock \showarticletitle{Information theoretic measures for clusterings
  comparison: Variants, properties, normalization and correction for chance}.
\newblock \bibinfo{journal}{\emph{The Journal of Machine Learning Research}}
  \bibinfo{volume}{11} (\bibinfo{year}{2010}), \bibinfo{pages}{2837--2854}.
\newblock


\bibitem[\protect\citeauthoryear{Wang, Xiao, Shao, and Wang}{Wang
  et~al\mbox{.}}{2014}]%
        {wxsw-icde-2014}
\bibfield{author}{\bibinfo{person}{Lu Wang}, \bibinfo{person}{Yanghua Xiao},
  \bibinfo{person}{Bin Shao}, {and} \bibinfo{person}{Haixun Wang}.}
  \bibinfo{year}{2014}\natexlab{}.
\newblock \showarticletitle{How to partition a billion-node graph}. In
  \bibinfo{booktitle}{\emph{{IEEE} 30th International Conference on Data
  Engineering, Chicago, {ICDE} 2014, IL, USA, March 31 - April 4, 2014}}.
  \bibinfo{pages}{568--579}.
\newblock


\bibitem[\protect\citeauthoryear{Wen, Qin, Zhang, Chang, and Lin}{Wen
  et~al\mbox{.}}{2017}]%
        {wqzcl-pvldb-2017}
\bibfield{author}{\bibinfo{person}{Dong Wen}, \bibinfo{person}{Lu Qin},
  \bibinfo{person}{Ying Zhang}, \bibinfo{person}{Lijun Chang}, {and}
  \bibinfo{person}{Xuemin Lin}.} \bibinfo{year}{2017}\natexlab{}.
\newblock \showarticletitle{Efficient Structural Graph Clustering: An
  Index-Based Approach}.
\newblock \bibinfo{journal}{\emph{{PVLDB}}} \bibinfo{volume}{11},
  \bibinfo{number}{3} (\bibinfo{year}{2017}), \bibinfo{pages}{243--255}.
\newblock


\bibitem[\protect\citeauthoryear{Wen, Qin, Zhang, Chang, and Lin}{Wen
  et~al\mbox{.}}{2019}]%
        {wqzcl-vldbj-2019}
\bibfield{author}{\bibinfo{person}{Dong Wen}, \bibinfo{person}{Lu Qin},
  \bibinfo{person}{Ying Zhang}, \bibinfo{person}{Lijun Chang}, {and}
  \bibinfo{person}{Xuemin Lin}.} \bibinfo{year}{2019}\natexlab{}.
\newblock \showarticletitle{Efficient structural graph clustering: an
  index-based approach}.
\newblock \bibinfo{journal}{\emph{{VLDB} J.}} \bibinfo{volume}{28},
  \bibinfo{number}{3} (\bibinfo{year}{2019}), \bibinfo{pages}{377--399}.
\newblock


\bibitem[\protect\citeauthoryear{Xu, Yuruk, Feng, and Schweiger}{Xu
  et~al\mbox{.}}{2007}]%
        {xyfs-kdd-2007}
\bibfield{author}{\bibinfo{person}{Xiaowei Xu}, \bibinfo{person}{Nurcan Yuruk},
  \bibinfo{person}{Zhidan Feng}, {and} \bibinfo{person}{Thomas A.~J.
  Schweiger}.} \bibinfo{year}{2007}\natexlab{}.
\newblock \showarticletitle{{SCAN:} a structural clustering algorithm for
  networks}. In \bibinfo{booktitle}{\emph{Proceedings of the 13th {ACM}
  {SIGKDD} International Conference on Knowledge Discovery and Data Mining, San
  Jose, California, USA, August 12-15, 2007}}. \bibinfo{pages}{824--833}.
\newblock


\bibitem[\protect\citeauthoryear{Yuruk, Xu, and Schweiger}{Yuruk
  et~al\mbox{.}}{2008a}]%
        {yxs-hicss-2008}
\bibfield{author}{\bibinfo{person}{Nurcan Yuruk}, \bibinfo{person}{Xiaowei Xu},
  {and} \bibinfo{person}{Thomas~AJ Schweiger}.}
  \bibinfo{year}{2008}\natexlab{a}.
\newblock \showarticletitle{On Structural Analysis of Large Networks}. In
  \bibinfo{booktitle}{\emph{Proceedings of the 41st Annual Hawaii International
  Conference on System Sciences (HICSS 2008)}}. IEEE,
  \bibinfo{pages}{143--143}.
\newblock


\bibitem[\protect\citeauthoryear{Yuruk, Xu, and Schweiger}{Yuruk
  et~al\mbox{.}}{2008b}]%
        {yxs-icss-2008}
\bibfield{author}{\bibinfo{person}{Nurcan Yuruk}, \bibinfo{person}{Xiaowei Xu},
  {and} \bibinfo{person}{Thomas A.~J. Schweiger}.}
  \bibinfo{year}{2008}\natexlab{b}.
\newblock \showarticletitle{On Structural Analysis of Large Networks}. In
  \bibinfo{booktitle}{\emph{Proceedings of the 41st Annual Hawaii International
  Conference on System Sciences (HICSS 2008)}}. \bibinfo{pages}{143}.
\newblock


\end{thebibliography}
\end{document}